# Measuring Investor Learning in Private Markets: A Sequential LLM–Bayesian Analysis of Expert Network Calls*

Yidong Chai[a], Yanguang Liu[b], Xuan Tian[c], Jiaheng Xie[d], and Yonghang Zhou[e]

## Abstract

We study investor learning and information acquisition in private markets using a large dataset of expert network calls. We develop a sequential Large Language Model (LLM)–Bayesian framework that treats expert interactions as sequential signals and recovers time-varying beliefs about firm success and associated uncertainty from unstructured conversations, providing a measurement system for how qualitative information is aggregated into investment expectations. We show that expert network calls contain decision-relevant information: a single call increases subsequent investment probability by 6.9 to 9.0 percentage points, while positive sentiment raises deal likelihood by 3.9 to 4.1 percentage points. Informativeness varies across topics and environments: discussions of technology adoption and customer acquisition increase deal probability by up to 14.7 percentage points, particularly in high-uncertainty settings. Information is asymmetric across horizons, with positive signals predicting short-term investment decisions and negative signals more informative about long-run firm performance. Consistent with a belief-based mechanism, investment decisions respond to inferred beliefs rather than raw signals. A one standard deviation increase in success belief raises deal probability by approximately 11 percentage points, while reductions in uncertainty further increase investment likelihood. Our framework improves capital allocation, increasing portfolio returns by 15.26% and F1 by 6.69%, with gains concentrated in the upper tail. Attention and ablation analyses show that conversational cues are particularly informative for technologically complex startups, young firms, diverse founding teams, and firms with low public visibility, where information frictions are severe.

---

* Tian acknowledges financial support from the National Natural Science Foundation of China (Grant No.72425002). Chai, Liu, Xie, and Zhou did not receive any financial support. We remain responsible for all remaining errors and omissions.
a City University of Hong Kong, yidong.chai@cityu.edu.hk.
b Martin Tuchman School of Management, New Jersey Institute of Technology, yl224@njit.edu.
c Guanghua School of Management, Peking University, China, tianx@gsm.pku.edu.cn.
d Alfred Lerner College of Business and Economics, University of Delaware, jxie@udel.edu.
e Hefei University of Technology, 2014213676@mail.hfut.edu.cn.

# Measuring Investor Learning in Private Markets: A Sequential LLM-Bayesian Analysis of Expert Network Calls*

**Abstract**

We study investor learning and information acquisition in private markets using a large dataset of expert network calls. We develop a sequential Large Language Model (LLM)–Bayesian framework that treats expert interactions as sequential signals and recovers time-varying beliefs about firm success and associated uncertainty from unstructured conversations, providing a measurement system for how qualitative information is aggregated into investment expectations. We show that expert network calls contain decision-relevant information: a single call increases subsequent investment probability by 6.9 to 9.0 percentage points, while positive sentiment raises deal likelihood by 3.9 to 4.1 percentage points. Informativeness varies across topics and environments: discussions of technology adoption and customer acquisition increase deal probability by up to 14.7 percentage points, particularly in high-uncertainty settings. Information is asymmetric across horizons, with positive signals predicting short-term investment decisions and negative signals more informative about long-run firm performance. Consistent with a belief-based mechanism, investment decisions respond to inferred beliefs rather than raw signals. A one standard deviation increase in success belief raises deal probability by approximately 11 percentage points, while reductions in uncertainty further increase investment likelihood. Our framework improves capital allocation, increasing portfolio returns by 15.26% and F1 by 6.69%, with gains concentrated in the upper tail. Attention and ablation analyses show that conversational cues are particularly informative for technologically complex startups, young firms, diverse founding teams, and firms with low public visibility, where information frictions are severe.

---

*

## 1. Introduction

How do investors form beliefs about firm quality when reliable financial information is unavailable? This question is central to entrepreneurial finance, where a small fraction of startups generate outsized returns while most fail, producing highly skewed payoffs and amplifying the costs of misallocation (Kerr et al., 2014). Unlike public firms, early-stage ventures lack audited financials, stable business models, and comparable benchmarks, forcing investors to operate under severe information asymmetry (Gompers & Lerner, 2001; Ewens & Marx, 2018). As a result, investment decisions rely heavily on qualitative, forward-looking signals and on updating beliefs as new information arrives. Yet how investors acquire, interpret, and aggregate such signals, and translate them into capital allocation, remains largely a "black box" (Kaplan & Strömberg, 2004; Bernstein et al., 2017; Liberti and Petersen, 2019).

This paper opens this "black box" by proposing a framework that measures investor belief formation from qualitative information and links it directly to capital allocation. We study investor learning in private markets using sequential expert network calls signals, investor-initiated consultations with domain specialists, and develop a measurement system that transforms unstructured conversations into time-varying beliefs about firm success and associated uncertainty. Our analysis is based on a large dataset of expert network calls transcripts covering startups over 2017–2024. While prior work models information acquisition and belief updating theoretically, we recover these dynamics directly from data to examine how qualitative signals translate into investment decisions. We address four related questions: what information investors seek from experts; how these signals are incorporated into beliefs; whether these beliefs drive investment decisions and performance outcomes; and when information acquisition is most valuable under uncertainty and limited disclosure. We study this process in the setting of expert network calls, which provide a unique window into investor-driven information acquisition.

Expert network calls provide a natural setting to study information acquisition under uncertainty. Through interactions with customers, competitors, consultants, and former executives, investors obtain experiential insights on product adoption, competitive positioning, and execution risk, dimensions largely absent from structured disclosures (Solomon & Soltes, 2015; Bernstein et al., 2017; Liberti & Petersen, 2019; Drake et al., 2012). These interactions generate sequential, investor-driven signals, allowing us to study how beliefs evolve as information accumulates. We show that expert network calls reflect a dynamic due diligence process: call activity spikes prior to investment decisions and is concentrated on younger, more uncertain firms. Conversations cover core dimensions of firm evaluation, including technology, demand, competition, and execution, and their content and tone vary systematically across expert types, producing heterogeneous signals. Together, these patterns define a setting in which signals arrive sequentially, differ in credibility, and must be aggregated, motivating a framework that models belief updating under uncertainty.

Despite their richness, expert network calls data pose key challenges: they are unstructured and sparse, hierarchical and sequential, and involve latent beliefs that evolve over time while only final outcomes are observed. To address these challenges, we develop a sequential LLM-Bayesian framework that integrates text-based signal extraction with probabilistic belief updating. LLMs extract structured representations from unstructured dialogue, while a Bayesian network with latent variables captures dependencies across interactions and models the evolution of beliefs and uncertainty. Treating expert network calls as sequential information signals, the framework produces time-varying estimates of firm success and associated uncertainty, serving as a measurement system for investor learning (see Section 3). We link these belief measures to investment outcomes using return-based metrics such as Return on Investment (ROI) and Multiple on Invested Capital (MOIC), aligning the framework with the asymmetric payoff structure of venture capital.

Empirically, modeling expert network calls as sequential information signals yields substantial improvements in both predictive accuracy and investment performance. Incorporating expert network calls signals enhances performance relative to fundamentals alone, indicating that these interactions contain economically meaningful soft information, such as product traction, competitive positioning, and managerial quality, not captured by structured data. However, models that treat these signals as static features cannot fully exploit this information, as they fail to capture how signals arrive and accumulate over time. Our framework instead models expert network calls as sequential signals and maps them into time-varying beliefs, enabling this information to be incorporated more effectively into investment decisions. It outperforms standard machine learning models, transformer-based architectures, and recent text-based approaches, suggesting that the gains do not arise from richer representations alone, but from capturing how information is dynamically incorporated into beliefs. Relative to the best-performing baselines, the model improves F1-score by up to 6.69% and increases portfolio-level ROI by 15.26%, with corresponding gains in MOIC. Taken together, these results suggest that combining informative qualitative signals with a sequential belief-updating structure enhances both measurement and economically meaningful investment outcomes in private markets.

We provide direct evidence that the model's performance is driven by its ability to capture the structure and dynamics of information acquisition. Ablation tests show that ignoring dependencies across expert signals, within conversations or across calls, leads to the largest deterioration in both predictive and economic performance, while removing belief updating also significantly reduces performance. These results indicate that the primary value of our framework lies in modeling how information is accumulated and incorporated into beliefs over time. Consistent with this mechanism, performance improves as additional expert network calls are incorporated, with accuracy increasing from 0.769 to 0.777 and ROI rising by over 32.5%, reflecting progressive refinement of beliefs as new information arrives. We further show that uncertainty declines leading up to investment events and stabilizes afterward, indicating that investors resolve uncertainty during due diligence rather than post-investment.

Beyond predictive performance, our framework delivers economically meaningful improvements in investment decisions. The model substantially enhances the ranking of investment opportunities, particularly in the upper tail where returns are concentrated. As investors focus on increasingly selective subsets of top-ranked firms, realized performance improves sharply, and the model outperforms benchmarks across all thresholds. This evidence indicates that the framework improves capital allocation by better identifying high-quality opportunities, rather than merely increasing predictive accuracy. These gains translate into tangible portfolio improvements: reallocating capital from benchmark-selected firms to those identified by our model leads to monotonic increases in both ROI and MOIC, indicating systematic identification of superior investments. The improvements are robust across a wide range of payoff and cost assumptions, suggesting that the economic value reflects a general enhancement in capital allocation efficiency rather than specific return realizations. Overall, the results show that improved measurement of beliefs from qualitative signals translates into superior investment outcomes.

To uncover the mechanism, we examine how the model weights qualitative signals in belief updating using attention scores. The model places the greatest weight on operationally informed sources such as customers and industry consultants, and on core uncertainty-related topics including regulation, customer acquisition, and technology. It also assigns asymmetric weights to signals: positive signals receive two to six times the weight of negative signals in technology-related domains, and attention to customer acquisition rises from 0.023 (negative) to 0.102 (positive), highlighting the importance of market traction. Importantly, these effects are state-dependent. Attention declines sharply, often by an order of magnitude, as firms mature, indicating that qualitative signals are most valuable when firms are young, less funded, and more opaque. These patterns provide a transparent view of the belief-updating mechanism, in which investors selectively weight signals based on source, content, and context.

Finally, we provide direct evidence linking information acquisition to belief formation and capital allocation. Using model-implied measures of investor beliefs and uncertainty, we show that expert network calls signals generate economically meaningful belief updates that vary systematically with the information environment: learning is stronger when prior uncertainty is high, in more opaque settings, and for signals from operational experts and core fundamentals. These belief dynamics translate into investment decisions. In reduced-form estimates, the occurrence of an expert network call increases deal probability by 6.9 to 9.0 percentage points, while positive sentiment raises it by 3.9 to 4.1 percentage points, with larger effects for discussions of technology and customer acquisition (up to 14.7 percentage points). In contrast, negative signals are more informative for long-run outcomes: discussions of regulatory and security challenges significantly reduce future success, while positive signals have limited long-term predictive power, consistent with investors responding to short-term validation but underweighting downside risk. Importantly, these effects operate through belief formation. Once model-implied beliefs and uncertainty are included, raw text signals lose explanatory power: a one standard deviation increase in success belief raises deal probability by approximately 11 percentage points, while higher uncertainty reduces it. Moreover, reductions in uncertainty, that is, learning, further increase investment likelihood, particularly when prior uncertainty is high. Together, these results show that expert network calls information affects capital allocation by shaping beliefs and resolving uncertainty, highlighting the economic value of learning in private markets.

We further show that the value of learning is state-dependent and shaped by the information environment. The model delivers the largest gains in settings with high uncertainty and fragmented information, such as technologically complex industries, early-stage firms, and low-visibility firms with limited disclosure. In these environments, where traditional data are less informative, improvements in belief measurement translate into the largest gains in both predictive accuracy and investment performance. These results highlight that the economic value of expert network calls information is greatest precisely when information frictions are most severe.

This paper contributes to the literature on information frictions in venture capital. A central role of venture capitalists is to overcome severe information asymmetries (Gompers and Lerner, 2001; Kaplan and Strömberg, 2004), yet observable predictors of startup success are often limited, especially at early stages (Ewens and Marx, 2018; Gompers et al., 2020). While recent work highlights the importance of qualitative information in mitigating these frictions (Bernstein et al., 2017; Howell, 2017; Guzman and Li, 2023), how such information is acquired during due diligence remains largely unobserved. We address this gap by introducing expert network calls as a novel data source and providing direct evidence on how investors acquire information from heterogeneous sources during the due diligence process.

We further contribute to the literature on soft information, information acquisition, and learning by providing an empirical framework that traces how qualitative signals are incorporated into investment decisions. Prior work shows that investor interactions and information networks shape outcomes and uses new data to uncover otherwise unobservable aspects of the investment process (Hochberg, Ljungqvist, and Lu, 2007; Solomon and Soltes, 2015; Liberti and Petersen, 2019; Drake et al., 2012; Cao et al., 2023), but largely relies on reduced-form relationships between signals and outcomes. In contrast, we use the content and structure of expert network calls transcripts to track how signals from heterogeneous sources are aggregated into beliefs and uncertainty over time. Linking to the theoretical literature on information acquisition and Bayesian learning (Sims, 2003; Veldkamp, 2006; Van Nieuwerburgh and Veldkamp, 2010; Pastor and Veronesi, 2009), we construct time-varying measures of beliefs and show that investment decisions respond to these inferred beliefs rather than raw signals, establishing a belief-based mechanism linking information to capital allocation.

Finally, we develop a measurement framework that integrates LLM-based signal extraction with sequential Bayesian updating to recover belief dynamics from unstructured text. Qualitative information is high-dimensional, noisy, and arrives sequentially, making it difficult to map textual signals into economically interpretable belief updates. Prior work either relies on handcrafted textual features (e.g., Blei, Ng, and Jordan, 2003; Mayew and Venkatachalam, 2012; Hobson et al. 2012; Larcker and Zakolyukina, 2012) or uses deep learning models that improve representation but do not capture belief dynamics or uncertainty (Vaswani et al., 2017; Beltagy et al., 2020; Zaheer et al., 2020; Yang et al., 2016; Wu et al., 2021). Bayesian approaches provide a principled framework for sequential updating (Carvalho and West, 2007; Pastor and Veronesi, 2009), but are difficult to apply to high-dimensional text. By combining LLM-based signal extraction with a sequential updating framework, we transform unstructured conversations into interpretable, time-varying measures of beliefs and uncertainty. More broadly, our approach enables the measurement of dynamic belief formation and uncertainty resolution from qualitative data, providing a general framework for studying learning from unstructured information.

## 2. Institutional Setting and Stylized Facts of Expert Network Calls

### 2.1 Expert Network Calls and Sample Construction

The expert network industry emerged in the early 2000s following regulatory reforms such as Regulation Fair Disclosure and the Global Analyst Research Settlement, which limited selective disclosure and led investors to seek alternative information sources. The industry has since grown to more than 100 firms, generating approximately $1.9 billion in revenue by 2021.

Expert network calls are client-initiated consultations in which institutional investors engage domain specialists to obtain perspectives on companies, industries, or technologies. These interactions typically take the form of structured question-and-answer exchanges, allowing investors to adapt inquiries in real time. Through this process, investors extract soft information on product positioning, competitive dynamics, and execution risk, dimensions rarely observable in traditional disclosures (Liberti and Petersen 2019).

A key feature of expert network calls is the diversity and independence of information sources. Experts include customers, competitors, consultants, former executives, and business partners of focal firms, each offering distinct perspectives on firm fundamentals. By synthesizing these heterogeneous viewpoints, investors triangulate firm quality and mitigate biases in managerial narratives (Drake et al. 2012). Because experts are external to the focal firm, they face fewer incentives to distort information, further enhancing credibility (Gompers & Lerner, 2001; Kaplan & Strömberg, 2004).

Expert network calls, therefore, provide a natural setting to study information acquisition under uncertainty. Through compliance-monitored, multi-turn conversations, investors elicit experiential insights largely absent from structured disclosures (Solomon & Soltes, 2015; Bernstein et al., 2017; Liberti & Petersen, 2019; Drake et al., 2012). Their use reflects endogenous information acquisition: investors intensify information gathering for firms that are younger, more uncertain, and economically important. As a result, expert network calls generate sequential, investor-driven signals, allowing us to study how beliefs evolve as information accumulates.

Our analysis uses a dataset of expert network calls transcripts matched to startup firms over time. To ensure compliance with securities regulations, expert networks implement safeguards including call recordings, transcript archiving, and participant pre-vetting, consistent with Section 204A of the Investment Advisers Act of 1940. These protocols enable systematic collection of conversational data while limiting the transmission of material non-public information.

Table 1 reports summary statistics on call characteristics, expert composition, and firm attributes. The sample spans 2017 to 2024, the earliest period with comprehensive transcript data. Conversations are information-rich: the average transcript contains approximately 5,880 words and 74 question–answer exchanges. The distribution of expert types highlights the nature of information sought, with customers (37.86%), consultants (31.90%), and former executives (19.10%) comprising the majority, reflecting a focus on market traction, technical implementation, and organizational capability.

To characterize covered firms, we combine transcript data with startup fundamentals from Crunchbase and VentureXpert. We restrict the sample to firms with more than 50 employees to focus on startups at a commercially meaningful stage (Sorenson 2021). Panel C of Table 1 shows that these firms have, on average, completed five financing rounds, are backed by approximately 14 institutional investors, and operate multiple products with substantial technological investment.

Finally, we examine firm characteristics associated with demand for expert network calls. Panel D of Table 1 compares firms that receive calls with those that do not. Consistent with uncertainty-driven information demand, firms receiving expert network calls are younger, indicating that investors seek additional signals when fundamentals are less established. At the same time, these firms have raised more external capital and attract greater institutional attention, as reflected in higher prior funding. Together, these patterns indicate that expert network calls are concentrated on firms that are both uncertain and economically important, consistent with endogenous information acquisition during due diligence.

### 2.2 Information Acquisition Around Investment Decisions

Expert network calls are actively initiated by investors and therefore reflect endogenous information acquisition decisions. To characterize how call activity evolves around investment events, we estimate the following event-study specification:

$$Call_{i,t} = \alpha + \sum_{s=-7}^{7} \beta_s \mathbf{1}(Deal)_{i,t+s} + \mu_i + \tau_t + \epsilon_{it}, \quad (1)$$

where $Call_{i,t}$ measures call activity for firm $i$ in quarter $t$, and $\mathbf{1}(Deal)_{i,t+s}$ indicates whether firm $i$ completes a deal in quarter t+s. The specification includes firm and quarter fixed effects.

Figure 1 plots the estimated coefficients $\beta_s$ and reveals a clear pattern: call frequency rises sharply prior to investment decisions and declines thereafter, indicating that expert consultations are concentrated around the decision process rather than routine monitoring.

These patterns suggest that investors dynamically gather signals rather than rely on a fixed information set. Expert network calls appear to serve as a mechanism for resolving uncertainty and refining beliefs about firm quality. More broadly, the evidence highlights two key features of the information environment: information arrives sequentially, and uncertainty evolves as new signals are acquired, motivating continued information gathering prior to capital allocation decisions.

Importantly, call activity is endogenously determined by investors. We interpret this as a central feature of the information acquisition process: investors allocate attention to firms where uncertainty is high, and the value of information is greatest. Accordingly, we view these patterns as reflecting equilibrium information acquisition and learning, rather than causal effects of individual signals.

### 2.3 The Structure and Content of Expert Network Calls

Expert network calls are rich, multi-turn dialogues that allow investors to probe specific dimensions of firm performance. Unlike static disclosures, these interactions enable follow-up questioning and scenario testing, generating detailed discussions of technology, customer demand, competitive dynamics, and unit economics (Gompers and Lerner 2001; Kaplan and Strömberg 2004).

To characterize their informational content, we extract topics and sentiment using an LLM-based approach that transforms unstructured dialogue into structured representations of topic relevance and tone. The extraction proceeds in three steps, topic generation, refinement, and assignment, implemented using GPT-4 models (Appendix A.I). The model identifies investment-relevant themes, consolidates similar topics, and assigns each conversation to topics and sentiment polarities (−2 to +2; Table A.3).

Panel A and Panel B of Table 2 show that conversations concentrate on core drivers of startup performance, including product development, market demand, and operational execution. The most frequent topics, Regulatory and Security Challenges (16.88%), Customer Acquisition and Pricing (15.46%), and AI and Machine Learning Platforms (14.91%), highlight the importance of risk, monetization, and technological capability. Investors also emphasize Data Management and Preparation (11.89%) and Industry-Specific AI Applications (12.01%), reflecting the role of data infrastructure and domain-specific deployment. Product-related topics such as Product Development and Innovation (8.87%) and Market Expansion and Strategy (7.68%) indicate a focus on scaling, while Digital Advertising Strategies (1.53%) and Corporate Culture and Management (1.75%) receive limited attention. Overall, expert network calls provide a forward-looking, multi-dimensional view of firm prospects.

Figure 2 shows that topic emphasis varies systematically across expert types, indicating that investors match questions to domain-specific expertise. Competitive and market expansion topics are concentrated among competitors and consultants, while Customer Acquisition and Pricing is primarily discussed with customers. Product and technology topics are directed to consultants and former executives, and Business Models and Revenue Streams to former executives. Regulatory and Security Challenges span multiple expert types but are most prominent among consultants and customers. These patterns indicate that investors combine complementary perspectives. In our framework, this heterogeneity implies that signals differ in informational content and reliability and are weighted differently in belief formation.[1]

Panel B of Table 2 reports sentiment at both transcript and topic levels. The average sentiment is positive (mean 0.55), with 63.36% positive and 8.22% negative observations, but dispersion is substantial (standard deviation 0.66), indicating both favorable and critical assessments within calls. Topic-level sentiment shows further heterogeneity: growth and technology discussions are most positive (e.g., Digital Advertising Strategies 0.74; AI and Machine Learning Platforms 0.70), while risk and organizational topics are more neutral or negative (e.g., Regulatory and Security Challenges 0.33; Corporate Culture and Management 0.14). Topics such as Competitive Landscape (0.44) and Customer Acquisition and Pricing (0.55) lie in between.

Figure 3 further illustrates heterogeneity across topics and expert types. Product and technology topics are consistently positive, while organizational and risk-related topics are more dispersed. For example, Corporate Culture and Management ranges from 0.61 (customers) to 0.09 (former executives), and some risk-related topics exhibit strongly negative sentiment for specific groups. Sentiment also varies by expert type: customers provide relatively positive and stable evaluations, consultants and former executives are more moderate, and competitors exhibit greater polarity. These patterns indicate systematic variation in informational content and credibility across sources, which informs belief updating in our framework.

Taken together, expert network calls signals are multi-dimensional, heterogeneous across sources, and arrive sequentially over time, creating a non-trivial aggregation problem. Static summaries or reduced-form approaches cannot capture these dynamics. To address this, we model investor learning as a sequential updating process and develop an LLM-Bayesian framework that maps expert signals into time-varying beliefs and uncertainty, allowing us to track how expectations evolve as new information arrives.

[1] Figure A.1 documents industry-level specialization in expert discourse, indicating that discussions reflect economically meaningful sectoral differences.

## 3. Sequential Learning from Expert Signals: An LLM-Bayesian Framework

### 3.1 Overview of Our Framework

Investors evaluating early-stage firms operate under substantial uncertainty and limited hard information. Unlike public firms, where standardized disclosures provide stable signals, startup evaluation relies on fragmented, qualitative, and forward-looking information (Gompers and Lerner, 2001; Kaplan and Strömberg, 2004; Ewens and Farre-Mensa, 2022). A key feature of this setting is that information arrives sequentially, as investors actively acquire signals over time rather than observing a fixed information set.

Expert network calls provide a natural setting to study this process. Each call generates a signal about firm fundamentals from domain specialists with heterogeneous perspectives. As investors consult multiple experts, signals differ in content, credibility, and direction, and may reinforce or contradict one another. The central economic problem is therefore the aggregation of dispersed and noisy information under uncertainty (Veldkamp, 2006; Van Nieuwerburgh and Veldkamp, 2010; Sims, 2003).

We model this process as sequential belief updating. In our framework, investor learning refers to the incorporation of qualitative expert signals into beliefs about firm prospects, which we measure through changes in model-implied beliefs and uncertainty. At time $t$, investors hold a belief about a firm's success probability (denoted as $r^t$), which is updated as new signals arrive. Updating depends on both signal content and source, reflecting differences in informational relevance and credibility. When uncertainty is high, the marginal value of information is greater, implying state-dependent learning. A Bayesian framework provides a natural approach: it accommodates sequential arrival, allows precision-weighted signals, and yields both posterior beliefs and uncertainty. Table A.1 summarizes the notation used throughout the paper.

Motivated by this setting, we develop an integrated LLM-Bayesian framework. The LLM extracts structured signals from unstructured expert network calls conversations, while the Bayesian component aggregates the extracted structured signals into beliefs over time. This framework allows us to recover time-varying beliefs and uncertainty from qualitative interactions.

In our framework, we formalize the learning environment at the level of expert interactions. For each firm, investors observe a sequence of expert network calls, each consisting of multiple question–answer pairs. Let $C^t = \{c^1, c^2, \dots, c^t\}$ denote the sequence of calls up to time $t$, where each call (e.g., $c^t$) contains $K^t$ question–answer pairs. Since firm characteristics $\boldsymbol{e}$ (e.g., age) also influence investor belief formation, we include them in our model. Whether the firm succeeds (denoted as $y \in [0,1]$) is unobserved until the outcome event at time $t^{\text{Event}}$, defined as IPO, M&A, or subsequent financing. We therefore dynamically estimate investors' beliefs in the firm's success using data observed at each time $t < t^{\text{Event}}$. Formally, $r^t = f_{\boldsymbol{\theta}}([C^t, \boldsymbol{e}]), \forall t$, and the estimated belief should be consistent with the observation. We define firm success using standard measures in the venture capital literature (Nanda and Rhodes-Kropf, 2013; Hegde and Tumlinson, 2014; Maarouf et al., 2025): a firm is successful if it achieves an IPO, acquisition, or external financing.

Constructing this framework presents three challenges. First, expert network call data are lengthy and unstructured, with sparse information, requiring scalable signal extraction. Second, the data are hierarchical and sequential, with interdependent signals across interactions and calls. Third, only the final outcome is observed, whereas investors' beliefs are latent and formed and updated dynamically over time. To address the first challenge, we extract economically relevant signals from text, summarizing key information from expert network calls conversations. For the second and third challenges, which are also where our novelty lies, we design a novel Bayesian network with latent distributed variables that dynamically model investors' success beliefs and associated uncertainty, while capturing dependencies among expert network calls, questions, and answers.

This framework recovers model-implied investor beliefs and their corresponding uncertainties, which evolve dynamically as new signals arrive. These beliefs not only reveal investors' expectations over time but also enable dynamic prediction of firm success. While our framework produces model-implied beliefs, we do not interpret them as directly observed subjective expectations. Instead, they are data-driven proxies inferred from qualitative signals, constructed using only information available up to each point in time and updated sequentially. They serve as intermediate economic objects summarizing how signals are aggregated into expectations. We later validate their economic relevance by showing that they explain subsequent investment decisions and outcomes and subsume the predictive content of raw text-based signals (Section 4).

### 3.2 LLM-Based Signal Extraction from Expert Network Calls

As discussed above, although expert network calls are information-rich, economically relevant content is often sparse and embedded in lengthy conversations. We therefore extract the most informative signals from expert network calls transcripts. In each call, questions and answers typically revolve around the firm's underlying state, such as operations, reputation, and products, with questions seeking to uncover this state and answers reflecting expert assessments. Because this state is unobserved, we summarize each interaction into a structured representation that captures its key informational content.

However, expert network calls are typically lengthy and exhibit a complex structure. Directly feeding the entire text into an LLM would exceed its context limit and result in an ultra-long inference time. Conversely, splitting the text into smaller segments and processing each separately would neglect the structural dependencies inherent in the expert network call data. To address this, we leverage an LLM (GPT

4) to extract key information from lengthy and complex call transcripts.[2] Based on the literature review in Section 1, we identify the key factors associated with successful private firm outcomes. These factors are then incorporated into domain-specific prompts designed to embed relevant expert knowledge (The detailed prompts are presented in Appendix A. III). Guided by these prompts, the LLM condenses each question–answer pair into a shorter, information-dense summary. Finally, following previous research (Koval et al. 2023), we adopt MPNet, a variant of SentenceTransformer (Reimers and Gurevych 2019), to obtain dense, context-sensitive representations of the summarized questions and answers. Formally, we denote the resulting representations of the question and answer in the $k$-th question–answer pair as $\boldsymbol{q}^{t,k}$ and $\boldsymbol{a}^{t,k}$, which serve as inputs to the Bayesian learning framework.

**3.3 Bayesian Learning from Dynamic Signals**

We model investor learning via a Bayesian network in which beliefs about firm quality are updated as new expert signals arrive. Investors do not observe fundamentals directly, but instead infer them from qualitative information obtained through expert interactions. The Bayesian network is constructed to capture the three main factors that influence investors' belief $r^t$ at time $t$ through a generative process. The Bayesian network is shown in Figure 4.

**3.3.1 Modeling of Investors' Beliefs and Observed Outcomes**

The investor's belief $r^t$ is influenced by three factors. First, investors' belief $r^t$ is formed based on their perception of the firm's current fundamental state, such as its operations, reputation, products, and other key attributes. Thus, we introduce $\boldsymbol{s}^t \in \mathbb{R}^V$ to denote the firm's state at time $t$, which encodes the informative aspects from the expert network calls. Given the persistence in the firm's fundamentals, we assume the state evolves smoothly over time: $\boldsymbol{s}^1 \sim \mathcal{N}(0, I)$, and for $t \geq 2$, $\boldsymbol{s}^t \sim \mathcal{N}(\boldsymbol{s}^{t-1}, I)$. Second, the belief $r^t$ is affected by historical states prior to the expert network call at time $t$, i.e., $\{\boldsymbol{s}^{t-1}, \dots, \boldsymbol{s}^t\}$. Hence, when estimating the company's success rate, historical states should be considered. For instance, even if the current state appears unfavorable (e.g., short-term operational setbacks or negative near-term signals), a strong history of positive fundamental states can still lead investors to maintain a positive belief. Third, the belief $r^t$ is also influenced by external factors such as firm size, market conditions, and other environmental attributes, collectively denoted as $\boldsymbol{e}$.

Hence, the belief $r^t$ should be determined by three factors: $\boldsymbol{s}^t, \{\boldsymbol{s}^{t-1}, \dots, \boldsymbol{s}^1\}$, and $\boldsymbol{e}$. However, the functional relationship that maps these factors to $r^t$ is complex and nonlinear. This complexity arises because each $\boldsymbol{s}^t$ is a high-dimensional vector, and the number of such vectors varies dynamically with $t$. To flexibly model this relationship, we use a neural network. Specifically, we apply an attention mechanism over the sequence of firm states to assign adaptive weights to historical information. The resulting context vector is concatenated with the external factor $\boldsymbol{e}$ and fed into a multilayer perceptron (MLP) to generate $r^t$: $r^t = \text{MLP}_1(\text{Attention}(\boldsymbol{s}^{t-1}, \dots, \boldsymbol{s}^1);\ \boldsymbol{e})$, if $t \geq 2$. At the initial time ($t = 1$), no historical state is available. Therefore, we estimate the belief directly based on the current state and the external factors: $r^t = \text{MLP}_2(\boldsymbol{s}^1;\ \boldsymbol{e})$, if $t = 1$.

[2] We use GPT-4 as a representative large language model with strong summarization and information extraction capabilities. Our setting requires only that the model reliably extract structured signals from complex transcripts; the specific LLM is not central to our contribution. Any sufficiently capable model would yield similar inputs, and newer models can be readily incorporated. The contribution of the paper lies in the economic measurement and sequential belief-updating framework, rather than in the underlying language model.

The observed outcome $y$ is generated based on the last belief $r^T$. Specifically, the belief $r^T$ is treated as the expectation of a Bernoulli distribution, which is used to generate the observed label: $y \sim \text{Bernolli}(r^T)$.

**3.3.2 Modeling of Each Expert Network Call**

Within each expert network call, the dialogue follows a sequence of question–answer pairs (e.g., "question 1–answer 1," "question 2–answer 2"). Accordingly, we model the conversation as an ordered sequence of such paired exchanges, each comprising a question and its corresponding answer.

For a given answer $\boldsymbol{a}^{t,k}$, it is generated in response to its paired question $\boldsymbol{q}^{t,k}$. Meanwhile, from the expert's behavioral perspective, neighboring answers are often dependent, as human communication exhibits continuity and memory. For example, if an expert feels that a previous answer was unclear, they may clarify it in the next response. Therefore, we need to model dependencies between the current answer and the previous answer $\boldsymbol{a}^{t,k-1}$. In addition, we model the dependency between the current answer and the previous question $\boldsymbol{q}^{t,k-1}$, since an expert may recall an unaddressed point from the prior question and incorporate it into the current answer. However, to balance model complexity and noise, we restrict the dependency to only the immediately preceding question and answer rather than modeling longer histories. Finally, each answer is also shaped by the firm's state $\boldsymbol{s}^t$, as a more promising state tends to elicit more positive expert responses. In summary, the generation of each answer depends on four factors: $\boldsymbol{s}^t, \boldsymbol{q}^{t,k-1}, \boldsymbol{a}^{t,k-1}$ and $\boldsymbol{q}^{t,k}$. Since $\boldsymbol{a}^{t,k}$ is a continuous vector, we assume it follows a multivariate Gaussian distribution, with its mean determined by these four factors: $\mu(\boldsymbol{a}^{t,k}) = \text{MLP}_3(\boldsymbol{s}^t, \boldsymbol{q}^{t,k-1}, \boldsymbol{a}^{t,k-1}, \boldsymbol{q}^{t,k}), \boldsymbol{a}^{t,k} \sim \mathcal{N}(\mu(\boldsymbol{a}^{t,k}), \Sigma)$, where $\Sigma$ is a hyperparameter. When $k = 1$, i.e., for the first answer in the expert network call, no historical questions or answers are available. Therefore, the mean simplifies $\mu(\boldsymbol{a}^{t,k}) = \text{MLP}_4(\boldsymbol{s}^t, \boldsymbol{q}^{t,k})$.

Similarly, each question arises from multiple factors. First, a question may be prompted by ambiguity in the previous question or by elements in the prior exchange that spark the investor's interest. Second, questions in an expert network call typically evolve sequentially rather than randomly, making adjacent questions highly dependent. Third, questions are usually centered on the firm's current state, which shapes the investor's curiosity. Accordingly, each question $\boldsymbol{q}^{t,k}$ is determined by three factors: the firm's state $\boldsymbol{s}^t$, the previous question $\boldsymbol{q}^{t,k-1}$, and the previous answer $\boldsymbol{a}^{t,k-1}$. Since questions are represented as continuous vectors, we assume a multivariate Gaussian distribution whose mean is given by: $\mu(\boldsymbol{q}^{t,k}) = \text{MLP}_5(\boldsymbol{s}^t, \boldsymbol{q}^{t,k-1}, \boldsymbol{a}^{t,k-1}), \boldsymbol{q}^{t,k} \sim \mathcal{N}(\mu(\boldsymbol{q}^{t,k}), \Sigma)$. When $k = 1$, i.e., for the first question in the expert network call, no prior question or answer is available. In this case, the mean simplifies to $\mu(\boldsymbol{q}^{t,k}) = \text{MLP}_6(\boldsymbol{s}^t)$.

### 3.4 The Learning Process of Our Framework

Our framework involves two components: an LLM and a Bayesian network. In the first component, no parameters need to be learned, as we utilize a pretrained LLM. Parameter learning occurs only in the second component, i.e., how to learn the parameters of the Bayesian network.

A key challenge in learning Bayesian networks lies in handling latent variables. In our model, $r^t$ and $\boldsymbol{s}^t$ for $t = 1, \dots, T$, which represent the investor's belief and summarize firm fundamentals over time, are latent and are collectively denoted as $r^{1:T}$ and $\boldsymbol{s}^{1:T}$. A natural approach is to first infer their distributions, then treat the latent variables as observed under these inferred distributions to construct the objective, and finally learn the model parameters by maximizing this objective. While the generative process specifies distributions for $r^t$ and $\boldsymbol{s}^t$, these are defined independently of observed data and thus serve as priors, denoted by $p^{\text{prior}}(r^{1:T}, \boldsymbol{s}^{1:T})$. Given observed data, the distributions should be updated to the posterior $p(r^{1:T}, \boldsymbol{s}^{1:T} | \boldsymbol{A}^{1:T}, \boldsymbol{Q}^{1:T}, \boldsymbol{e}, y)$, where the input to the variational distribution consists of the observed variables: the environmental factors $\boldsymbol{e}$, all extracted answer vectors in the expert network call (denoted as $\boldsymbol{A}^{1:T}$), all extracted question vectors in the expert network call (denoted as $\boldsymbol{Q}^{1:T}$) and the final outcome $y$.

However, the true posterior is generally intractable. To address this, we adopt variational inference and introduce a tractable variational distribution to approximate it. Since $r^t$ is deterministically determined by $\boldsymbol{s}^{1:t}$ and the observed variable $\boldsymbol{e}$, we only need a variational distribution over $\boldsymbol{s}^{1:t}$, denoted as $q(\boldsymbol{s}^{1:T})$, to approximate the true posterior $p^{\text{post}}(\boldsymbol{s}^{1:T})$, while incorporating the prior $p^{\text{prior}}(\boldsymbol{s}^{1:T})$. The posterior of $r^t, \forall t$, then follows directly. Since the variational distribution is tractable, its variance can be computed to reflect the uncertainty of the updated belief. Technical details of the learning process are provided in the Appendix A.II.

Importantly, this estimation procedure yields a sequence of model-implied beliefs $\{r^t\}_{t=1}^{T}$ that evolve as new expert signals arrive. These beliefs integrate prior information with newly acquired signals, capturing a dynamic learning process for investors. Because they are constructed using only information available up to each point in time, they can be interpreted as forward-looking expectations and used to study subsequent investment decisions and outcomes.

### 3.5 Novelty of Our Framework

Investors evaluating early-stage firms face substantial uncertainty and limited hard information.

Expert network calls offer a window into belief updating, yet modeling such data poses three challenges: unstructured, information-sparse transcripts; hierarchical sequential dependencies across calls and dialogue turns; and latent, dynamically evolving beliefs with only terminal outcomes observed.

Our framework addresses all three challenges. First, we introduce an LLM-based extraction module that converts unstructured expert dialogue into dense vector representations. Using domain-specific prompts and a sentence transformer, the LLM distills each question–answer pair into a semantically rich representation. Our novelty lies in the latter two. Second, we design a novel Bayesian network that explicitly models dependencies of the data. Investors' beliefs depend on the firm's states and external factors, while the firm's states evolve smoothly over time. Meanwhile, each question–answer pair is generated conditionally on the firm's state, the prior question, and the prior answer, consistent with the natural flow of expert conversations. Third, we provide a fully Bayesian solution to latent dynamic belief updating. Using variational inference, we recover posterior distributions over time-varying beliefs $r^t$ (using only information up to $t$) and probabilistic (quantifying uncertainty). These beliefs aggregate dispersed signals in a theoretically grounded manner and are validated to explain subsequent investment decisions.

Together, these innovations transform unstructured sequential expert dialogues into dynamic, interpretable measures of investor beliefs, addressing a core challenge in early-stage entrepreneurial finance.

### 3.6 Evaluation Setting

We evaluate model performance along both statistical and economic dimensions. For statistical accuracy, we report *Accuracy, Precision, F1-score, Weighted F1, Macro F1,* and *Area Under the ROC Curve (AUC).*[3]

To assess economic relevance, we construct portfolio-based performance measures. Each correctly identified successful startup generates a final investment value ($FIV_{TP}$), defined as the post-success valuation following financing events such as acquisitions, IPOs, or subsequent financing. Each selected startup incurs an investment cost (*IC*), approximated as the last pre-success valuation plus 10% to capture screening and monitoring costs (Gompers and Lerner 2004; Metrick and Yasuda 2021). To account for missed high-value opportunities, we introduce an opportunity cost (*OC*), defined as the gap between a top-decile post-success valuation benchmark and the average pre-success valuation, consistent with private equity benchmarking practices (Harris et al. 2014; Cambridge Associates 2018; Institutional Limited

[3] AUC measures how well the model ranks firms by investment quality; a higher AUC implies a higher probability that successful firms are ranked above unsuccessful ones, improving investment selection. The F1-score summarizes the trade-off between identifying successful firms (recall) and avoiding unsuccessful ones (precision), capturing the overall effectiveness of investment screening.

Partners Association 2019). Because startup-level valuations are not always publicly available, we approximate these quantities using benchmark values from Crunchbase and VentureXpert.[4]

Let *TP*, *FP*, and *FN* denote true positives, false positives, and false negatives.[5] The net investment gain is calculated as the value of correctly predicted successful startups ($TP \times FIV_{TP}$) minus the total investment cost for all selected startups ($(TP + FP) \times IC$) and an additional penalty for missed successful startups ($FN \times OC$). For companies that were non-successful, we conservatively assign a final investment value of zero ($FIV_{FP}$). This Return on Investment (*ROI*) formulation explicitly penalizes both false positives and false negatives, aligning predictive accuracy with the asymmetric payoff structure of venture capital. Formally, *ROI* is defined as the ratio of net investment gain to total investment cost (Gompers and Lerner 2004, Metrick and Yasuda 2021, Maarouf et al. 2025):

$$ROI = \frac{TP \times FIV_{TP} + FP \times FIV_{FP} - (TP + FP) \times IC - FN \times OC}{(TP + FP) \times IC} \times 100 \qquad (2)$$

As robustness checks, we report another complementary metric widely used in private equity and venture capital: the Multiple on Invested Capital (*MOIC*). *MOIC*, the ratio of total realized and unrealized value to invested capital, captures gross value creation without time adjustment (Cumming 2009, Korteweg and Sorensen 2023). Alongside *ROI*, this metric provides a comprehensive view of model-driven portfolio performance. *ROI* remains our primary measure, reflecting net economic gain while accounting for missed high-value opportunities, whereas *MOIC* offers time-agnostic and profitability-focused perspectives, respectively (Cumming 2009, Ljungqvist 2024). Together, these metrics ensure that our evaluation captures both statistical accuracy and the economic relevance of predictive performance for institutional investors (Harris et al. 2014, KPMG International 2016).

$$MOIC = \frac{TP \times FIV_{TP} + FP \times FIV_{FP} - FN \times OC}{(TP + FP) \times IC} \qquad (3)$$

## 4. Empirical Results

### 4.1 Benchmark Performance and Predictive Accuracy

To evaluate our framework, we compare its performance against a broad set of benchmark models. While standard metrics assess predictive accuracy, our objective is broader: to test whether modeling expert network calls as sequential information signals improves the measurement of firm-level beliefs and yields economically meaningful implications for investment decisions. Accordingly, prediction serves as a validation of belief measurement rather than the primary objective.

---

[4] In our dataset, the valuation of a startup after a success event (i. e., initial public offerings, funding, acquisitions) is, on average, $248.44 million. The investment cost is, on average, $10.24 million. The opportunity cost is, on average, $198.81 million.

[5] TP represents the number of correctly predicted successful startups, FP represents the number of startups incorrectly predicted as successful, and FN represents the number of missed successful startups.

We consider three groups of benchmarks. The first includes standard machine learning methods commonly used in startup success prediction, support vector machines (SVM), logistic regression (LR), Naïve Bayes (NB), and multilayer perceptrons (MLP), which rely on engineered inputs constructed from (i) topic distributions extracted using Latent Dirichlet Allocation (LDA), (ii) the number of expert network calls in the prior 24 months, and (iii) firm-level fundamentals from Crunchbase. The second group consists of transformer-based models for long-text classification, including hierarchical architectures (HAT; Hu et al. 2021, Hi-Transformer; Wu et al. 2021) and sparse-attention models (Longformer; Beltagy et al. 2020, BigBird; Zaheer et al. 2021), augmented with firm-level fundamentals. The third group includes recent text-based startup prediction approaches (Maarouf et al. 2025; Katsafados et al. 2024; Guzman and Li 2023), which use textual data but do not model sequential learning. All models are finetuned, and results are averaged over ten runs.

Panel A isolates the role of feature construction. As shown in Table 3 Panel A, performance improves as additional inputs are incorporated, progressing from firm fundamentals (fv) to combinations of textual features (topic) and call volume (#call). Incorporating expert network calls signals further enhances performance relative to fundamentals alone, with higher AUC indicating improved ranking of firms by investment quality and better identification of high-return opportunities (Figure A.3). This evidence shows that expert network calls contain economically meaningful soft information, such as product traction, competitive positioning, and managerial quality, that is not captured by structured data. However, models that treat these signals as static features cannot fully exploit this information, as they fail to capture how signals arrive and accumulate over time. In contrast, our framework models expert network calls as sequential signals and maps them into time-varying beliefs, allowing the information embedded in expert interactions to be more effectively incorporated into investment decisions. Relative to the best baseline (MLP with topic, fv, and #call), it improves the F1-score by 7.08% and delivers higher ROI (723 vs. 640) and MOIC (8.23 vs. 7.40).[6] These results indicate that while expert network calls data provide valuable information, fully realizing their economic value requires modeling their temporal structure.

Panel B isolates the role of text representation. As reported in Table 3 Panel B, our framework consistently outperforms transformer-based models designed for long-text classification. Hierarchical models perform better than sparse-attention models, likely because they better align with the multi-call, multi-turn structure of expert conversations, whereas sparse-attention models struggle to capture long-range dependencies under practical constraints. Despite their ability to process complex and lengthy text, these models do not capture how information arrives and is incorporated over time. Relative to the best-performing transformer (HAT; Hu et al. 2021), our framework improves F1-score by 6.69% and increases ROI by 95.74 (a 15.26% relative gain).[7] These results show that improvements are not driven by more sophisticated text representations, but by modeling the dynamic arrival of information and its impact on belief formation.

[6] The percentage improvement is computed as the relative increase in F1-score over the best-performing baseline (MLP using textual topic features (topic), firm-level fundamentals (fv), and call volume (#call)), $(0.590/0.551 - 1) \times 100\% = 7.08\%$.

Panel C isolates the contribution relative to existing text-based approaches. As shown in Table 3 Panel C, our model substantially outperforms recent startup prediction methods that combine textual and structured inputs. While Katsafados et al. (2024) performs best among these benchmarks, our framework delivers an additional 11.74% improvement in F1-score and a 65.16% increase in ROI.[8] Unlike these approaches, which treat text as static input, our framework models how qualitative signals are sequentially incorporated into beliefs. This comparison shows that the gains are substantial relative to the existing literature and arise from introducing a belief-based, dynamic learning structure.

Taken together, Panels A–C isolate the source of the performance gains. Panel A shows that incorporating expert network calls data improves performance, but these gains are not fully realized without modeling how information accumulates over time. Panel B shows the gains are not driven by more advanced text representations, and Panel C shows they are substantial relative to existing approaches rather than merely incremental. Across all comparisons, the advantage of our framework arises from modeling expert network calls as sequential information signals and mapping them into time-varying beliefs. This evidence indicates that the key contribution lies in capturing how information accumulates and is incorporated into beliefs over time, rather than in feature engineering or text representation.

We further assess robustness through sensitivity analysis of key hyperparameters (Table A.4). Performance remains stable across specifications, with optimal results at a learning rate of $1 \times 10^{-5}$, a dropout rate of 0.15, a latent state dimension of 512, and a constraint weight of 0.0001. These findings indicate that results are not driven by parameter choices but reflect the model's ability to capture the structure of expert signals and belief updating.

**4.2 Sequential Learning and Model Validation**

We examine the mechanisms underlying model performance by evaluating how different components contribute to learning from expert signals and how beliefs evolve over time. The results show that the gains documented in Section 4.1 arise from capturing the structure and dynamics of information acquisition in private markets.

[7] The percentage improvement is computed as the relative increase in F1-score over the best-performing baseline (HAT), $(0.590/0.553 - 1) \times 100\% = 6.69\%$. The increase in ROI is measured as both an absolute difference, $723.350 - 627.61 = 95.74$, and a relative change, $(723.350/627.61 - 1) \times 100\% = 15.26\%$

[8] The reported percentage improvements are computed relative to the best-performing baseline (Katsafados et al. (2024)). The increase in F1-score is calculated as $(0.590/0.528 - 1) \times 100\% = 11.74\%$. The increase in ROI is measured as both an absolute difference, $723.350 - 437.973 = 285.38$, and a relative change, $(723.350/437.973 - 1) \times 100\% = 65.16\%$.

We begin with ablation tests. Table 4 reports variants that remove key components, alongside both statistical metrics (Accuracy, Precision, F1-score, Weighted F1, Macro F1, AUC) and economic measures (ROI and MOIC). Modeling dependencies across expert signals is critical. A static aggregation benchmark (Row (1), *No Signal Dependence*), which treats calls as independent, performs worst across all metrics, indicating substantial information loss when interaction structure is ignored. Introducing partial structure, either across calls (Row (2)) or within calls (Row (3)), yields modest improvements but remains well below the full model. These results show that both within-call conversational structure and cross-call information accumulation are essential for extracting meaningful signals.

Sequential belief updating is also central. Removing this component (Row (4), *No Belief Updating*) leads to a consistent decline in both predictive and economic performance. While this variant captures static relationships, it fails to reflect how information is incorporated over time. From a finance perspective, this result captures a core feature of investor behavior: beliefs are revised sequentially as new information arrives.

Signal extraction quality further matters. Replacing the LLM-based extraction with simpler representations (Row (5), *No LLM Signal Extraction*) reduces performance across all metrics, indicating that expert network calls transcripts contain rich but noisy information that must be effectively summarized to recover economically meaningful variation.

Taken together, the ablation results show that dependency structure, sequential updating, and signal extraction all contribute to performance. The largest deterioration occurs when dependencies are removed, indicating that the primary value of the framework lies in capturing how information is structured, accumulated, and incorporated into beliefs over time.

We next examine the dynamics of learning. As shown in Table 5, both statistical and economic performance improve as additional expert network calls are incorporated. Accuracy increases from 0.769 to 0.777, while ROI rises by over 32.5%, indicating that posterior beliefs are progressively refined as new signals arrive. This pattern closely mirrors the iterative due-diligence process used by venture investors.

These results reflect the role of information acquisition under uncertainty. Private firms operate with limited disclosure and severe information asymmetry (Gompers and Lerner 2001; Kaplan and Strömberg 2004), and information arrives sporadically rather than continuously. By treating expert network calls as sequential signals, the framework allows beliefs about firm quality to evolve over time as new information on product traction, competition, and execution becomes available.

To further examine belief dynamics, Figure 5 analyzes investor uncertainty around investment events using an event-study specification. Uncertainty, measured as the dispersion of model-implied beliefs (Appendix Table A.2), declines as the deal approaches and stabilizes afterward. This pattern indicates that uncertainty is resolved during the pre-investment phase, consistent with a sequential learning process in which investors actively acquire information before committing capital. The post-deal stabilization suggests that the most relevant uncertainty is resolved during due diligence.

Overall, these results show that the framework transforms fragmented qualitative signals into a dynamic process of belief updating. By aligning model updating with the timing of real-world information arrival, it provides an economically grounded representation of investor learning in private markets.

**4.3 Economic Relevance: Implications for Investment Decisions**

We next examine whether improved belief measurement translates into economically meaningful gains in investment decisions.

Figure 6 evaluates ranking performance. As selectivity increases (i.e., focusing on smaller fractions of top-ranked firms), realized performance improves across all panels, with ROI and MOIC highest in the top tail of the distribution. This pattern reflects the concentration of returns in a small subset of investments. Within each selectivity threshold, our model consistently outperforms benchmark methods, indicating that it more effectively identifies high-quality opportunities. These results show that the framework improves capital allocation primarily through better ordering of investment opportunities, rather than through uniform improvements in average predictive accuracy.

Figure 7 evaluates the implications for portfolio allocation. Starting from portfolios constructed using benchmark rankings, we progressively replace selected firms with those identified by our model. Portfolio performance increases monotonically with the replacement fraction for both ROI and MOIC. This pattern indicates that firms selected by our model systematically outperform those chosen by alternative approaches, and that the gains extend beyond a small subset of investments to improve overall portfolio outcomes. Together, these results provide direct evidence that improved belief measurement leads to more efficient capital allocation and higher realized returns.

Figure A.4 evaluates robustness to payoff and cost assumptions. Across a wide range of investment costs (IC) and payoff levels ($FIV_{TP}$), our model consistently outperforms both HAT- and SVM-based portfolios. The performance advantage is strongest in economically favorable regions (low costs and high payoffs) but remains positive across most of the parameter space. This indicates that the gains are not driven by a specific payoff specification, but reflect a general improvement in selecting high-quality investments.

Overall, these results show that modeling expert network calls as sequential signals improves the measurement of investor beliefs, which in turn translates into better ranking, more efficient capital allocation, and higher investment returns.

### 4.4 Information Processing: What Signals Does the Model Learn?

To interpret the mechanisms behind the LLM-Bayesian's predictive success, we combine qualitative case evidence and quantitative attention analysis, linking conversational patterns to model inference and investment outcomes. This analysis identifies which signals matter, how they are incorporated into beliefs, and why this process improves investment decisions.

#### 4.4.1 Exemplary Transcript by Expert Type

We begin by examining representative excerpts from expert network consultations across five expert categories, customers, industry consultants, former executives, competitors, and partners (Table 6). These transcripts illustrate how the LLM-Bayesian framework learns economically meaningful signals from dialogue. Each excerpt aligns the model-identified sentiment with the startup's realized outcome, allowing direct comparison between evaluative tone and ex-post success.

Customer discussions reveal that language about product functionality and user experience strongly differentiates outcomes. Successful firms are described with concrete and confident expressions, while failed ones draw skepticism. These patterns indicate that product–market fit and demand-side validation are key signals in early-stage investment. Industry consultants emphasize technological feasibility and scalability. Positive discussions highlight measurable improvements, whereas negative ones focus on adoption challenges. These signals directly inform beliefs about growth potential and execution risk. Former executives provide insights into organizational capability, while competitors and partners offer external validation regarding market positioning and trust. Together, these perspectives triangulate firm quality and reduce uncertainty through independent sources.

Overall, expert discourse encodes multi-dimensional soft information, product, technology, management, and competition, that is not captured in standard datasets. The model works by transforming these qualitative signals into structured latent representations that evolve with new information.

#### 4.4.2 Exemplary Transcript by Attention Scores

Building on the qualitative insights from transcripts, we next quantify how the LLM-Bayesian framework weighs different sources of information during prediction. In our setting, the attention score captures the relative weight assigned to a signal in belief formation, so higher attention indicates a larger influence on the model-implied posterior belief about firm success. Figures 8–10 show how the LLM-Bayesian framework allocates informational importance across expert types, topics, sentiment orientations, and firm characteristics. Our attention-based approach provides a dynamic and interpretable view of how the model assigns informational importance within a hierarchical and sequential structure. While conventional interpretability methods identify which linguistic or topical features are most strongly associated with predicted outcomes, they typically operate on static models trained on fixed data snapshots. By contrast, the LLM-Bayesian framework assigns attention weights endogenously within a hierarchical and sequential structure, allowing informational importance to evolve as new information arrives, an essential property in settings such as startup financing, where information is sparse and continuously updated.

We begin with Figure 8, which examines the distribution of attention across expert types and discussion topics. Panel (a) shows that customers and industry consultants receive the largest weights, indicating that the model places greatest value on voices with firsthand operational and market knowledge. Former executives also contribute meaningfully, reflecting insights into managerial and organizational execution, while competitors and partners play more peripheral roles. Panel (b) highlights that discussions on regulatory challenges, customer acquisition, and industry-specific AI applications dominate the model's learned focus, aligning closely with the uncertainty dimensions most salient in early-stage investment, compliance risk, scalability, and technological viability. The correspondence between model attention and investor intuition suggests that the LLM-Bayesian framework learns to extract information hierarchically, consistent with how investors prioritize diverse but economically material signals.

Having identified the most influential sources and topics, Figure 9 examines how attention varies with sentiment orientation. The model assigns the strongest weights to neutral and positive tones when discussing regulatory or technology-related topics, indicating that optimism in technically complex or compliance-heavy domains carries greater predictive value. Attention to customer acquisition increases markedly with positive sentiment (0.102), implying that expressions of confidence in market traction are particularly informative about future success. Conversely, negative tone in these same domains reduces attention weights (0.023), suggesting that the model learns to discount pessimistic views when uncertainty remains high. Across technology-related topics, positive signals consistently receive two to six times the attention of negative signals, highlighting the model's emphasis on signals related to feasibility and execution. This pattern mirrors investor behavior during venture screening, where signals associated with positive assessments receive greater weight in updating beliefs about firm quality, while negative signals have more limited influence on posterior beliefs.

We next turn to Figure 10, which examines how attention evolves over the firm lifecycle. A clear pattern emerges: attention to key topics declines sharply as firms mature and the information environment becomes more transparent. For example, attention to regulatory discussions falls from approximately 0.010–0.011 in early funding rounds to near zero in later rounds, from about 0.017 at low funding levels to below 0.002 as firms become more capitalized, and from roughly 0.006 in early-stage firms to below 0.001 in later stages. Customer acquisition and technology-related discussions exhibit similar dynamics, with attention peaking in early rounds (around 0.011–0.012 and 0.011, respectively) and declining to negligible levels as firms progress. These economically large declines, often by an order of magnitude, indicate that qualitative expert signals receive substantially greater weight when firms are young, less funded, and more opaque. As firms mature and observable performance measures become available, these signals are systematically downweighted in belief updating. Overall, the evidence is consistent with an information substitution mechanism, in which qualitative signals play a central role under high uncertainty but lose marginal value as more reliable, quantitative information becomes available.

Taken together, the attention patterns reveal that the LLM-Bayesian approach provides an economically interpretable decomposition of how investors learn from expert interactions over time. Rather than offering a static explanation of feature importance, the model dynamically tracks which information sources and tones are most predictive given the firm's stage and conversational context. This dynamic interpretability bridges the gap between linguistic explainability and financial reasoning, capturing how investors update their beliefs in environments characterized by limited disclosure, evolving uncertainty, and high stakes.

**4.5 Belief Updating and Capital Allocation: The Economic Value of Information**

Building on Section 4.4, which characterizes how the model extracts and processes information from expert network calls, we next examine how these belief dynamics translate into economically meaningful outcomes. In particular, we study how expert network calls signals induce changes in model-implied beliefs and uncertainty, our measures of investor learning, and how these changes ultimately guide capital allocation decisions.

Table 7 examines how expert network calls signals are transformed into economic outcomes through a belief-updating mechanism. Specifically, we analyze how new information from expert interactions reduces investor uncertainty and shapes beliefs that guide investment decisions. The dependent variable is the reduction in uncertainty, measured as$\Delta \text{Uncertainty}_{it} = 100 \times (\text{Uncertainty}_{i,t-1} - \text{Uncertainty}_{it})$, so larger values indicate greater belief updating. Because both belief and uncertainty are model-implied quantities (see Appendix Table A.2 for formal definition), the LLM-Bayesian framework serves as a measurement system that maps qualitative signals into changes in investor beliefs.

Panel A provides evidence on state-dependent learning by linking information signals to belief updating.[9] The positive and highly significant coefficient on prior uncertainty indicates that belief revisions are larger when initial uncertainty is high, consistent with a higher marginal value of information. We measure *Opacity* using a composite index of firm youth and product scope, capturing the complexity of the firm's information environment: younger firms have fewer verifiable signals, while broader product scope increases the difficulty of assessing performance and growth prospects. The interaction between prior uncertainty and opacity is positive and significant, indicating that signals induce larger belief revisions in more opaque environments where public information is limited. At the same time, the negative coefficient on opacity implies that uncertainty is, on average, more difficult to resolve for such firms. Overall, these findings show that expert network calls signals are most consequential when information frictions are severe.

[9] Given the call-level structure of the data, including firm fixed effects substantially reduces the sample size and variation. Therefore, we include time fixed effects and focus on within-call dynamics, while complementing this analysis with firm fixed effects at the deal level in Table 8.

Panel B shows that the informational value of signals varies across expert types. Calls involving *Operational Expert*, such as customers, consultants, and former executives, lead to significantly larger reductions in uncertainty, particularly when prior uncertainty is high. This finding suggests that investors place greater weight on high-quality, experience-based information sources. In our framework, this reflects the role of source credibility in shaping belief updating.

Panel C demonstrates that belief updating also depends on signal content. Calls focusing on technology adoption topics generate significantly larger reductions in uncertainty, consistent with these discussions capturing hard-to-observe fundamentals such as feasibility and scalability. This indicates that signals tied to core economic drivers have greater influence on belief formation than more general information.

Taken together, Table 7 shows that expert network calls information systematically reduces uncertainty and drives belief updating in economically meaningful ways. The magnitude of belief updating varies with prior uncertainty, information opacity, and the quality and content of signals, leading to more precise and better-informed investor beliefs.

Table 8 links belief formation to capital allocation using specifications with firm and quarter fixed effects, so identification comes from within-firm variation over time. Panel A shows that expert network calls information is strongly associated with subsequent investment outcomes. The occurrence of a call increases the probability of a financing event by 6.9 to 9.0 percentage points, indicating that expert consultations contain information beyond observable firm characteristics. The content of these interactions is also informative: positive sentiment raises investment likelihood by 3.9 to 4.1 percentage points, with substantially larger effects for discussions of technology adoption and customer acquisition (10.8 and 13.2 percentage points, rising to 11.4 and 14.7 percentage points when sentiment is positive). These economically large magnitudes suggest that expert network calls signals, particularly those related to execution and market traction, play an important role in shaping investment decisions. Appendix Table A.6 provides additional insight into the informational content of expert signals across horizons. While positive signals primarily predict short-term deal outcomes, negative signals, particularly discussions of regulatory and security risks, are more informative about long-run performance. The coefficient on negative risk-related discussions is economically meaningful, corresponding to a decline of approximately 0.02 standard deviations in future success (mean = 0.23, standard deviation = 0.64).[10] This asymmetry suggests that expert concerns capture forward-looking downside risk, whereas positive signals mainly reflect short-term validation.

However, these reduced-form relationships do not reveal how signals are incorporated into beliefs. Panel B establishes that model-implied beliefs and uncertainty serve as the primary channels through which expert network calls information affects investment decisions. Using the summary statistics in Appendix Table A.5, a one standard deviation increase in *Investor Success Belief* (0.052) corresponds to approximately a 0.11 increase in the probability of a financing event, while a one standard deviation increase in *Uncertainty* (0.008) reduces deal probability by about 0.01.[11] These effects are economically meaningful and precisely estimated. Once beliefs and uncertainty are included, the coefficients on baseline text-based signals become small and statistically insignificant, indicating that their predictive power operates through belief formation rather than directly influencing outcomes. Columns (3) and (4) confirm the robustness of this mechanism. The negative and significant interaction term in column (4) shows that the effect of beliefs is attenuated when uncertainty is high, suggesting that investors place less weight on signals when confidence is limited. Importantly, beliefs are constructed using only information available up to time $t$, and we evaluate their ability to predict subsequent investment outcomes. This temporal separation mitigates concerns of mechanical reweighting of text features and instead supports an interpretation in which the model recovers data-driven proxies for investor beliefs that govern capital allocation decisions.

Panel C examines the economic value of learning by linking changes in uncertainty to investment decisions. While Table 7 studies how expert network calls signals generate learning, here we test whether this learning translates into capital allocation outcomes. The negative coefficient on *ΔUncertainty* indicates that greater reductions in uncertainty increase the likelihood of investment, with a stronger effect when prior uncertainty is high. Unlike Panel B, which focuses on belief levels, these results show that the process of updating beliefs itself has direct economic value.

Taken together, Table 8 provides a direct link between information acquisition and economic outcomes. Signals from expert network calls are aggregated into beliefs and uncertainty measures, and these belief states in turn shape investment decisions. Overall, Tables 7 and 8 show that the LLM-Bayesian framework functions as a measurement system for investor learning, translating qualitative signals into structured belief updates that guide capital allocation and improve investment efficiency in private markets.

[10] The economic magnitude is obtained by dividing the estimated coefficient (−0.014) by the standard deviation of the dependent variable (0.643), implying an effect of approximately −0.022, or about 0.02 standard deviations in future success.
[11] Economic magnitudes are based on one standard deviation changes. Using SDs of 0.052 for Investor Success Belief and 0.008 for Uncertainty (Appendix Table A.5), the implied effects are 0.11 (= 2.20 × 0.052) and −0.01 (= −1.42 × 0.008) on deal probability.

**4.6 Heterogeneity: Information Environments and Variation in the Value of Learning**

To examine how the value of learning varies across information environments, we analyze heterogeneity in predictive and economic performance across four firm dimensions that proxy for information frictions: industry complexity, firm age, public visibility, and founder diversity. These dimensions capture differences in the availability of structured information and the role of qualitative signals in belief formation. Because the framework maps expert network calls transcripts into beliefs and uncertainty, cross-sectional variation in performance provides indirect evidence on when qualitative information is more informative.

Table 9 shows a consistent pattern: the model delivers the largest gains in environments where information is fragmented and difficult to interpret. In Panel A, performance gains are strongest in high-complexity industries such as AI, software, and biotechnology. In these technologically dense sectors, expert discussions highlight interoperability challenges and technical risks that are difficult to capture using structured data. Accordingly, the model achieves higher Precision (0.518 vs. 0.313) and larger ΔROI (+106.260 vs. +8.070), consistent with qualitative signals playing a larger role in belief updating when structured information is limited.

Panel B shows similar effects across the firm lifecycle. Young firms (less than six years old) exhibit higher predictive and economic gains (Precision = 0.561; ΔROI = 268.300) than mature firms. Because early-stage startups lack long operating histories, expert commentary provides forward-looking assessments of scalability and execution risk, increasing reliance on qualitative signals in belief formation.

A similar pattern appears in Panel C, where low-visibility firms with younger web domains yield ΔROI = 336.920, more than twice that of high-visibility peers. When digital footprints are sparse, expert network calls serve as a primary source of information, increasing the value of learning from qualitative signals.

Panel D shows stronger gains for firms led by diverse founding teams (Precision = 0.605; ΔROI = 203.120). One interpretation is that greater diversity is associated with more varied and complex information, consistent with qualitative signals playing a larger role in belief formation in these environments.

Overall, the model's advantage increases with information opacity. Across complex, young, low-visibility, and diverse firms, it delivers larger gains in predictive accuracy and portfolio returns by transforming unstructured expert discourse into actionable signals. These patterns highlight how the economic value of information aggregation varies across environments: when hard information is limited and difficult to interpret, investors rely more heavily on qualitative signals, and the ability to structure and integrate these signals becomes more valuable. In this sense, the model provides a measurement framework for investor learning, showing where qualitative information is most consequential for belief formation and capital allocation.

## 5. Discussion and Conclusion

We study investor learning and information acquisition in private markets using sequential expert network calls signals. In settings with limited disclosure and high uncertainty, investors rely on fragmented qualitative information that arrives over time. We develop a sequential LLM-Bayesian framework that converts expert conversations into time-varying beliefs and uncertainty, providing a measurement system for how information enters investment decisions. Specifically, we combine LLM-based signal extraction with sequential Bayesian updating to model how investors process unstructured dialogue. By capturing dependencies across calls and the temporal structure of information arrival, the framework moves beyond static text-based approaches and yields interpretable measures of beliefs and uncertainty alongside improved investment performance.

Our analysis yields three insights. First, learning is state-dependent: belief updating is strongest when prior uncertainty is high and when firms are more opaque. Second, information is heterogeneous: signals from operational experts and core fundamentals generate larger belief revisions. Third, information is asymmetric across horizons: positive signals primarily support short-term investment decisions, while negative signals, especially those related to regulatory and security risks, are more informative about long-run outcomes. Belief formation is the key channel linking information to capital allocation. Model-implied beliefs and uncertainty, rather than raw signals, drive investment decisions, and reductions in uncertainty increase investment likelihood. Improved belief measurement leads to better ranking and more efficient capital allocation, particularly in high-friction environments.

These findings contribute to entrepreneurial finance and information economics by providing direct evidence on how investors aggregate dispersed qualitative signals in real time. More broadly, the paper shows how machine learning can be used for economic measurement, enabling the study of belief dynamics and information aggregation when key variables are unobservable. Finally, the framework has practical implications. By converting unstructured expert dialogue into interpretable belief and uncertainty measures, it supports due diligence and screening in data-sparse environments and improves capital allocation when traditional information is limited.

**References**

Beltagy I, Peters ME, Cohan A (2020) Longformer: The Long-Document Transformer. Preprint, submitted December 2, http://arxiv.org/abs/2004.05150.

Bernstein S, Korteweg A, Laws K (2017) Attracting early-stage investors: evidence from a randomized field experiment. *J. Finance* 72(2):509–538.

Blei DM, Ng AY, Jordan MI, Lafferty J (2003) Latent dirichlet allocation. *J. Mach. Learn. Res.* 3(4/5):993.

Borchert P, Coussement K, De Caigny A, De Weerdt J (2023) Extending business failure prediction models with textual website content using deep learning. *Eur. J. Oper. Res.* 306(1):348–357.

Bowen DE, Frésard L, Hoberg G (2023) Rapidly evolving technologies and startup exits. *Manage. Sci.* 69(2):940–967.

Bushee BJ, Matsumoto DA, Miller GS (2003) Open versus closed conference calls: the determinants and effects of broadening access to disclosure. *J. Account. Econ.* 34(1–3):149–180.

Cambridge Associates (2018) *US Venture Capital Index and Selected Benchmark Statistics* (Cambridge Associates).

Cao S, Green TC, Lei L (Gillian), Zhang S (2023) Expert network calls. Preprint, submitted May 15, https://papers.ssrn.com/abstract=4280865.

Carvalho CM, West M (2007) Dynamic matrix-variate graphical models. *Bayesian Anal.* 2(1):69–97.

Cumming D (2009) *Private equity: fund types, risks and returns, and regulation* (John Wiley and Sons).

Drake MS, Roulstone DT, Thornock JR (2012) Investor information demand: evidence from Google searches around earnings announcements. *J. Account. Res.* 50(4):1001–1040.

Ewens M, Marx M (2018) Founder replacement and startup performance. *Rev. Financ. Stud.* 31(4):1532–1565.

Ewens, M., and Farre-Mensa, J. (2022). Private or public equity? The evolving entrepreneurial finance landscape. *Annu. Rev. Financ. Econ.* 14(1), 271-293.

Farboodi, M., Matray, A., Veldkamp, L., & Venkateswaran, V. (2022). Where has all the data gone?. *The Rev. Financ. Stud. 35*(7), 3101-3138.

Gans JS, Stern S, Wu J (2019) Foundations of entrepreneurial strategy. *Strateg. Manag. J.* 40(5):736–756.

Giglio S, Maggiori M, Stroebel J, Utkus S (2021) Five facts about beliefs and portfolios. *Am. Econ. Rev.* 111(5):1481–1522.

Gompers P, Lerner J (2001) The venture capital revolution. *J. Econ. Perspect.* 15(2):145–168.

Gompers PA, Gornall W, Kaplan SN, Strebulaev IA (2020) How do venture capitalists make decisions? *J. Financ. Econ.* 135(1):169–190.

Gompers PA, Lerner J (2004) *The venture capital cycle* (MIT Press).

Gruber TR (1995) Toward principles for the design of ontologies used for knowledge sharing? *Int. J. Hum.-Comput. Stud.* 43(5–6):907–928.

Guzman J, Li A (2023) Measuring founding strategy. *Manage. Sci.* 69(1):101–118.

Hansen, L. P., & Sargent, T. J. (2011). Robustness.

Harris RS, Jenkinson T, Kaplan SN (2014) Private equity performance: what do we know? *J. Finance* 69(5):1851–1882.

Hegde D, Tumlinson J (2014) Does social proximity enhance business partnerships? Theory and evidence from ethnicity's role in U.S. venture capital. *Manage. Sci.* 60(9):2355–2380.

Hevner, A. R., March, S. T., Park, J., and Ram, S. (2004). Design science in information systems research. MIS quarterly, 75-105.

Hobson JL, Mayew WJ, Venkatachalam M (2012) Analyzing speech to detect financial misreporting. *J. Account. Res.* 50(2):349–392.

Hochberg YV, Ljungqvist A, Lu Y (2007) Whom you know matters: venture capital networks and investment performance. *J. Finance* 62(1):251–301.

Howell ST (2017) Financing innovation: evidence from R&D grants. *Am. Econ. Rev.* 107(4):1136–1164.

Hu Y, Chen P, Liu T, Gao J, Sun Y, Yin B (2021) Hierarchical Attention Transformer Networks for Long Document Classification. *2021 International Joint Conference on Neural Networks (IJCNN)* 1–7

Huang AH, Wang H, Yang Y (2023) FinBERT: a large language model for extracting information from financial text. *Contemp. Account. Res.* 40(2):806–841.

Institutional Limited Partners Association (2019) ILPA Principles 3.0: Fostering Transparency, Governance and Alignment of Interests for General and Limited Partners. *Institutional Limited Partners Association*.

Kadhim AI (2019) Survey on supervised machine learning techniques for automatic text classification. *Artif Intell Rev* 52(1):273–292.

Kaplan SN, Schoar A (2005) Private equity performance: returns, persistence, and capital flows. *J. Finance* 60(4):1791–1823.

Kaplan SN, Strömberg P (2004) Characteristics, contracts, and actions: evidence from venture capitalist analyses. *J. Finance* 59(5):2177–2210.

Kaplan SN, Strömberg P (2009) Leveraged buyouts and private equity. *J. Econ. Perspect.* 23(1):121–146.

Katsafados AG, Leledakis GN, Pyrgiotakis EG, Androutsopoulos I, Fergadiotis M (2024) Machine learning in bank merger prediction: a text-based approach. *Eur. J. Oper. Res.* 312(2):783–797.

Ke ZT, Kelly B, Xiu D (2019) Predicting returns with text data. Preprint, submitted August, http://www.nber.org/papers/w26186.pdf.

Kerr WR, Nanda R, Rhodes-Kropf M (2014) Entrepreneurship as experimentation. *J. Econ. Perspect.* 28(3):25–48.

Korteweg AG, Sorensen M (2023) Performance measures in private equity. Preprint, submitted October 7, https://papers.ssrn.com/abstract=4595553.

KPMG International (2016) *Evaluating private equity's performance*

R. Koval, N. Andrews, X. Yan (2023) Forecasting earnings surprises from conference call transcripts. in: Findings of the Association for Computational Linguistics (ACL), pp. 8197–8209.

Larcker DF, Zakolyukina AA (2012) Detecting deceptive discussions in conference calls. *J. Account. Res.* 50(2):495–540.

Lerner, J., and Nanda, R. (2020). Venture capital's role in financing innovation: What we know and how much we still need to learn. Journal of Economic Perspectives, 34(3), 237-261.

Li, K., Mai, F., Shen, R., & Yan, X. (2021). Measuring corporate culture using machine learning. *The Rev. Financ. Stud. 34*(7), 3265-3315.

Liberti JM, Petersen MA (2019) Information: hard and soft. *Rev. Corp. Finance Stud.* 8(1):1–41.

Liu J, Jia M (2025) Financial distress prediction with annual reports-based deep textual feature extraction: A hybrid approach. *Inform. Sci.* 686:121318.

Ljungqvist A (2024) The economics of private equity: a critical review. Preprint, submitted February 12, https://papers.ssrn.com/abstract=4723171.

Loughran, T., & McDonald, B. (2011). When is a liability not a liability? Textual analysis, dictionaries, and 10-Ks. *J. Finance 66*(1), 35-65.

Maarouf A, Feuerriegel S, Pröllochs N (2025) A fused large language model for predicting startup success. *Eur. J. Oper. Res.* 322(1):198–214.

Mayew WJ, Venkatachalam M (2012) The power of voice: managerial affective states and future firm performance. *J. Finance* 67(1):1–43.

Metrick A, Yasuda A (2021) *Venture capital and the finance of innovation* (John Wiley & Sons).

Miao Y, Yu L, Blunsom P (2016) Neural variational inference for text processing. *Proceedings of The 33rd International Conference on Machine Learning* (PMLR), 1727–1736

Nanda R, Rhodes-Kropf M (2013) Investment cycles and startup innovation. *J. Financ. Econ.* 110(2):403–418.

Pastor L, Veronesi P (2009) Learning in financial markets. *Annu. Rev. Financ. Econ.* 1(1):361–381.

Popoola G, Abdullah KK, Fuhnwi GS, Agbaje J (2024) Sentiment analysis of financial news data using TF-IDF and machine learning algorithms. *2024 IEEE 3rd International Conference on AI in Cybersecurity (ICAIC)* 1–6

Puri M, Zarutskie R (2012) On the life cycle dynamics of venture-capital- and non-venture-capital-financed firms. *J. Finance* 67(6):2247–2293.

Reimers N, Gurevych I (2019) Sentence-BERT: sentence embeddings using siamese BERT-networks. *Proceedings of the 2019 Conference on Empirical Methods in Natural Language Processing and the 9th International Joint Conference on Natural Language Processing (EMNLP-IJCNLP)* (Association for Computational Linguistics, Hong Kong, China), 3980–3990

Sims, C. A. (2003). *Implications of rational inattention*. *J. Monetary Econ.* 5**0**(3), 665–690.

Solomon D, Soltes E (2015) What Are We Meeting For? The Consequences of Private Meetings with Investors. *J. Law Econ.* 58(2):325–355.

Sorenson, O., Dahl, M. S., Canales, R., & Burton, M. D. (2021). Do startup employees earn more in the long run?. *Organ. Sci.* 32(3), 587-604.

Van Nieuwerburgh, S., & Veldkamp, L. (2010). Information acquisition and under-diversification. *The Rev. Econom. Stud. 77*(2), 779-805.

Vaswani A, Shazeer N, Parmar N, Uszkoreit J, Jones L, Gomez AN, Kaiser Ł, Polosukhin I (2017) Attention is all you need. *Adv. Neural Inform. Process. Syst.* (Curran Associates, Inc.), 261–272

Veldkamp, L. L. (2006). Media frenzies in markets for financial information. *Amer. Econ. Rev. 96*(3), 577–601.

Wang H, Yeung DY (2021) A survey on bayesian deep learning. *ACM Comput. Surv.* 53(5):1–37.

Wu C, Wu F, Qi T, Huang Y (2021) Hi-transformer: hierarchical interactive transformer for efficient and effective long document modeling. *Proceedings of the 59th Annual Meeting of the Association for Computational Linguistics and the 11th International Joint Conference on Natural Language Processing (Volume 2: Short Papers)* (Association for Computational Linguistics, Online), 848–853

Yang Z, Yang D, Dyer C, He X, Smola A, Hovy E (2016) Hierarchical attention networks for document classification. Knight K, Nenkova A, Rambow O, eds. *Proceedings of the 2016 Conference of the North American Chapter of the Association for Computational Linguistics: Human Language Technologies* (Association for Computational Linguistics, San Diego, California), 1480–1489

Zaheer M, Guruganesh G, Dubey A, Ainslie J, Alberti C, Ontanon S, Pham P, et al. (2020) Big bird: transformers for longer sequences. *Proceedings of the 34th International Conference on Neural Information Processing Systems* NIPS '20. (Curran Associates Inc., Red Hook, NY, USA), 17283–17297

Zaheer M, Guruganesh G, Dubey A, Ainslie J, Alberti C, Ontanon S, Pham P, et al. (2021) Big bird: transformers for longer sequences. Preprint, submitted January 8, http://arxiv.org/abs/2007.14062.

Zhang H, Shafiq MO (2024) Survey of transformers and towards ensemble learning using transformers for natural language processing. *J Big Data* 11(1):25.

Zhang Y, Liu T, Li W (2024) Corporate fraud detection based on linguistic readability vector: Application to financial companies in China. *Int. Rev. Financ. Anal.* 95:103405.

Zhu Y, Liu X, Sheng ORL (2025) Post-earnings-announcement drift prediction: leveraging postevent investor responses with multitask learning. *Inf. Syst. Res.*:isre.2022.0358.

Zimmermann J, Champagne LE, Dickens JM, Hazen BT (2024) Approaches to improve preprocessing for Latent Dirichlet Allocation topic modeling. *Decis. Support Syst.* 185:114310.

**Figure 1: Information Acquisition Intensifies around Deal Quarters**

This figure presents event-time estimates of the change in the number of expert network calls around investment deals. The horizontal axis shows event quarters relative to the deal quarter (quarter 0), and the vertical axis reports coefficient estimates from a regression of the number of expert network calls on event-quarter indicators. The vertical error bars represent 95% confidence intervals. The vertical dashed line marks the deal quarter.

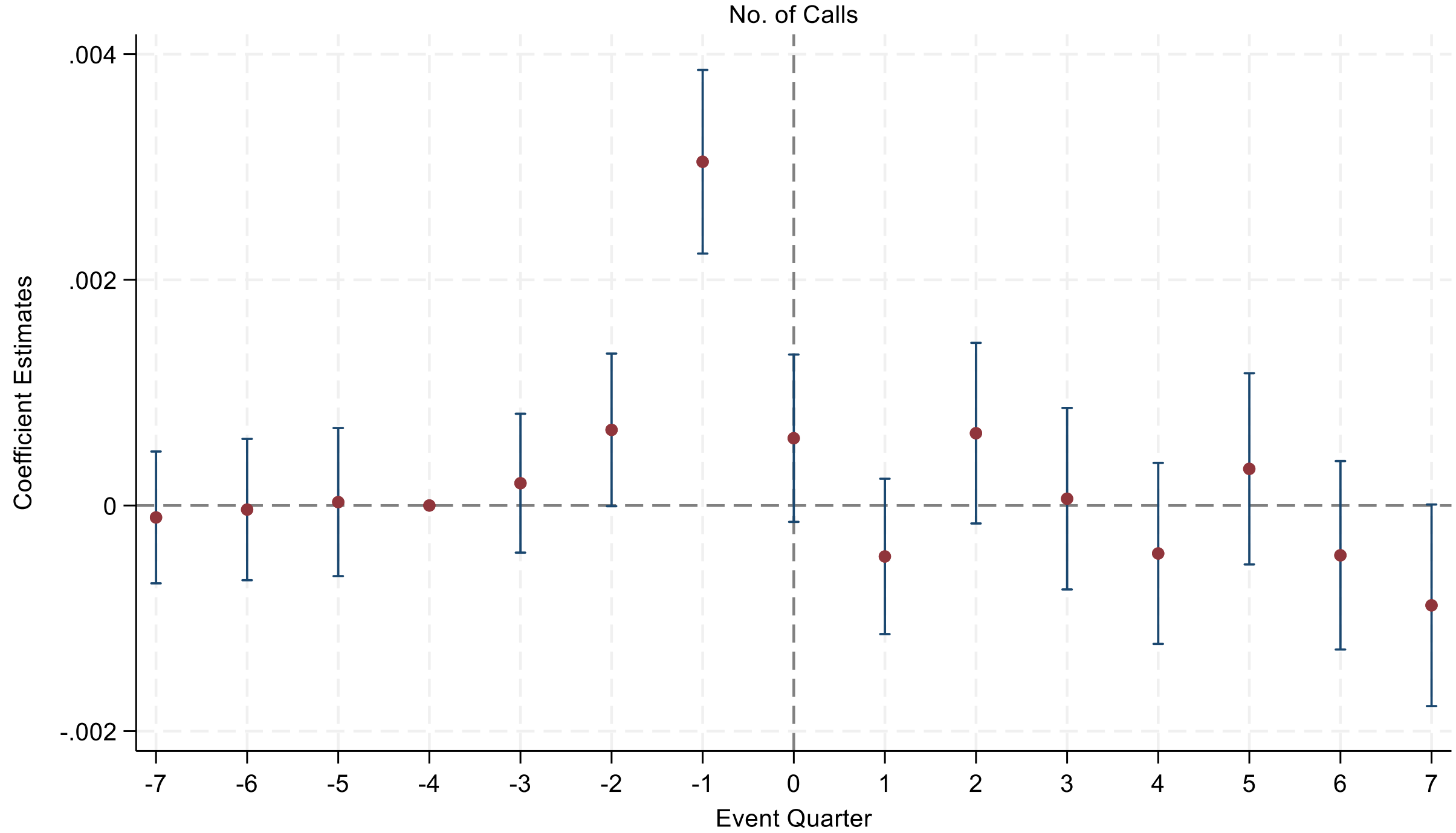

**Figure 2: Expert–Topic Knowledge Flows**

This figure visualizes the distribution of discussion topics across expert types in the expert network call dataset. Each flow represents the proportion of calls linking a given expert type, *Competitor*, *Customer*, *Former executive*, *Industry consultant*, or *Partner*, to one of the top refined topics identified in the topic-model analysis. Flow width reflects the relative frequency of topic mentions within each expert group. The visualization highlights how different expert roles emphasize distinct domains, such as strategy, product development, regulation, and data analytics.

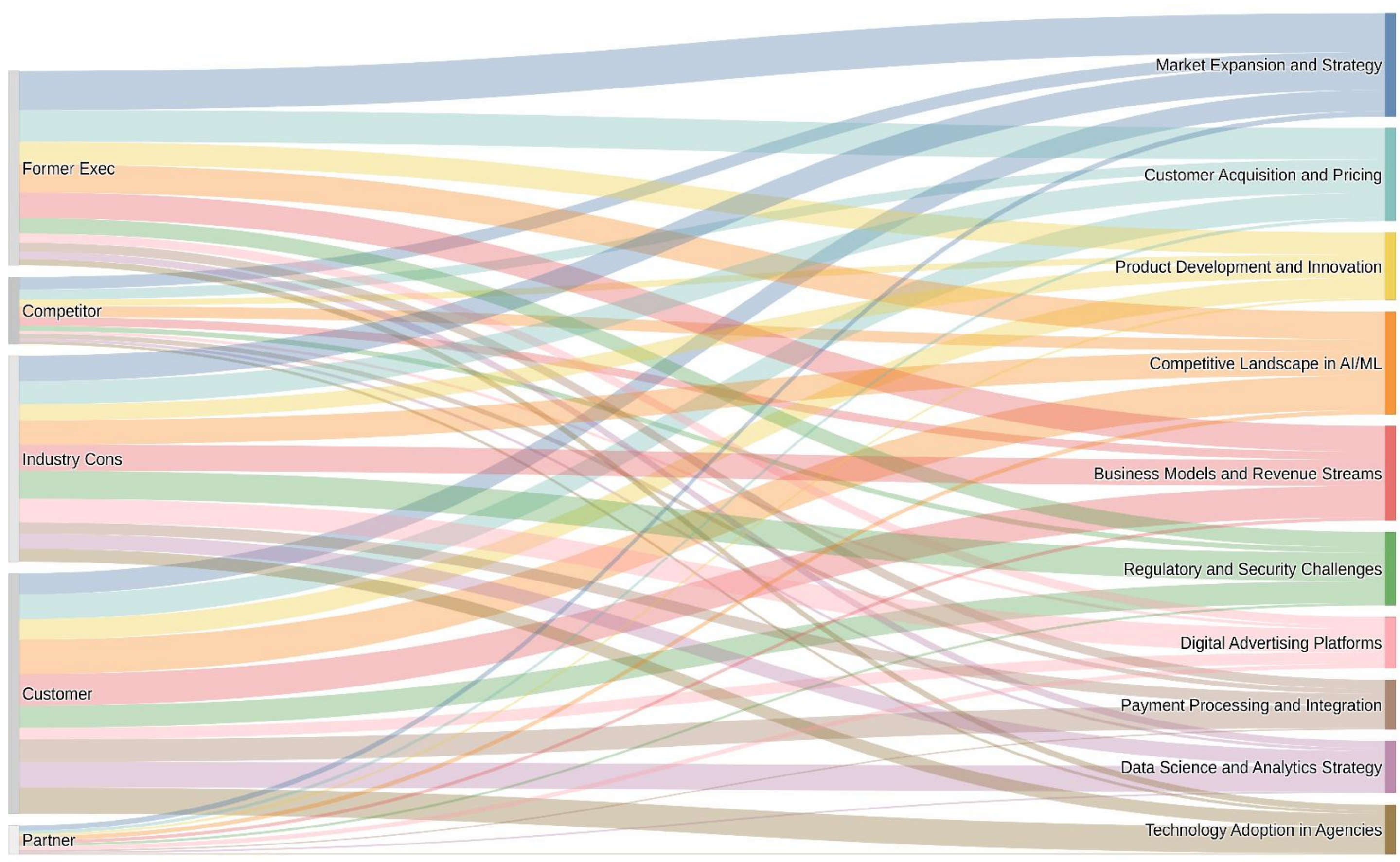

**Figure 3: Sentiment by Topic and Expert Type**

This heatmap reports the average sentiment score for each combination of discussion topic and expert type derived from expert–client consultation transcripts. Sentiment scores range from −2 (negative tone) to +2 (positive tone), with colors shifting from red (pessimistic outlook) to blue and green (optimistic outlook). Each cell reflects the mean sentiment expressed by a given expert group when discussing a specific topic.

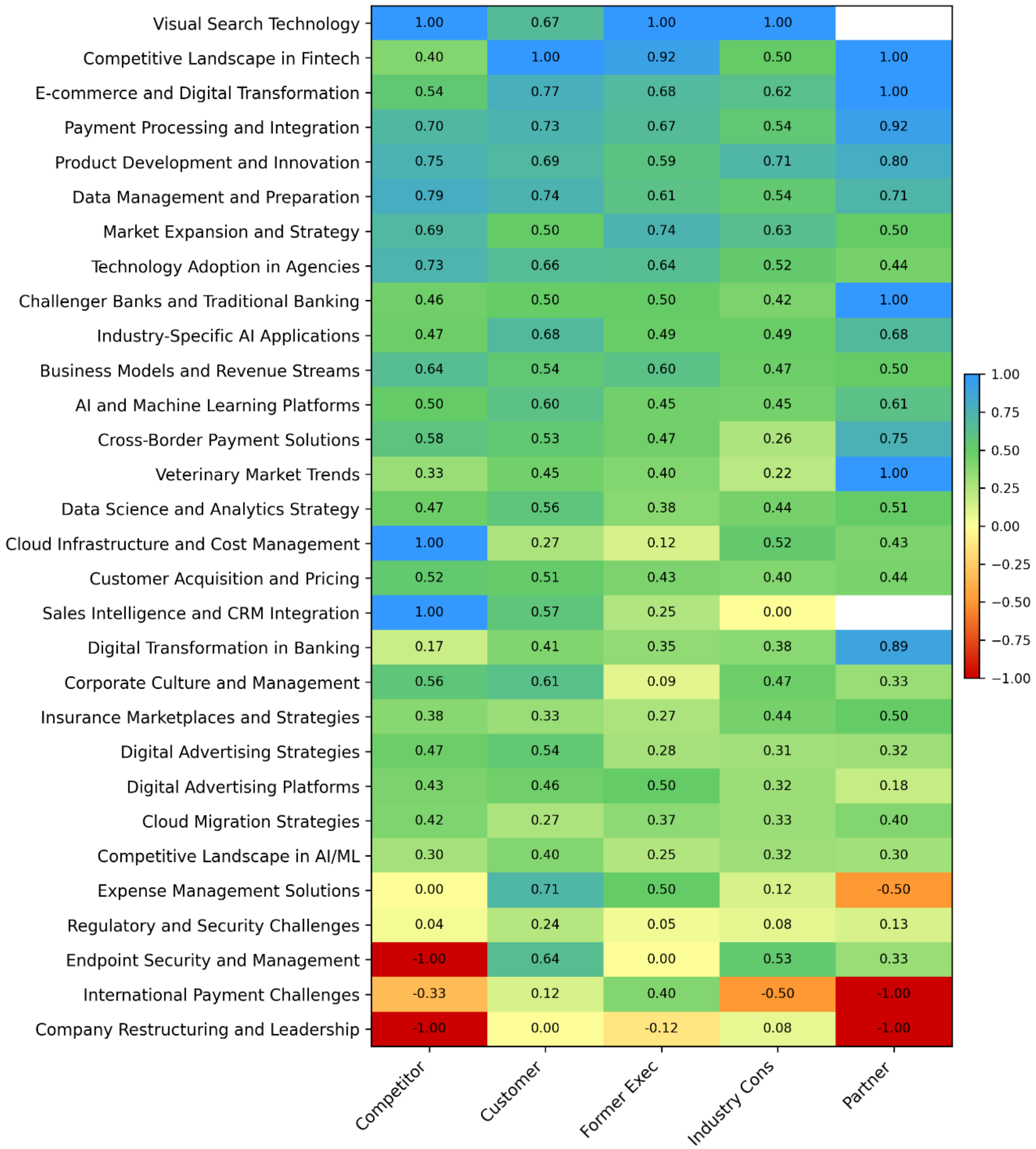

**Figure 4: The Generation Process of Our Proposed Bayesian Network**

This figure illustrates the sequential dependency structure of our Bayesian network, which models how conversational exchanges across expert network calls inform latent estimates of a firm's success rate. Each call $t$ produces question–answer pairs $(q^{t,k}, a^{t,k})$, summarized into a latent conversational state $s^t$. The state evolves through the sequence of calls, influencing the latent success variable $r^t$ and updating the firm-level success belief $y^{t+1}$. This dynamic generative process enables the model to infer and update the underlying success probability after each new expert network call. All notations are reported in Table A.1.

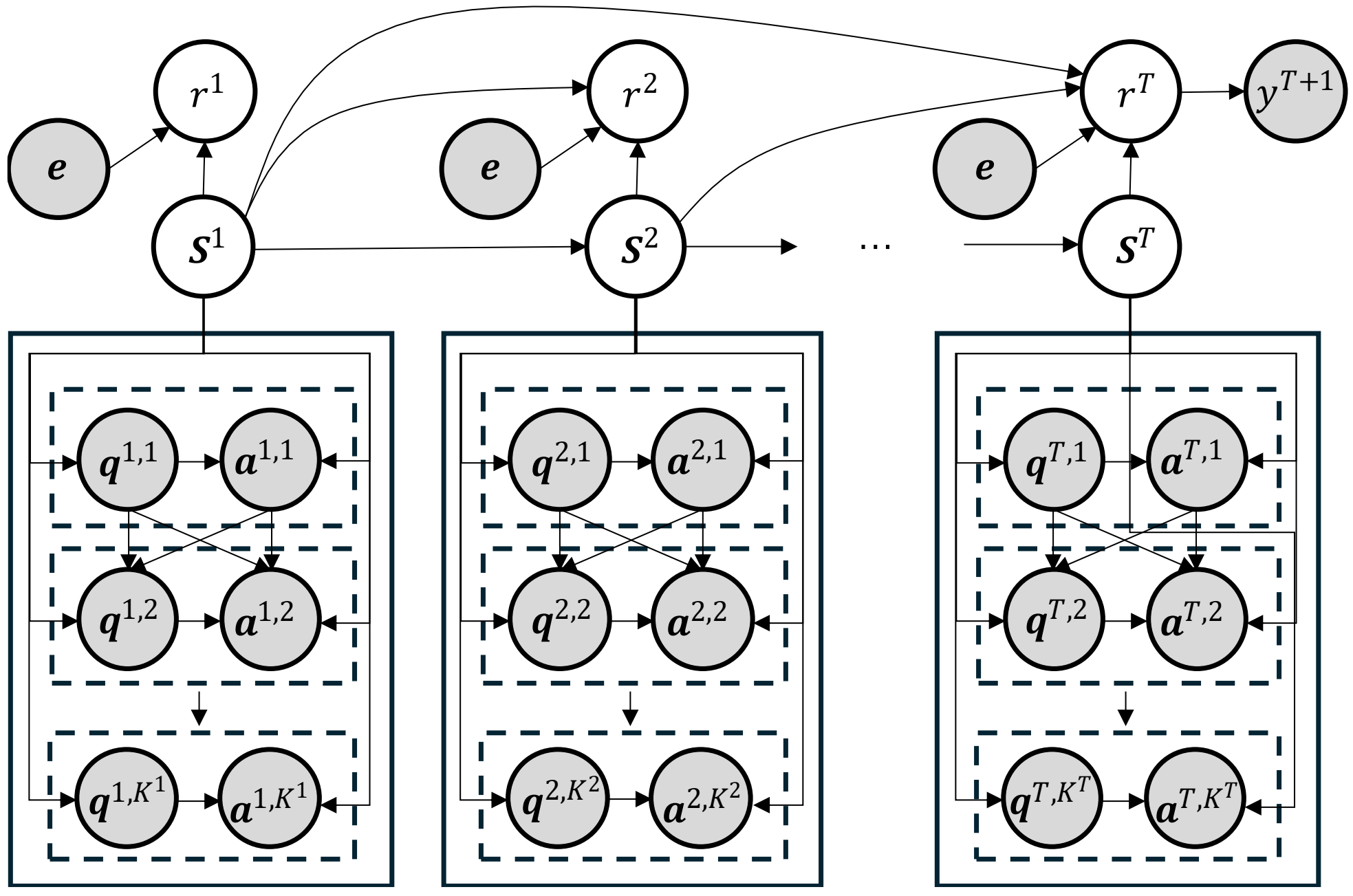

**Figure 5: Investor Uncertainty around Deal Quarters**

This figure presents event-time estimates of investor uncertainty around investment deals, as measured from expert network calls data. The horizontal axis shows event quarters relative to the deal quarter (quarter 0), and the vertical axis reports coefficient estimates from regressions of the uncertainty measure on event-quarter indicators. *Uncertainty* refers to the model-implied dispersion in beliefs about firm success (see Appendix Table A.2 for formal definition). Vertical error bars represent 95% confidence intervals, and the dashed vertical line marks the deal quarter.

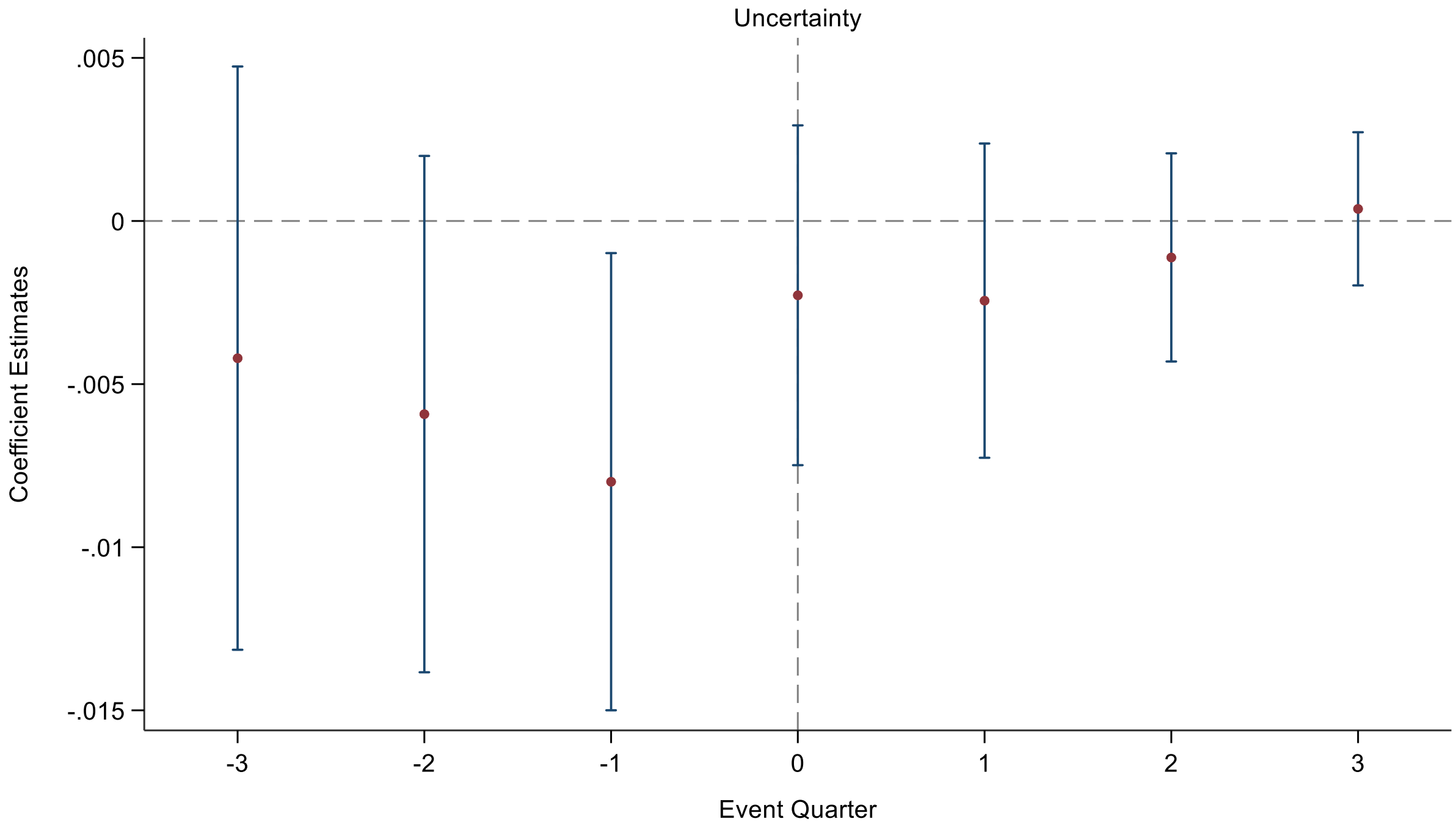

**Figure 6: Model Performance in the Top Tail of Investment Opportunities**

This figure compares the performance of our model with benchmark approaches in identifying high-quality investment opportunities. The horizontal axis ranks firms by predicted success probability and reports the fraction of top-ranked firms selected (e.g., 0.2 corresponds to the top 20% of firms and 0.4 corresponds to the top 40%), while the vertical axis shows realized investment outcomes. Panel A reports success rates, Panel B reports return on investment (ROI), and Panel C reports multiples of invested capital (MOIC).

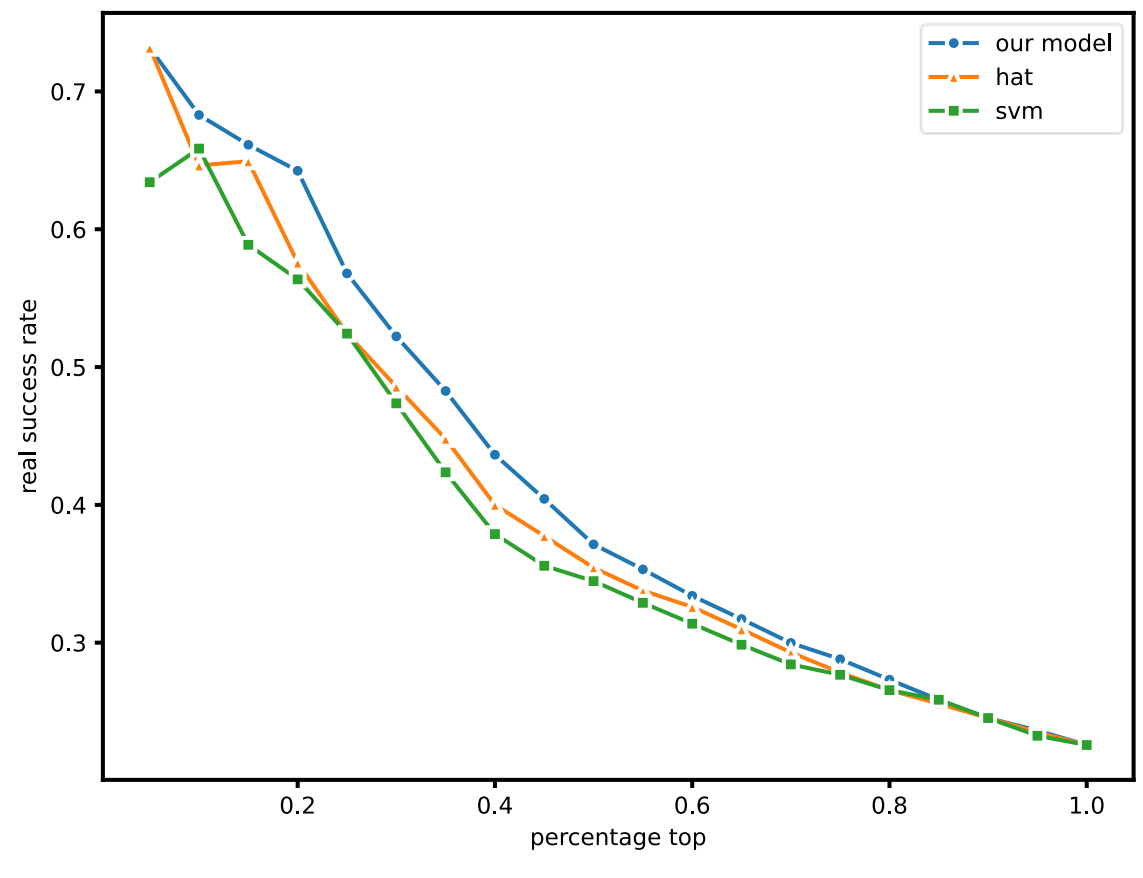

Panel A: Success Rate

Panel B: ROI

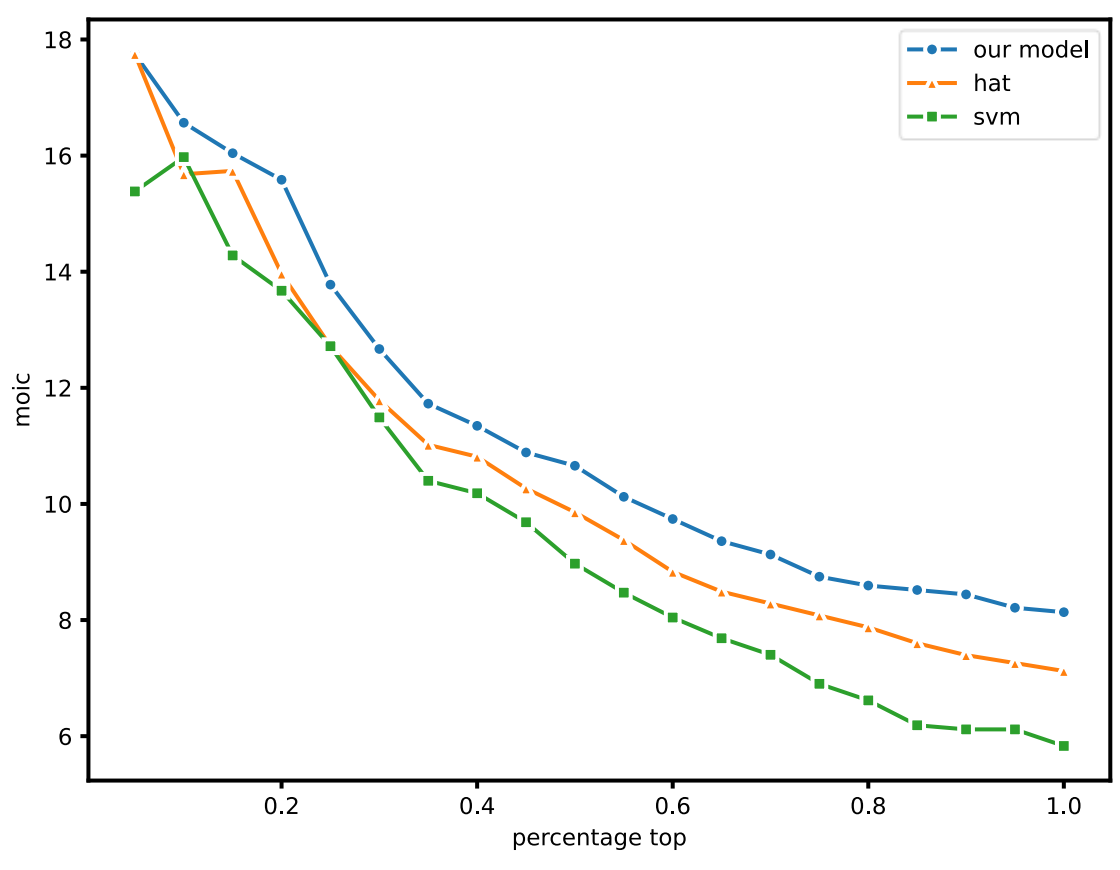

Panel C: MOIC

**Figure 7: Improving Portfolio Performance through Model-Based Reallocation**

This figure evaluates the economic gains from reallocating investment portfolios using our model. Starting from benchmark portfolios constructed using alternative selection methods (HAT and SVM), we sequentially replace a fraction of firms with those selected by our model based on predicted performance. The horizontal axis reports the fraction of firms in the benchmark portfolio that are replaced (e.g., 0.2 corresponds to replacing 20% of firms), while the vertical axis reports the realized return on investment (ROI) in Panel A and multiple of invested capital (MOIC) in Panel B. As the replacement fraction increases, portfolio performance improves monotonically, indicating that firms selected by our model systematically outperform those selected by benchmark methods. The gains are economically meaningful and persist across both benchmark portfolios. These results demonstrate that the model generates value through improved capital allocation rather than merely improving predictive accuracy, as replacing lower-quality investments with model-selected firms leads to higher realized returns.

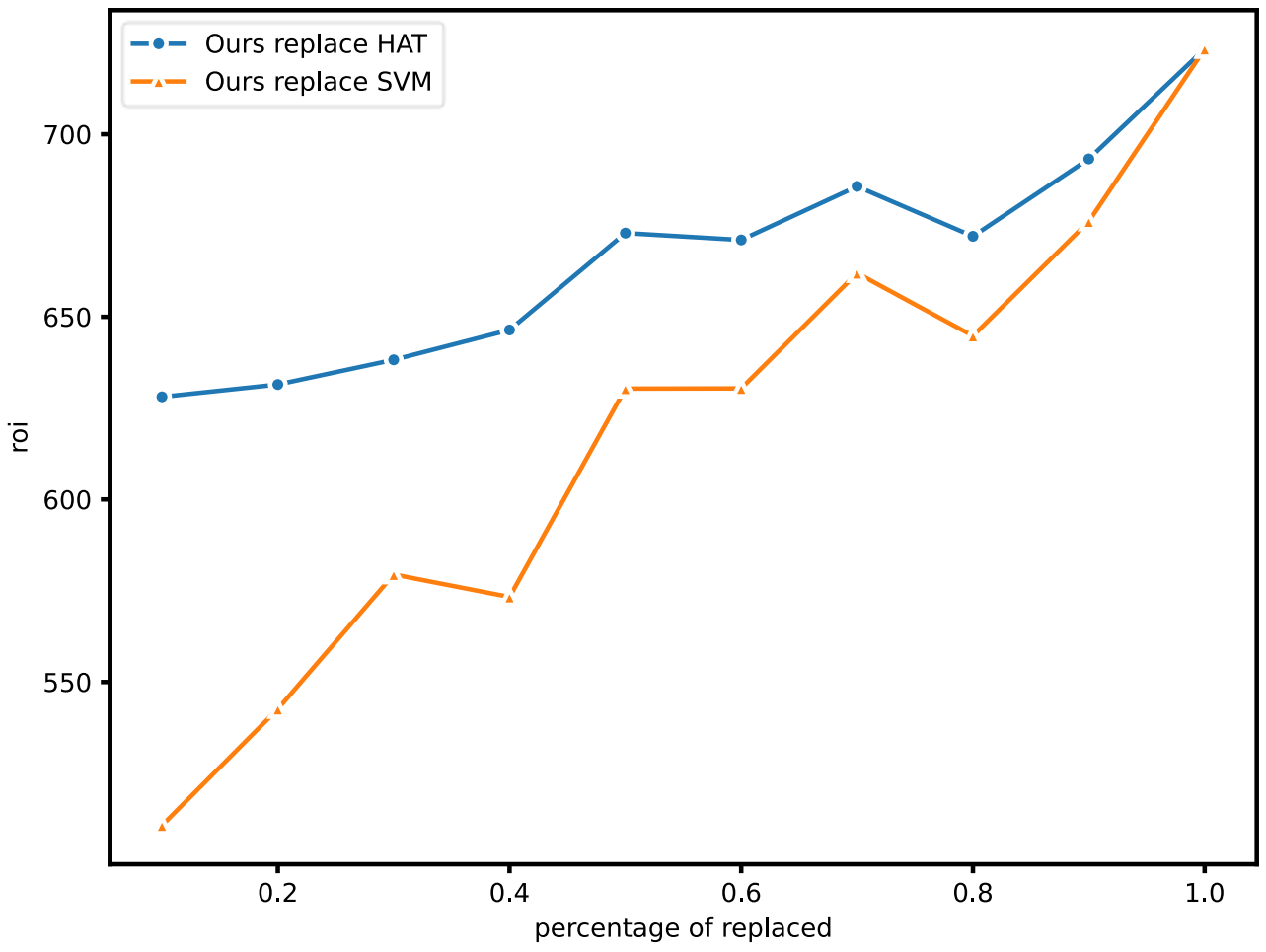

Panel A: ROI

Panel B: MOIC

**Figure 8: Attention Score by Expert Type and Topics**

This figure summarizes how attention is distributed across expert types and discussion topics. Panel (a) shows the average attention score by expert type, *Customers*, *Industry consultants*, *Former executives*, *Competitors*, and *Partners*. Panel (b) reports the distribution of attention across the most discussed topics, including *Regulatory and Security Challenges*, *Customer Acquisition and Pricing*, and *Industry-Specific AI Applications* as illustrative examples.

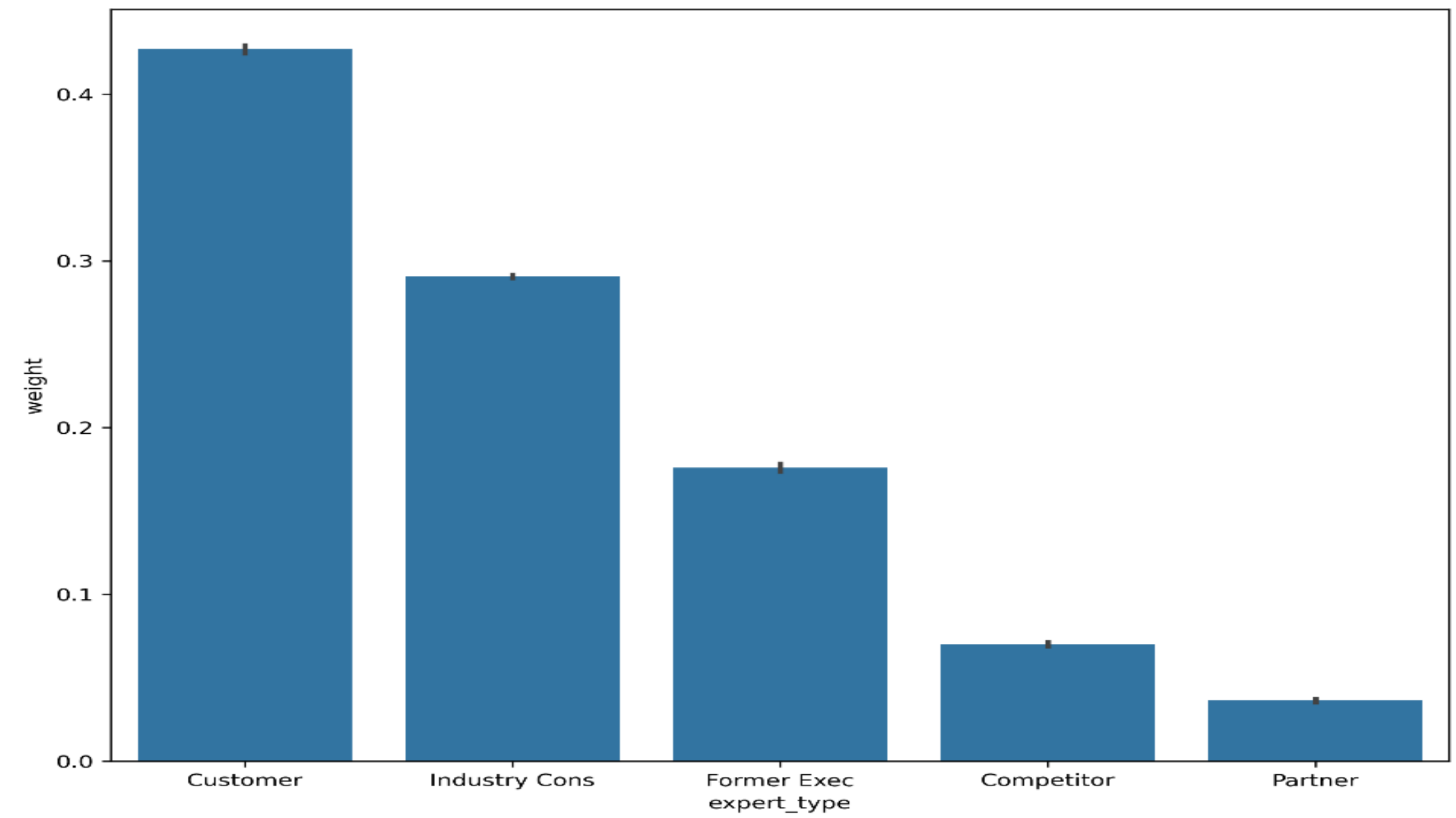

(a) Attention Score by Expert Type

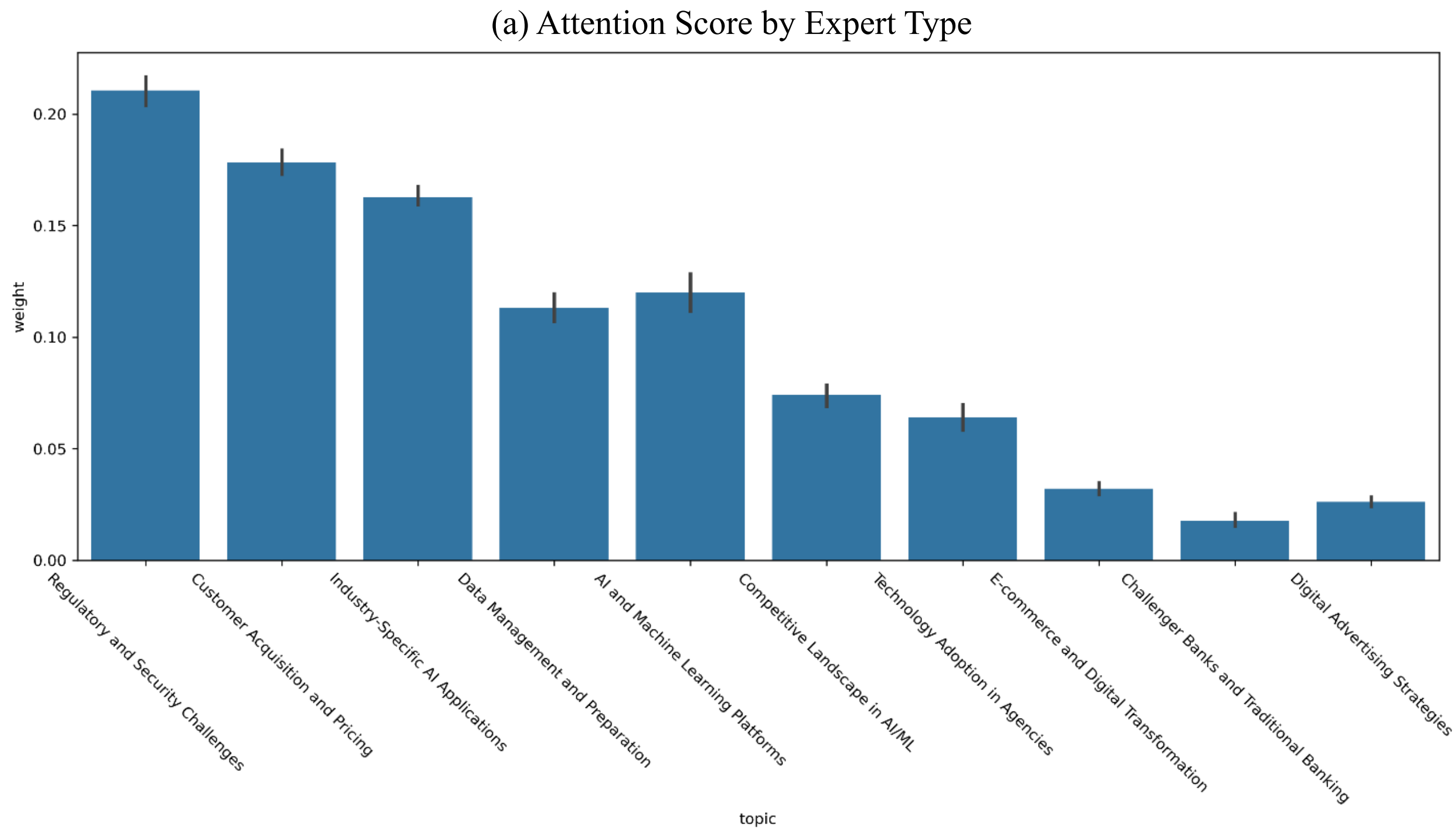

(b) Attention Score by Topics

**Figure 9: Attention Score by Topic and Sentiment**

This figure visualizes how expert attention varies across major discussion topics and sentiment orientations. Each cell represents the average model-derived attention score for a given topic–sentiment pair, with warmer colors indicating higher emphasis.

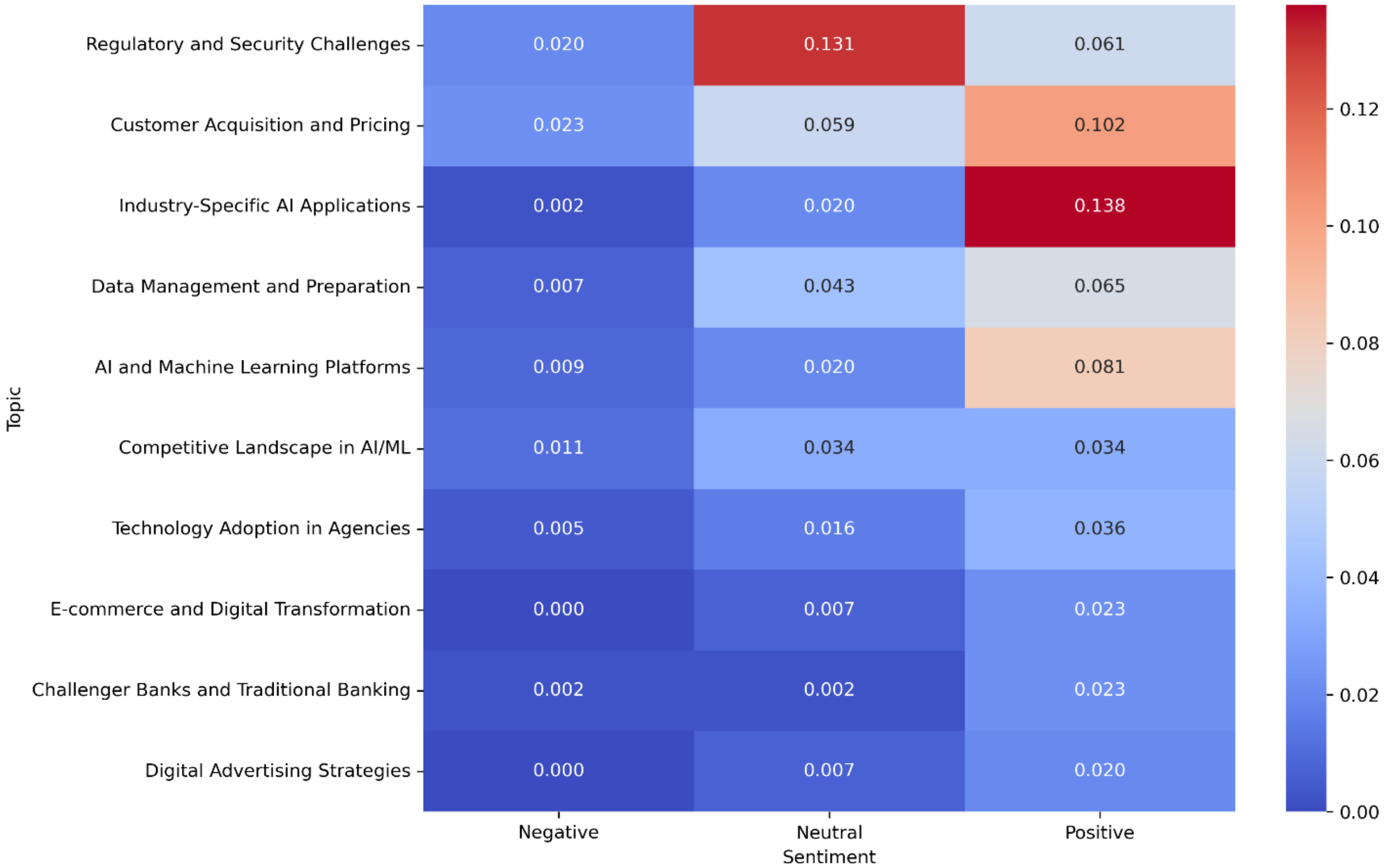

**Figure 10: Attention Scores for Each Topic versus Firm Characteristics**

This figure shows attention scores to key topics, (a) *Regulatory and Security Challenges*, (b) *Customer Acquisition and Pricing*, and (c) *Industry-Specific AI Applications*, varies with firm characteristics. For each topic, attention scores are plotted against *Funding rounds*, *Funding amount*, and *Stages*. Firm stage is defined based on the most recent funding round, ranging from early-stage, expansion, later stage, exit. Blue lines are attention score. Shaded areas represent 95% confidence intervals.

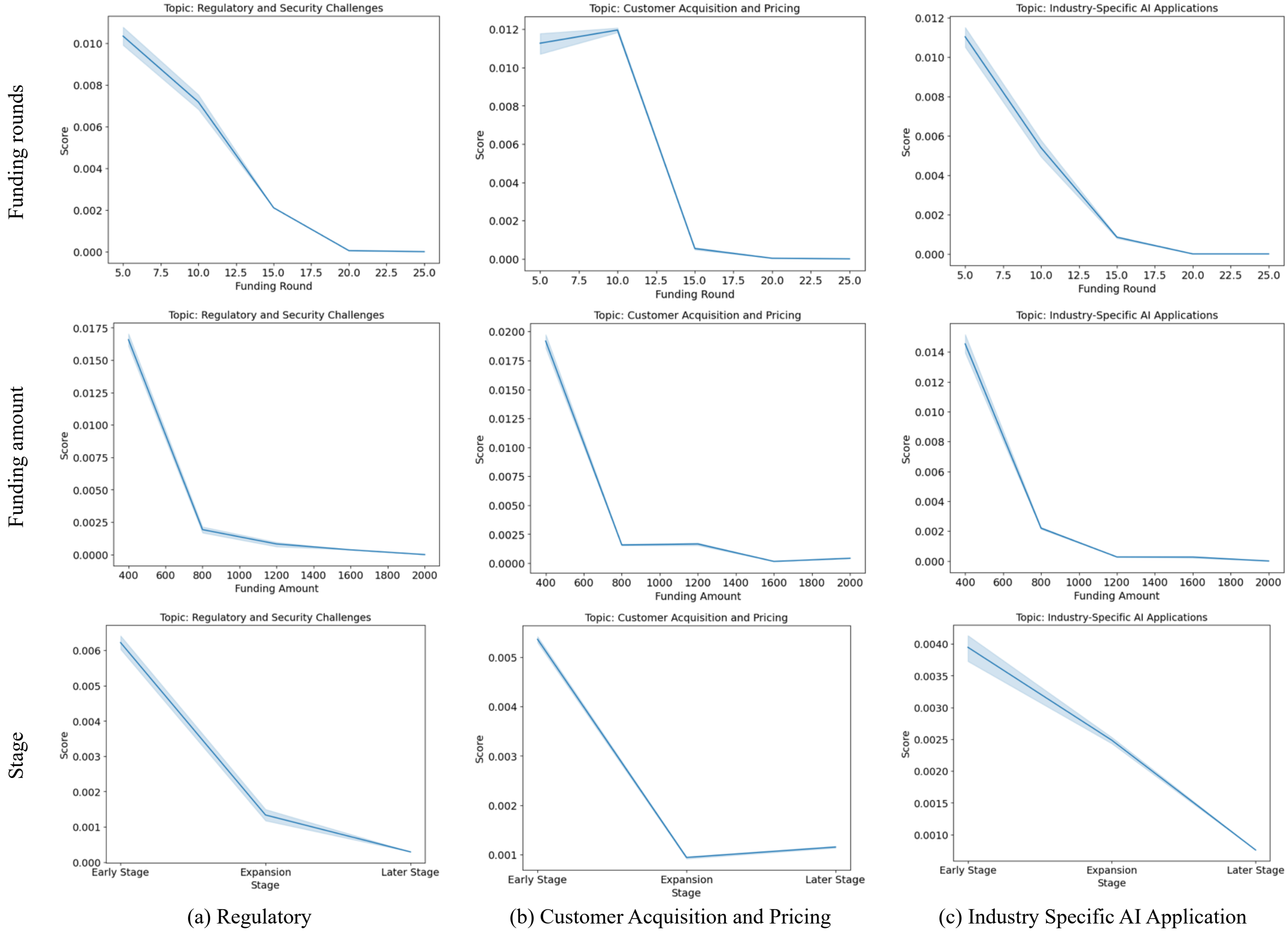

(a) Regulatory (b) Customer Acquisition and Pricing (c) Industry Specific AI Application

**Table 1: Summary Statistics**

This table presents descriptive statistics for the sample of expert network calls and firms included in the analysis. Panel A summarizes *Call Characteristics*, including *Number of words per transcript*, *Number of QA per transcript*, *Number of calls*, and *Number of words per transcript after gpt summary*. Panel B reports the *Expert Types Distribution* by their relationship to the target company. Panel C provides Firm Characteristics, including *Age*, *Founders count*, *Number of investment rounds*, *Raised funding*, *Investor count*, *Active products count*, and *IT spending*. Panel D compares firm characteristics between firms that ever received expert network calls during 2017-2024 versus those that never received calls. All variable definitions are reported in Table A.2.

Panel A: Call Characteristics

| Transcript | Min | p25 | p50 | p75 | Max | Mean |
|---|---|---|---|---|---|---|
| Number of words per transcript | 767 | 1330.323 | 1610.450 | 1730.935 | 16011 | 5879.730 |
| Number of QA per transcript | 2 | 31 | 59 | 111 | 333 | 73.726 |
| Number of calls | 1 | 1 | 2 | 5 | 5 | 2.769 |
| Number of words per transcript after gpt summary | 93 | 391.968 | 489.675 | 560.968 | 8808 | 2053.146 |

Panel B: Expert Types Distribution

| Expert Type | Percent |
|---|---|
| Competitor | 7.995 |
| Customer | 37.857 |
| Former exec | 19.095 |
| Industry cons | 31.899 |
| Partner | 3.155 |

Panel C: Firm Characteristics

| Variables | N | Min | p25 | p50 | p75 | Max | mean |
|---|---|---|---|---|---|---|---|
| Age (months) | 2214 | 7.033 | 73.475 | 103.667 | 148.592 | 279.367 | 117.099 |
| Founders count | 2214 | 1 | 2 | 2 | 3 | 11 | 2.394 |
| Number of investment rounds | 2214 | 1 | 4 | 5 | 7 | 40 | 5.796 |
| Raised funding (millions) | 2214 | 0.000 | 73 | 193.984 | 308.650 | 28500 | 435.384 |
| Investor count | 2214 | 1 | 7 | 11 | 18 | 124 | 14.138 |
| Active products count | 2214 | 2 | 21 | 27.365 | 37 | 139 | 30.979 |
| IT spending (millions) | 2214 | 0.000 | 2.748 | 11.429 | 30.014 | 33912.320 | 60.432 |

Panel D: Firm Characteristics of Expert Network Call Recipients

| Variable | Calls | No Calls | Diff. |
|---|---|---|---|
| Age (months) | 115.97 | 148.26 | 32.29*** |
| Raised funding (million) | 411.11 | 26.82 | -384.29*** |

**Table 2: Top 10 Transcript Topics by ChatGPT**

This table summarizes the ten most frequently discussed thematic categories in expert network transcripts, with concise descriptions generated through a LLM-based topic modeling pipeline. Each topic represents a distinct area of focus within expert discussions. Panel A presents topics and descriptions. Panel B presents summary statistics for sentiment measures in expert network calls. The first row shows overall sentiment statistics across all calls. Subsequent rows show topic-specific sentiment patterns for the 10 main topics identified through Topic analysis. Mean and standard deviation refer to the underlying continuous sentiment scores. Frequency shows the percentage of calls discussing each topic. Sentiment distribution shows the percentage of calls with negative, neutral, and positive sentiment for each topic. N shows the number of non-missing observations for each topic or overall sentiment. Overall sentiment frequency is 100% as it covers all calls, while individual topics appear in a subset of calls. Topic sentiment categories reflect the distribution of opinion valence across expert network calls.

Panel A: Topics and Descriptions

| Rank | Topic | Description |
|---|---|---|
| 1 | AI and Machine Learning Platforms | Examination of AI/ML platforms for data preparation, model building, and deployment, including the impact of low-code solutions on reducing the need for skilled data scientists. |
| 2 | Data Management and Preparation | Challenges and solutions in data ingestion, normalization, and preparation, highlighting tools and platforms that facilitate these processes. |
| 3 | Industry-Specific AI Applications | Use cases of AI across various industries, focusing on applications like customer churn analysis and healthcare diagnostics. |
| 4 | Competitive Landscape in AI/ML | Analysis of the competitive dynamics among AI/ML vendors and the influence of major cloud providers on market dynamics. |
| 5 | Regulatory and Security Challenges | Discussion on regulatory and security considerations affecting the adoption of AI/ML tools, particularly in sensitive sectors. |
| 6 | Digital Advertising Strategies | Exploration of strategies and platforms used for digital advertising, including the impact of external factors like COVID-19 on budgets and platform-specific insights. |
| 7 | Corporate Culture and Management | Analysis of corporate culture and management styles across tech companies, focusing on leadership, execution speed, and strategic vision. |
| 8 | Product Development and Innovation | Insights into product development strategies, including the role of acquisitions and leadership in driving innovation and market positioning. |
| 9 | Market Expansion and Strategy | Discussion on strategies for market expansion, including international growth, competitive dynamics, and customer acquisition. |
| 10 | Customer Acquisition and Pricing | Analysis of customer acquisition channels and pricing strategies, emphasizing the importance of market-specific pricing models. |

Panel B: Topic and Sentiment Summary Statistics

| Topic | Mean | SD | Freq% | Neg% | Neut% | Pos% |
|---|---|---|---|---|---|---|
| Overall Sentiment | 0.55 | 0.66 | 100.00 | 8.22 | 28.42 | 63.36 |
| T1 AI and Machine Learning Platforms | 0.70 | 0.61 | 14.91 | 6.67 | 17.29 | 76.04 |
| T2 Data Management and Preparation | 0.58 | 0.60 | 11.89 | 5.90 | 30.59 | 63.51 |
| T3 Industry-Specific AI Applications | 0.68 | 0.59 | 12.01 | 5.71 | 20.52 | 73.78 |
| T4 Competitive Landscape in AI/ML | 0.44 | 0.67 | 9.00 | 9.24 | 37.86 | 52.90 |
| T5 Regulatory and Security Challenges | 0.33 | 0.71 | 16.88 | 11.98 | 41.55 | 46.47 |
| T6 Digital Advertising Strategies | 0.74 | 0.55 | 1.53 | 5.32 | 14.89 | 79.79 |
| T7 Corporate Culture and Management | 0.14 | 0.81 | 1.75 | 23.36 | 39.25 | 37.38 |
| T8 Product Development and Innovation | 0.64 | 0.63 | 8.87 | 7.35 | 21.69 | 70.96 |
| T9 Market Expansion and Strategy | 0.57 | 0.65 | 7.68 | 7.86 | 26.54 | 65.61 |
| T10 Customer Acquisition and Pricing | 0.55 | 0.66 | 15.46 | 8.02 | 28.69 | 63.29 |

**Table 3: Predictive performance comparison across benchmark models**

This table compares the predictive performance of the proposed sequential learning framework with a range of benchmark models commonly used for startup success prediction. Panels A–C compare benchmarks that differ in feature construction, text representation, and modeling approach. Panel A compares the LLM-Bayesian framework with traditional machine learning and deep learning classifiers, Support Vector Machine (SVM), Logistic Regression (LR), Naïve Bayes (NB), and Multi-Layer Perceptron (MLP), using various combinations of engineered textual features (*topic*), call volume (*#call*), structured firm variables (*fv*), and expert network call transcripts (*text*). Panel B compares the LLM-Bayesian framework's predictive and economic performance with representative transformer-based long-text classification models applied to expert network call transcripts combined with structured firm variables (*fv+ text*). Benchmarked models include HAT, Hi-transformer, Longformer, and BigBirds, which differ in their hierarchical and sparse attention mechanisms. Panel C compares the LLM-Bayesian model's predictive and investment performance with recent text-based approaches to startup success prediction using comparable Crunchbase-derived inputs. Reported benchmarks include Maarouf et al. (2025), Katsafados et al. (2024), and Guzman and Li (2023), which combine textual and structured firm-level features (*fv+text*) or use text alone. Performance is evaluated using predictive metrics, *Accuracy*, *Precision*, *F1-score*, *Weighted F1*, *Macro F1*, and *AUC*. Economic metrics include *ROI* (Return on Investment) and *MOIC* (*Multiple on Invested Capital*), reflecting realized investor performance under model-guided portfolio allocation. All variable definitions are reported in Table A.2.

Panel A: Static Feature Models (Machine Learning Benchmarks)

| Baselines | Features | Accuracy | Precision | F1-score | Weighted F1 | Macro F1 | AUC | ROI | MOIC |
|---|---|---|---|---|---|---|---|---|---|
| SVM | topic | 0.630 | 0.337 | 0.446 | 0.660 | 0.584 | 0.641 | 382.391 | 4.824 |
| | #call | 0.698 | 0.391 | 0.476 | 0.717 | 0.632 | 0.666 | 358.147 | 4.581 |
| | fv | 0.601 | 0.340 | 0.480 | 0.632 | 0.578 | 0.677 | 577.222 | 6.772 |
| | topic + #call | 0.716 | 0.411 | 0.487 | 0.732 | 0.645 | 0.674 | 358.040 | 4.580 |
| | topic + fv | 0.649 | 0.369 | 0.501 | 0.678 | 0.615 | 0.695 | 592.482 | 6.925 |
| | topic + fv + #call | 0.759 | 0.473 | 0.536 | 0.769 | 0.686 | 0.709 | 480.816 | 5.808 |
| LR | topic | 0.637 | 0.333 | 0.431 | 0.665 | 0.582 | 0.627 | 290.574 | 3.906 |
| | #call | 0.752 | 0.461 | 0.507 | 0.752 | 0.461 | 0.686 | 327.497 | 4.275 |
| | fv | 0.641 | 0.358 | 0.484 | 0.670 | 0.604 | 0.679 | 533.876 | 6.339 |
| | topic + #call | 0.749 | 0.455 | 0.504 | 0.758 | 0.668 | 0.684 | 321.945 | 4.219 |
| | topic + fv | 0.672 | 0.386 | 0.513 | 0.699 | 0.633 | 0.705 | 603.926 | 7.039 |
| | topic + fv + #call | 0.731 | 0.437 | 0.530 | 0.748 | 0.670 | 0.710 | 546.179 | 6.462 |
| NB | topic | 0.632 | 0.337 | 0.444 | 0.662 | 0.585 | 0.639 | 366.127 | 4.661 |
| | #call | 0.714 | 0.409 | 0.487 | 0.730 | 0.644 | 0.674 | 367.291 | 4.673 |
| | fv | 0.545 | 0.301 | 0.433 | 0.578 | 0.526 | 0.624 | 454.551 | 5.546 |
| | topic + #call | 0.651 | 0.347 | 0.443 | 0.706 | 0.622 | 0.638 | 352.885 | 4.529 |
| | topic + fv | 0.634 | 0.336 | 0.440 | 0.664 | 0.585 | 0.635 | 349.309 | 4.493 |
| | topic + fv + #call | 0.685 | 0.389 | 0.500 | 0.709 | 0.635 | 0.690 | 518.688 | 6.187 |
| MLP | topic | 0.618 | 0.325 | 0.432 | 0.649 | 0.572 | 0.627 | 341.661 | 4.417 |
| | #call | 0.688 | 0.388 | 0.489 | 0.711 | 0.632 | 0.679 | 455.437 | 5.554 |
| | fv | 0.677 | 0.381 | 0.490 | 0.702 | 0.627 | 0.681 | 489.012 | 5.890 |
| | topic + #call | 0.689 | 0.391 | 0.496 | 0.712 | 0.636 | 0.685 | 487.499 | 5.875 |
| | topic + fv | 0.686 | 0.393 | 0.508 | 0.710 | 0.639 | 0.697 | 533.264 | 6.333 |
| | topic + fv + #call | 0.737 | 0.448 | 0.551 | 0.754 | 0.682 | 0.729 | 639.874 | 7.399 |
| **Ours** | fv+text | **0.777** | **0.505** | **0.590** | **0.789** | **0.718** | **0.753** | **723.350** | **8.230** |

Panel B: Deep Text Representation Models (Transformer Benchmarks)

| Models | Features | Accuracy | Precision | F1-score | Weighted F1 | Macro F1 | AUC | ROI | MOIC |
|---|---|---|---|---|---|---|---|---|---|
| HAT | fv + text | 0.745 | 0.458 | 0.553 | 0.761 | 0.688 | 0.729 | 627.610 | 7.280 |
| Hi-transformer | fv + text | 0.756 | 0.469 | 0.532 | 0.767 | 0.683 | 0.706 | 462.850 | 5.630 |
| Longformer | fv + text | 0.660 | 0.376 | 0.504 | 0.688 | 0.623 | 0.697 | 585.300 | 6.850 |
| Bigbirds | fv + text | 0.645 | 0.366 | 0.498 | 0.674 | 0.612 | 0.693 | 601.810 | 7.120 |
| **Ours** | fv + text | **0.777** | **0.505** | **0.590** | **0.789** | **0.718** | **0.753** | **723.350** | **8.230** |

Panel C: Existing Text-Based Startup Prediction Approaches

| | | Accuracy | Precision | F1-score | Weighted F1 | Macro F1 | AUC | ROI | MOIC |
|---|---|---|---|---|---|---|---|---|---|
| Maarouf et al. (2025) | fv+text | 0.746 | 0.452 | 0.511 | 0.757 | 0.670 | 0.690 | 376.917 | 4.769 |
| Katsafados et al. (2024) | fv+text | 0.757 | 0.471 | 0.528 | 0.767 | 0.683 | 0.702 | 437.973 | 5.380 |
| Guzman and Li (2023) | text | 0.598 | 0.318 | 0.434 | 0.631 | 0.561 | 0.628 | 385.065 | 4.851 |
| **Ours** | fv+text | **0.777** | **0.505** | **0.590** | **0.789** | **0.718** | **0.753** | **723.350** | **8.230** |

**Table 4: Architecture Validation: Sequential Learning from Expert Signals**

This table evaluates the contribution of different components of the proposed LLM-Bayesian framework for interpreting expert network calls signals and updating beliefs about startup success. Each row reports results from a model variant that removes or isolates a key component of the sequential learning framework. Predictive metrics include *Accuracy*, *Precision*, *F1-score*, *Weighted F1*, *Macro F1*, and *AUC*. Economic metrics include *ROI* (*Return on Investment*) and *MOIC* (*Multiple on Invested Capital*), reflecting realized investor performance under model-guided portfolio allocation. Row (1) *No Signal Dependence (Static Aggregation)* treats expert network calls as independent signals and aggregates their information without modeling relationships across calls. This corresponds to a static signal aggregation approach commonly used in standard machine learning models. Row (2) *Sequential Learning Only* captures temporal ordering across expert network calls and updates beliefs sequentially as new signals arrive but does not model richer structure within individual calls. Row (3) *Within-Call Signal Structure Only* captures relationships among signals within each expert network call but ignores the sequential dependence across calls. Row (4) *No Belief Updating* removes the latent belief updating component that dynamically updates the estimated success probability as new expert signals arrive. This specification evaluates the importance of modeling investor learning through sequential Bayesian updating. Row (5) *No LLM Signal Extraction* removes the LLM-based knowledge extraction component and replaces it with simpler signal representations, thereby isolating the contribution of AI-based interpretation of expert network calls content. All variable definitions are reported in Table A.2.

| Variants | Accuracy | Precision | F1-score | Weighted F1 | Macro F1 | AUC | ROI | MOIC |
|---|---|---|---|---|---|---|---|---|
| (1) No Signal Dependence (Static Aggregation) | 0.627 | 0.315 | 0.402 | 0.655 | 0.566 | 0.601 | 171.360 | 2.714 |
| (2) Sequential Learning Only | 0.593 | 0.315 | 0.420 | 0.623 | 0.551 | 0.614 | 310.260 | 4.103 |
| (3) Within-Call Signal Structure Only | 0.601 | 0.307 | 0.408 | 0.633 | 0.553 | 0.604 | 262.820 | 3.628 |
| (4) No Belief Updating | 0.768 | 0.490 | 0.580 | 0.781 | 0.710 | 0.747 | 699.249 | 7.993 |
| (5) No LLM Signal Extraction | 0.756 | 0.472 | 0.562 | 0.770 | 0.696 | 0.734 | 640.722 | 7.407 |
| **Ours** | **0.777** | **0.505** | **0.590** | **0.789** | **0.718** | **0.753** | **723.350** | **8.230** |

**Table 5: Dynamic Updating of Predictive Performance with Successive Expert Network Calls**

This table reports how the LLM-Bayesian model's predictive and economic performance evolves as additional expert network calls are incorporated into the training sequence. Each row represents the model's performance using transcripts from only the first call (*First call*), the first half of calls (*50% calls*), and all available calls (*All calls*) for each firm. Predictive metrics include *Accuracy*, *Precision*, *F1-score*, *Weighted F1*, *Macro F1*, and *AUC*. Economic metrics include *ROI* (*Return on Investment*) and *MOIC* (*Multiple on Invested Capital*), reflecting realized portfolio performance under model-guided investment. All variable definitions are reported in Table A.2.

| | Accuracy | Precision | F1-score | Weighted F1 | Macro F1 | AUC | ROI | MOIC |
|---|---|---|---|---|---|---|---|---|
| First call | 0.769 | 0.491 | 0.541 | 0.7772 | 0.695 | 0.710 | 461.568 | 5.616 |
| 50% calls | 0.777 | 0.504 | 0.560 | 0.785 | 0.7051 | 0.724 | 546.000 | 6.46 |
| All calls | **0.777** | **0.505** | **0.590** | **0.789** | **0.718** | **0.753** | **723.350** | **8.230** |

**Table 6: Exemplary Transcript by Expert Type**

This table presents representative excerpts from expert network call transcripts illustrating how experts of different types discuss firms and their prospects. Each excerpt includes an anonymized question–answer exchange highlighting the expert's evaluative tone and qualitative reasoning. Positive (blue) and negative (red) segments indicate the LLM-identified sentiment associated with each statement. To preserve confidentiality, firm names mentioned during the calls are replaced with the placeholder "ABC". All variable definitions are reported in Table A.2.

| Expert type | Textual self-description | Outcome |
|---|---|---|
| Customer | Q: So why would someone need ABC then? A:..The integration with the Tray was flunky and personally, I would not recommend them.. | non-succcessful |
| Customer | Q: That makes sense. How is what they're doing unique? A: ABC, ... it has a lot of artificial intelligence built in, and it gives you the ability and a powerful dashboard to create rules and policies... | successful |
| Industry cons | Q: could you please give us a quick overview of your experience in this space? A:..it's money spent and screening people that ultimately don't enroll or even worse, enrolled and then drop out prematurely... ..That was a no for us really out of the gate ... | non-succcessful |
| Industry cons | Q: But how likely is it to actually be available at scale? A: ...this could get 50% to 100% improvement in energy density over state-of-the-art cells today...And that's an obviously huge marketing point for the technology in the startups... | successful |
| Former exec | Q: How does the engineering team compare with other companies you may have worked out ...? A: ... the pressure to get people in the absence of those professionals in the market forced the company to make some compromises in the sense that not always the best ones were hired. ..So maintaining them, it's very difficult. | non-succcessful |
| Former exec | Q: So from your experience at ABC, how did you guys frame yourself in the market? A: ABC goes one step beyond… | successful |
| Competitor | Q:.Were there any sort of real credible competitors or folks that you saw as being potential challengers in this segment? A:...But all of them are slower to the market...No one was effective. | non-succcessful |
| Competitor | Q: we are interested in the PIM space generally and are looking to get your perspective on some of the players. A: ...I mean this PIM business has definitely exploded...It's exciting... | successful |
| Partner | Q: is there any barrier to entry in entering the cloud cost optimization space? A:...Not necessarily… | non-succcessful |
| Partner | Q: And then maybe just turning to kind of your experience with ABC or …how satisfied were you? A: ...Very pleased. I would absolutely buy them again for an organization that I'm with now or will be in the future...I absolutely love working with them... | successful |

**Table 7: Marginal Value of Information from Expert Network Calls**

This table examines the marginal value of information generated by expert network calls during the due diligence process. The dependent variable is the change in investor uncertainty following each call, measured as $\Delta Uncertainty(\%) = 100 \times (Uncertainty_{t-1} - Uncertainty_{t})$, where $\mathrm{Uncertainty_t}$ represents the investor's posterior uncertainty and $\mathrm{Uncertainty_{t-1}}$ is the prior uncertainty before the call. A larger value of $\Delta Uncertainty(\%)$ therefore indicates a greater reduction in uncertainty and stronger belief updating. Panel A examines how the marginal value of information varies with the firm's information environment. *Opacity* measures the degree to which the firm's information environment is difficult for investors to interpret. Panel B examines heterogeneity in learning across expert types. We construct an indicator variable *Operation Expert*, which equals one if the expert is a customer, industry consultant, or former executive, and zero otherwise. Panel C examines heterogeneity in learning across discussion topics. *Technology Adoption Topic* is an indicator for calls that focus on technology adoption and implementation issues (Topic 1, 2, 3, and 4). Control variables include *Age(years)*, *Founders count*, *Number of investment rounds*, *Raised funding (millions)*, *Investor count*, *Active products count*, and *IT spending (millions)*. Standard errors are clustered by firm. *, **, *** denote significance at 10%, 5%, 1% levels respectively. All variable definitions are reported in Table A.2.

Panel A: Information Environment

| | ΔUncertainty(%) | | | |
|---|---|---|---|---|
| | (1) | (2) | (3) | (4) |
| $\mathrm{Uncertainty}_{t-1}$ | 7.235*** | 7.250*** | 7.028*** | 7.113*** |
| | (7.45) | (7.43) | (7.51) | (7.57) |
| Opacity | | 0.002 | -1.367** | -1.409** |
| | | (0.58) | (-1.98) | (-2.04) |
| $\mathrm{Uncertainty}_{t-1}$×Opacity | | | 1.991** | 2.041** |
| | | | (1.99) | (2.05) |
| Controls | N | N | N | Y |
| Quarter FE | Y | Y | Y | Y |
| Observations | 4,255 | 4,255 | 4,255 | 4,255 |
| Adjusted R-squared | 0.035 | 0.035 | 0.037 | 0.037 |

Panel B: Expert Type

| | ΔUncertainty(%) | |
|---|---|---|
| | (1) | (2) |
| $Uncertainty_{t-1}$ | 2.478** | 2.585** |
| | (2.38) | (2.41) |
| Operation Expert | -3.740*** | -3.737*** |
| | (-3.59) | (-3.58) |
| $Uncertainty_{t-1}$× Operation Expert | 5.428*** | 5.425*** |
| | (3.59) | (3.58) |
| Controls | N | Y |
| Quarter FE | Y | Y |
| Observations | 4,255 | 4,255 |
| Adjusted R-squared | 0.037 | 0.037 |

Panel C: Topic-Level Information

| | ΔUncertainty(%) | |
|---|---|---|
| | (1) | (2) |
| $Uncertainty_{t-1}$ | 7.246*** | 7.348*** |
| | (7.48) | (7.49) |
| Technology Adoption Topic | 0.018** | 0.018** |
| | (2.08) | (2.02) |
| Controls | N | Y |
| Quarter FE | Y | Y |
| Observations | 4,255 | 4,255 |
| Adjusted R-squared | 0.036 | 0.036 |

**Table 8: Expert Network Call Information, Belief Updating, and Capital Allocation**

This table examines how information extracted from expert network calls shapes investor belief formation and capital allocation decisions. The dependent variable $Deal_{t+1}$ is a binary indicator equal to one if a startup receives a financing event in quarter $t+1$. *Call* indicates if an expert consultation occurred at t. *Positive Sentiment* captures whether the call had positive sentiment. *Technology Adoption Topic* is an indicator for calls that focus on technology adoption and implementation issues (Topic 1, 2, 3, and 4) and *Technology Adoption Topic Positive* captures whether *Technology Adoption Topic* had positive sentiment. *Customer Acquisition Topic* is an indicator for calls that focus on Topic 10 and *Customer Acquisition Topic Positive* captures whether *Customer Acquisition Topic* had positive sentiment. *Overall Sentiment* is the overall sentiment score. We construct model-implied measures of investor beliefs and uncertainty based on expert network call transcripts. $Investor\ Success\ Belief_{it}$, represents investors' updated assessment of the likelihood of startup success following expert information aggregation, while uncertainty, $Uncertainty_{it}$, represents the investor's posterior uncertainty. We also measure learning as the change in uncertainty, $\Delta Uncertainty(\%) = 100 \times (Uncertainty_{t-1} - Uncertainty_t)$, where larger values indicate greater reductions in uncertainty and stronger information acquisition. Panel A examines the relationship between baseline text-based signals from expert network calls and subsequent deal outcomes, including call occurrence, sentiment, and topic indicators. Panel B investigates how model-implied beliefs and uncertainty map into investment decisions, and whether these variables subsume the explanatory power of traditional text-based measures. Panel C examines the state-dependent value of information by interacting uncertainty reduction with prior uncertainty, testing whether learning is more valuable when initial uncertainty is high. Control variables include *Age* and the logarithm of one *Raised funding (millions)*. Standard errors are clustered by firm. *, **, *** denote significance at 10%, 5%, 1% levels respectively. All variable definitions are reported in Table A.2.

Panel A: Baseline Text Signals and Deal Outcomes

| | $Deal_{t+1}$ | | | |
|---|---|---|---|---|
| | (1) | (2) | (3) | (4) |
| Call | 0.069*** | 0.090*** | | |
| | (4.22) | (5.36) | | |
| Positive Sentiment | 0.039** | 0.041** | | |
| | (2.01) | (2.05) | | |
| Technology Adoption Topic | | | 0.108*** | |
| | | | (8.23) | |
| Customer Acquisition Topic | | | 0.132*** | |
| | | | (5.30) | |
| Technology Adoption Topic Positive | | | | 0.114*** |
| | | | | (7.54) |
| Customer Acquisition Topic Positive | | | | 0.147*** |
| | | | | (4.83) |
| Observations | 2,504,490 | 2,097,472 | 2,097,472 | 2,097,472 |
| Adjusted R-squared | 0.018 | 0.023 | 0.023 | 0.023 |
| Controls | N | Y | Y | Y |
| Firm FE | Y | Y | Y | Y |
| Quarter FE | Y | Y | Y | Y |

Panel B: Model-Implied Beliefs, Uncertainty, and Investment Decisions

| | $Deal_{t+1}$ | | | |
|---|---|---|---|---|
| | (1) | (2) | (3) | (4) |
| Overall Sentiment | 0.049*** | 0.020 | | |
| | (3.20) | (1.29) | | |
| Technology Adoption Topic | 0.081*** | 0.009 | | |
| | (5.04) | (0.47) | | |
| Customer Acquisition Topic | 0.115*** | 0.034 | | |
| | (4.48) | (1.26) | | |
| Investor Success Belief | | 2.195*** | 2.223*** | 6.625*** |
| | | (11.11) | (11.23) | (3.52) |
| Uncertainty | | -1.421*** | -1.408*** | -1.345*** |
| | | (-10.25) | (-10.18) | (-9.73) |
| Investor Success Belief× Uncertainty | | | | -6.530** |
| | | | | (-2.43) |
| Observations | 2,097,472 | 2,097,472 | 2,097,472 | 2,097,472 |
| Adjusted R-squared | 0.023 | 0.023 | 0.023 | 0.023 |
| Controls | Y | Y | Y | Y |
| Firm FE | Y | Y | Y | Y |
| Quarter FE | Y | Y | Y | Y |

Panel C: State-Dependent Learning and the Value of Information

| | $Deal_{t+1}$ | |
|---|---|---|
| | (1) | (2) |
| Investor Success Belief | 2.221*** | 2.221*** |
| | (11.23) | (11.23) |
| ΔUncertainty(%) | -2.505** | -2.519** |
| | (-2.23) | (-2.24) |
| Uncertainty | -1.404*** | -1.420*** |
| | (-10.15) | (-10.27) |
| ΔUncertainty(%)×Uncertainty | 3.720** | 3.741** |
| | (2.26) | (2.27) |
| Overall Sentiment | | 0.021 |
| | | (1.39) |
| Observations | 2,097,472 | 2,097,472 |
| Adjusted R-squared | 0.023 | 0.023 |
| Controls | Y | Y |
| Firm FE | Y | Y |
| Quarter FE | Y | Y |

**Table 9: Heterogeneity Analysis of Model Performance Across Firm Subgroups**

This table evaluates whether the LLM-Bayesian model's incremental predictive and economic performance varies across key firm subgroups. *Precision* measures the fraction of correctly predicted successful startups, while *ROI* (*Return on Investment*) and *MOIC* (*Multiple on Invested Capital*) reflect realized portfolio profitability under model-guided investment. *ΔROI* and *ΔMOIC* denote improvements relative to a baseline model. Panel A partitions firms by *Industry complexity*, classified through a two-stage text procedure combining keyword heuristics and LLM validation. Industries referencing advanced technologies (e.g., AI, software, biotech) are labeled *High*, while traditional sectors are labeled *Low*. Panel B partitions firms by *Firm age*, defining *Young* firms as those less than six years old, and *Mature* firms as those six years or older. Panel C partitions firms by *Public visibility*, proxied by domain age. Domains younger than six years are labeled *Low*, while older domains are labeled *High*. Panel D partitions firms by *Founder diversity*, capturing gender and ethnic minority representation. All variable definitions are reported in Table A.2.

| | | LLM-Bayesian | | | | |
|---|---|---|---|---|---|---|
| | | Precision | ROI | MOIC | ΔROI vs Baseline | ΔMOIC vs Baseline |
| Panel A: Industry complexity | | | | | | |
| (1) | High | 0.518 | 774.770 | 8.748 | 106.260 | 1.063 |
| (2) | Low | 0.313 | 29.220 | 1.290 | 8.070 | 0.079 |
| Panel B: Firm age | | | | | | |
| (3) | Young | 0.561 | 976.710 | 10.767 | 268.300 | 2.683 |
| (4) | Mature | 0.495 | 678.320 | 7.783 | 105.430 | 1.054 |
| Panel C: Public visibility | | | | | | |
| (5) | Low | 0.600 | 1085.790 | 11.858 | 336.920 | 3.369 |
| (6) | High | 0.504 | 703.660 | 8.037 | 152.420 | 1.524 |
| Panel D: Founder diversity | | | | | | |
| (7) | High | 0.605 | 915.300 | 10.153 | 203.120 | 2.031 |
| (8) | Low | 0.497 | 579.060 | 6.791 | 32.320 | 0.323 |

# Internet Appendix for "Measuring Investor Learning in Private Markets: A Sequential LLM-Bayesian Analysis of Expert Network Calls"
# (Not to be published)

## Appendix A. I: ChatGPT-Based Topic Modeling Pipeline for Expert Network Call Transcripts

In this appendix, we outline the technical details and prompt design underlying our three-step ChatGPT-based topic modeling framework applied to expert network call transcripts. The process is designed to enable large language models (LLMs) to identify, refine, and evaluate thematic structures within each transcript while preserving conversational context and ensuring topic consistency across the corpus.

### Step 1. Topic Generation

In the first step, ChatGPT is instructed to generate initial topics for each transcript individually. Each transcript is provided as input text, and the model outputs a list of concise thematic labels and corresponding summaries that capture the main discussion themes. The prompt is designed to balance interpretability and coverage by constraining the model to return no more than 8–10 core topics per transcript.

### Prompt Example (Step 1: Topic Generation)

Task Description
As part of a research initiative in entrepreneurial finance, we examine how early-stage investors approach their decision-making. Our dataset consists of transcripts from consultation calls between prospective early-stage investors (hereafter referred to as "Clients") and domain experts who possess deep knowledge of the companies under discussion. The transcripts will be analyzed individually to extract key thematic information.
Your Task
For the transcript below, identify the primary, top-level topics discussed. Focus on broadly relevant themes that capture the call's main focus and the clients' priorities.
Instructions
- Identify ONLY the primary, high-level topics central to this call.
- Avoid overly technical or company-specific topics.
- Keep labels concise and include a 1–2 sentence description for each topic.

This step yields an initial topic set for each transcript. The outputs are cached for subsequent refinement to ensure computational efficiency and reproducibility.

### Step 2. Topic Refinement and Global Topic Alignment

In the second step, all transcript-level topics are aggregated to identify the global top-30 themes across the entire dataset. The purpose is to merge semantically overlapping or synonymous topics (e.g., "AI adoption in finance" and "AI-driven financial innovation") and to construct a unified taxonomy of discussion categories.

ChatGPT is used to cluster, rename, and refine the topics for global consistency. The process ensures that topics assigned in later stages map to a stable and interpretable set of categories.

**Prompt Example (Step 2: Topic Refinement)**

Task Description
As part of an academic research project in entrepreneurial finance, we aim to better understand the key themes that shape how early-stage investors make decisions. To support this goal, we have compiled transcripts from consultation calls between investors and domain experts who are familiar with the companies. For each call, we have extracted a set of preliminary topic labels.
Your Task
Your job is to **clean and refine** these topic lists by identifying and merging topics that are duplicates, near-duplicates, or semantically overlapping. You should also adjust titles and descriptions for clarity and generality. If the provided topics are already clear and distinct, return "None" to indicate that no refinement is required.
This step represents the **topic refinement stage** in a multi-step topic modeling workflow.
Guidelines for merging topics:
- Merge topics that describe the same underlying idea or process, even if worded differently or from different industries.
- When one topic is more specific and another is broader, keep the broader one.
- Use concise, general titles (no company or product names).
- Avoid vague names like "Merged Topic."
- Write definitions that describe the general concept, not company-specific details.
Illustrative Merging Examples
- Customer Support Platforms + Customer Service Software →
Customer Service Infrastructure: Software systems for managing tickets, live chat, and client communication.
- Leadership Pipeline Development + Internal Talent Acceleration →
Leadership Development Programs: Strategies to identify, train, and promote internal talent.
- Nike Digital Campaigns + Adidas Online Marketing Strategy →
Digital Marketing Strategies: How firms use online platforms and partnerships to engage consumers.

This refinement step yields a stable taxonomy, which we use to assign consistent topic labels back to each transcript.

**Step 3. Topic Assignment and Sentiment Evaluation**

Finally, we re-evaluate each transcript and assign it to one or more of the refined topics identified in Step 2. ChatGPT receives both the transcript content and the refined topic list as context and determines which topics are most relevant to the discussion. The model also provides a sentiment evaluation (positive, neutral, or negative) for each topic occurrence within the transcript.

**Prompt Example (Step 3: Topic Assignment and Sentiment Evaluation)**

Task Description

Analyze the expert–client conversation to identify key discussion topics from the list below. Use only the exact topic names. For each relevant topic, assign a sentiment score (−2 to +2) and give short supporting evidence.
Sentiment Scale
+2 Strongly Positive — clear enthusiasm, major benefits
+1 Positive — favorable tone
0 Neutral — factual or balanced
−1 Negative — some concerns
−2 Strongly Negative — serious criticism or problems
Examples
(+2) "The Asia partnership doubled client volume." → Strongly Positive (+2); rapid regional growth.
(−2) "System outages caused cancellations and refunds." → Strongly Negative (−2); repeated failures, lost trust.
(0) "The firm publishes quarterly ESG updates." → Neutral (0); factual reporting.

This step generates the final per-transcript topic-sentiment dataset used for downstream quantitative analysis, including topic frequency distribution, co-occurrence patterns, and sentiment correlation with firm outcomes.

## Appendix A. II: Technical Details of the Sequential Bayesian Learning Framework

### 1 Overview of the Learning Process

A key challenge in learning Bayesian networks arises from the presence of latent variables. Specifically, our Bayesian network involves latent variables $r^t$ and $\boldsymbol{s}^t$ for $t = 1, \dots, T$. We denote them as $r^{1:T}$ and $\boldsymbol{s}^{1:T}$. A straightforward way would be to infer their distributions, treat the latent variables as observed according to these inferred distributions to obtain the objective, and then learn the model parameters in the usual manner to maximize the obtained objective. In the generative process, we have specified assumptions about how $r^t$ and $\boldsymbol{s}^t$ are distributed, but these assumptions are made without considering the observed data and thus serve only as priors. We denote the prior distribution of the latent variable as $p^{\text{prior}}(r^{1:T}, \boldsymbol{s}^{1:T})$. The actual distributions of $r^t$ and $\boldsymbol{s}^t$ must be updated based on the observed data to obtain the posterior distributions, which is denoted as $p(r^{1:T}, \boldsymbol{s}^{1:T} | \boldsymbol{A}^{1:T}, \boldsymbol{Q}^{1:T}, \boldsymbol{e}, y)$. However, the true posterior of the latent variables is generally intractable. To address this, we adopt a variational inference approach, introducing a variational distribution to approximate the true posterior distribution of the latent variables. Note that $r^t$ is deterministically determined given $\boldsymbol{s}^{1:t}$ and the observed variable $\boldsymbol{e}$. Therefore, we only need to introduce a variational distribution for $\boldsymbol{s}^{1:t}$, denoted as $q(\boldsymbol{s}^{1:T})$, to approximate the true posterior $p^{\text{post}}(\boldsymbol{s}^{1:T})$, while taking into account the prior distribution $p^{\text{prior}}(\boldsymbol{s}^{1:T})$. The variational distribution should have desirable properties to allow for tractable computation. Next, we introduce our variational distributions.

### 2 Variational Distribution and Inference Networks

Since we hope to infer the distribution of latent variables based on the observed data, the input to the variational distribution consists of the observed variables: the environmental factors $\boldsymbol{e}$, all extracted answer vectors in the expert network call (denoted as $\boldsymbol{A}^{1:T}$), all extracted question vectors in the expert network call (denoted as $\boldsymbol{Q}^{1:T}$) and the final outcome $y$. Then, the variation distribution $q(\boldsymbol{s}^{1:T})$ can be written as $q(\boldsymbol{s}^{1:T} | \boldsymbol{A}^{1:T}, \boldsymbol{Q}^{1:T}, \boldsymbol{e}, y)$.

Given the dependence relationships in the Bayesian network, we have

$$p^{\text{prior}}(\boldsymbol{s}^{1:T}) = p^{\text{prior}}(\boldsymbol{s}^1) \prod_{t=2}^{T} p^{\text{prior}}(\boldsymbol{s}^t | \boldsymbol{s}^{t-1}) \qquad (\text{A}.1)$$

Hence, we also follow the same dependency structure to design our variational distribution $q$:

$$q(\boldsymbol{s}^{1:T}) = q(\boldsymbol{s}^1) \prod_{t=2}^{T} q(\boldsymbol{s}^t | \boldsymbol{s}^{t-1}) \qquad (\text{A}.2)$$

We next introduce each factor of the variational distribution. First, for $\boldsymbol{s}^1$, it directly depends on $\boldsymbol{Q}^1$, $\boldsymbol{A}^1$ and $y$. Hence, $q(\boldsymbol{s}^1)$ should be determined by these three factors. However, during the testing phase, the inference network must estimate the latent status while $y$ remains unobserved. Therefore, it is inappropriate to use $y$ as an input to the variational distribution. To address this issue, we design a variational distribution $q(\boldsymbol{s}^1)$ that takes $\boldsymbol{Q}^1$ and $\boldsymbol{A}^1$ as input, while being constrained by $y$ during training. In this way, the determination of $q(\boldsymbol{s}^1)$ considers $y$, while still allowing $q(\boldsymbol{s}^1)$ to be inferred during the testing phase without requiring $y$.

Given that $\boldsymbol{s}^1$ is a continuous vector, we assume it follows a Gaussian distribution, whose mean value $\mu_q(\boldsymbol{s}^1)$ needs to be inferred. Leveraging the strength of neural networks in complex inference tasks, we use a neural network to estimate this mean. Furthermore, considering the structural characteristics of $\boldsymbol{Q}^1$ and $\boldsymbol{A}^1$, the inference network must effectively capture their sequential dependencies for improved predictive accuracy. To this end, we employ an attention mechanism to analyze the sequence of question-answer pairs within the conversation. Formally,

$$\mu_q(\boldsymbol{s}^1) = \text{MLP}\left[\text{Attention}\left((\boldsymbol{a}^{1,1}, \boldsymbol{q}^{1,1}), (\boldsymbol{a}^{1,2}, \boldsymbol{q}^{1,2}), \ldots, \left(\boldsymbol{a}^{1,K^1-1}, \boldsymbol{q}^{1,K^1-1}\right)\right)\right] \quad (\text{A}.3)$$

To enhance the nonlinearity of the inference network, we introduce an MLP following the attention mechanism. The overall inference network, comprising both the attention module and the MLP, is denoted as $\text{MLP}_1^{\text{Inf}}$ for simplicity. Accordingly, the variational distribution $q(\boldsymbol{s}^1)$ is defined as:

$$q(\boldsymbol{s}^1) = \mathcal{N}\left(\mu_q(\boldsymbol{s}^1), \Sigma_2\right) \quad (\text{A}.4)$$

where $\Sigma_2$ is a hyperparameter. A similar process is applied for $\boldsymbol{s}^t$ with $t \geq 2$. However, for these later statuses, $\boldsymbol{s}^t$ is also directly influenced by the previous status $\boldsymbol{s}^{t-1}$, whose variational distribution has already been inferred. To incorporate this information, we input the expectation of the previous distribution, $\mu_q(\boldsymbol{s}^{t-1})$, into the variational distribution for $\boldsymbol{s}^t$. The same attention mechanism is applied to capture the sequential dependencies:

$$\mu_q(\boldsymbol{s}^t) = \text{MLP}\left[\text{Attention}\left((\boldsymbol{a}^{1,1}, \boldsymbol{q}^{1,1}), (\boldsymbol{a}^{1,2}, \boldsymbol{q}^{1,2}), \ldots, \left(\boldsymbol{a}^{1,K^t-1}, \boldsymbol{q}^{1,K^t-1}\right)\right), \mu_q(\boldsymbol{s}^{t-1})\right] \quad (\text{A}.5)$$

$$q(\boldsymbol{s}^t) = \mathcal{N}\left(\mu_q(\boldsymbol{s}^t), \Sigma_2\right) \quad (\text{A}.6)$$

The network described above, which includes both the attention mechanism and the MLP, is denoted as $\text{MLP}_2^{\text{Inf}}$ for simplicity. Since the neural networks $\text{MLP}_1^{\text{Inf}}$ and $\text{MLP}_2^{\text{Inf}}$ are used for inference, they are called inference networks. The generation of $r^t$ is a deterministic process conditioned on $\boldsymbol{s}^t$ and $\boldsymbol{e}$. Therefore, once the distribution $q(\boldsymbol{s}^t)$ is obtained, the variational distribution $q(r^t)$ can be derived straightforwardly using $\text{MLP}_1$. Specifically, we first sample a latent variable $\tilde{\boldsymbol{s}}^t$ from $q(\boldsymbol{s}^t)$ and then input $\tilde{\boldsymbol{s}}^t$ into $\text{MLP}_1$ to obtain $\tilde{r}^t$, which is an instance of $q(r^t)$. Meanwhile, during training, we require $\tilde{r}^t$ to be close to the observed outcome $y$, thereby imposing a constraint on $s^t$ using $y$. Specifically, this constraint is stronger when the time gap between $\tilde{r}^t$ and $y$ is smaller. Let $t^t$ denote the time gap from $\tilde{r}^t$ to $y$. Then, the constraint can be formulated as:

$$C = \exp(-t^t) \cdot \text{CrossEntropy}(\tilde{r}^t, y) \quad (A.7)$$

This constraint is used to guide the training of $q(\boldsymbol{s}^t)$. In this way, $q(\boldsymbol{s}^t)$ is influenced by $y$ during training, while still remaining inferable during the testing phase without requiring $y$ as input.

**3 Deriving the Learning Objective**

Our goal is to maximize the model's fit to the observed variables, i.e., to maximize the likelihood of generating the observed data. The log-likelihood can be expressed as follows (details are provided in the appendix):

$$\log p(\boldsymbol{A}^{1:T}, \boldsymbol{Q}^{1:T}, \boldsymbol{e}, y) = \mathbb{E}_q[\log p(\boldsymbol{s}^{1:T}, \boldsymbol{A}^{1:T}, \boldsymbol{Q}^{1:T}, \boldsymbol{e}, y) - \log q(\boldsymbol{s}^{1:T})] + KL(q||p^{\text{post}}) (A.8)$$

where $KL$ denotes the Kullback–Leibler divergence. The equation above can be further derived as:

$$\log p(\boldsymbol{A}^{1:T}, \boldsymbol{Q}^{1:T}, \boldsymbol{e}, y) - KL = \mathbb{E}_q[\log p(\boldsymbol{s}^{1:T}, \boldsymbol{A}^{1:T}, \boldsymbol{Q}^{1:T}, \boldsymbol{e}, y) - \log q(\boldsymbol{s}^{1:T})] \quad (A.9)$$

Since the KL divergence measures the difference between the variational distribution and the true posterior, it should be minimized to ensure a high-quality approximation. For the likelihood of the observed variables, $\log p(\boldsymbol{A}^{1:T}, \boldsymbol{Q}^{1:T}, \boldsymbol{e}, y)$, the value is constant with respect to the model parameters and the variational distribution, as the observed variables are given. Therefore, minimizing the KL divergence is equivalent to maximizing the right-hand side of the equation. Moreover, since the KL divergence is non-negative, the right-hand side provides a lower bound on the log-likelihood of the observed variables. We denote this lower bound as $\mathcal{L}^q$, which can be further transformed as follows:

$$\begin{aligned}
\mathcal{L}^q &= \mathbb{E}_q[\log p(\boldsymbol{s}^{1:T}, \boldsymbol{A}^{1:T}, \boldsymbol{Q}^{1:T}, \boldsymbol{e}, y) - \log q(\boldsymbol{s}^{1:T})] \\
&= E_q[logp(\boldsymbol{A}^{1:T}, \boldsymbol{Q}^{1:T}, \boldsymbol{e}, y|\boldsymbol{s}^{1:T})p(\boldsymbol{s}^{1:T}) - logq(\boldsymbol{s}^{1:T})] \\
&= E_q[logp(\boldsymbol{A}^{1:T}, \boldsymbol{Q}^{1:T}, \boldsymbol{e}, y|\boldsymbol{s}^{1:T}) + logp(\boldsymbol{s}^{1:T}) - logq(\boldsymbol{s}^{1:T})] \\
&= E_q[logp(\boldsymbol{A}^{1:T}, \boldsymbol{Q}^{1:T}, \boldsymbol{e}, y|\boldsymbol{s}^{1:T})] + E_q[logp(\boldsymbol{s}^{1:T}) - logq(\boldsymbol{s}^{1:T})] \\
&= E_q[logp(\boldsymbol{A}^{1:T}, \boldsymbol{Q}^{1:T}, \boldsymbol{e}, y|\boldsymbol{s}^{1:T})] - KL(q(\boldsymbol{s}^{1:T})||logp(\boldsymbol{s}^{1:T})) \quad (A.10)
\end{aligned}$$

The lower bound $\mathcal{L}^q$ is our learning objective. It depends on the parameters of $\text{MLP}_1$ to $\text{MLP}_6$ as well as the introduced inference networks $\text{MLP}_1^{\text{Inf}}$ and $\text{MLP}_2^{\text{Inf}}$. Hence, their parameters are updated to maximize $\mathcal{L}^q$.

**4 Computation of the Learning Objective**

For the term $\log p(\boldsymbol{A}^{1:T}, \boldsymbol{Q}^{1:T}, \boldsymbol{e}, y|\boldsymbol{s}^{1:T})$ on the right-hand side of $\mathcal{L}^q$, we can decompose it according to the dependency structure of the Bayesian network as follows:

$$\begin{aligned}
&\log p(\boldsymbol{A}^{1:T}, \boldsymbol{Q}^{1:T}, \boldsymbol{e}, y|\boldsymbol{s}^{1:T}) \\
&\quad = \sum_{t=1}^{T} [\log p(\boldsymbol{q}^{t,1}|\boldsymbol{s}^t)] \\
&\quad + \sum_{t=1}^{T} \sum_{k=2}^{K^t-1} \left[\log p\left(\boldsymbol{q}^{t,k}|\boldsymbol{a}^{t,k-1}, \boldsymbol{q}^{t,k-1}, \boldsymbol{s}^t\right) + \log p\left(\boldsymbol{a}^{t,k}|\boldsymbol{a}^{t,k-1}, \boldsymbol{q}^{t,k-1}, \boldsymbol{q}^{t,k}, \boldsymbol{s}^t\right)\right] \\
&\quad + \log p(\boldsymbol{e}) + \log p(y|\boldsymbol{s}^{1:T}, \boldsymbol{e}) \quad (A.11)
\end{aligned}$$

For the first two terms on the right-hand side in Equation (A.11), their values can be computed using the generation networks (i.e., $\mathrm{MLP}_2$ to $\mathrm{MLP}_6$). The third term, $\log p(\boldsymbol{e})$ is constant with respect to our objective $\mathcal{L}^q$ and can therefore be ignored. For the fourth term, $p(y|\boldsymbol{s}^{1:T}, \boldsymbol{e})$, we use the same neural network as the generation network (i.e., $\mathrm{MLP}_1$) to obtain $r^T$ and then compute the probability of $y$ based on Bernoulli distribution. In this way, Equation (A.11) can be computed. To estimate its expectation in Equation (A.10), we adopt the Monte Carlo method: we draw samples of the latent variables $\boldsymbol{s}^1, \ldots, \boldsymbol{s}^T$ from the variational distribution $q$, and then compute the value of each sample to approximate the expectation.

Next, we focus on the second term on the right-hand side of Equation (A.10), i.e., $\mathrm{KL}(q(\boldsymbol{s}^{1:T})||p(\boldsymbol{s}^{1:T})^{\text{prior}})$. Based on the factorizations in Equations (A.1) and (A.2), it can be expressed as follows:

$$\mathrm{KL}\left(q(\boldsymbol{s}^{1:T}) \| p(\boldsymbol{s}^{1:T})^{\text{prior}}\right) = \sum_{t=1}^{T} \mathrm{KL}\left(q(\boldsymbol{s}^t) \| p^{\text{prior}}(\boldsymbol{s}^t)\right) \tag{A.12}$$

For each prior distribution $p^{\text{prior}}(\boldsymbol{s}^t)$, they are Gaussian distributions as mentioned before (i,e., $\boldsymbol{s}^t = \mathcal{N}(\boldsymbol{s}^{t-1}, \boldsymbol{I})$ if $t \geq 2$; $\boldsymbol{s}^t = \mathcal{N}(0, \boldsymbol{I})$ if $t = 1$). Since the variational distributions $q(\boldsymbol{s}^t)$ are also Gaussian, the KL divergence between each prior and variational distribution can be computed in closed form using the standard formula for the KL divergence between two Gaussians. For instance, when $t \geq 2$, $\mathrm{KL}\left(q(\boldsymbol{s}^t) \| p^{\text{prior}}(\boldsymbol{s}^t)\right)$ can be computed as:

$$\mathrm{KL}\left(q(\boldsymbol{s}^t) \| p^{\text{prior}}(\boldsymbol{s}^t)\right) = \frac{1}{2}\left[\mathrm{tr}(\Sigma_2) + \left(\mu_q(\boldsymbol{s}^t) - \boldsymbol{s}^{t-1}\right)^{\mathrm{T}}\left(\mu_q(\boldsymbol{s}^t) - \boldsymbol{s}^{t-1}\right) - d + \log\frac{|\boldsymbol{I}|}{|\Sigma_2|}\right] \tag{A.13}$$

where $d$ denotes the dimension of $\boldsymbol{s}^t$.

Since both components of the second term on the right-hand side of Equation (A.10) is computable, we can compute the value of $\mathcal{L}^q$ to update parameters.

**5 Iterative Training Process**

The above computation of $\mathcal{L}^q$ relies on instantiated samples of $\boldsymbol{s}^t$ and $r^t$ for all $t$. The quality of the variational distributions determines how well these samples approximate the true posterior, thereby affecting the KL divergence term. Meanwhile, the model parameters influence the first term of $\mathcal{L}^q$. Consequently, our optimization proceeds in two steps.

First, we fix the generative model parameters (i.e., the parameters of $\mathrm{MLP}_1$ to $\mathrm{MLP}_6$ , denoted as $\boldsymbol{\Phi}$ collectively) and update only the parameters of the inference networks $\mathrm{MLP}_1^{\mathrm{Inf}}$ and $\mathrm{MLP}_2^{\mathrm{Inf}}$ (denoted as $\boldsymbol{\Theta}$ collectively). Since a constraint was introduced in Equation (A.7) to train the inference network, it is incorporated into $\mathcal{L}^q$ via a weighted sum. We optimize this objective using the Adam optimizer. Since the Adam optimizer is a gradient-based algorithm that updates the model parameters by minimizing the given objective, we take the negative of $\mathcal{L}^q$ and combine it with the constraint term for optimization. We denote the parameters at the $m$-th round as $\boldsymbol{\Phi}^m$ and $\boldsymbol{\Theta}^m$. Then,

$$\boldsymbol{\Phi}^{m+1} \leftarrow \text{Adam}(-\mathcal{L}^q + wC, \boldsymbol{\Phi}^m, \boldsymbol{\Theta}^m)$$

where $w$ is the relative weight of the constraint term. In this way, we reduce the gap between the variational distribution and the true posterior, yielding high-quality samples of the latent variables $\boldsymbol{s}^t, r^t$ for all $t$, which allows us to compute $\mathcal{L}^q$.

Second, we freeze the inference networks (i.e., $\boldsymbol{\Theta}^{m+1}$) and update the generative model parameters to maximize $\mathcal{L}^q$, thereby improving the fit of the Bayesian network to the data. Formally,

$$\boldsymbol{\Theta}^{m+1} \leftarrow \text{Adam}(-\mathcal{L}^q, \boldsymbol{\Phi}^{m+1}, \boldsymbol{\Theta}^m)$$

The above two-step training procedure corresponds to the Expectation-Maximization (EM) algorithm, a classic algorithm for training Bayesian networks. We alternate between the two steps until convergence.

**6 Predicting A New Sample**

After the parameters have been learned, our LLM-Bayesian network method can be used to predict a company's success rate. At the first expert network call, we first apply the LLM to analyze its conservation and extract the dense vector to represent the textual content. We then apply the trained $\text{MLP}_1^{\text{Inf}}$ to analyze the extracted vector to infer the latent status distribution $\boldsymbol{s}_{\text{test}}^1 \sim \mathcal{N}\left(\mu_q(\boldsymbol{s}_{\text{test}}^1), \Sigma_2\right)$. For stability, instead of sampling from the Gaussian, we use the mean $\mathcal{N}\left(\mu_q(\boldsymbol{s}_{\text{test}}^1), \Sigma_2\right)$ to compute the success rate via $\text{MLP}_1$. The predicted success is determined by comparing $r$ with a threshold. For subsequent expert network calls (i.e., $t \geq 2$), given the inferred status $\mu_q(\boldsymbol{s}_{\text{test}}^{t-1})$, we use $\text{MLP}_2^{\text{Inf}}$ to infer the current status $\mu_q(\boldsymbol{s}_{\text{test}}^t)$. All inferred statuses up to $t$ are then input to $\text{MLP}_2$ to compute $r$ which is compared with the threshold to predict the company's success. In this way, we can not only infer the company's underlying success rate but also quantify the uncertainty and sequentially update the belief about the company's success.

| **Generative Process of the LLM-Bayesian Framework** |
|---|
| **for** $t = 1,2, \ldots, T$**:** |
|     **if** $t = 1$: |
|         generate $\boldsymbol{s}^1 \sim \mathcal{N}(\boldsymbol{0}, \boldsymbol{I})$ |
|         generate $r^1 = \text{MLP}_1(\boldsymbol{s}^1, \boldsymbol{e})$ |
|     **else**: |
|         generate $\boldsymbol{s}^t \sim \mathcal{N}(\boldsymbol{s}^{t-1}, \boldsymbol{I})$ |
|         generate $r^t = \text{MLP}_2(\boldsymbol{s}^{t-1}, \ldots, \boldsymbol{s}^1, \boldsymbol{e})$ |
|     **end if** |
|     generate $y = \text{Bernolli}(r^T)$ |

**for** $k = 1, \dots, K^t$:

  **if** $k = 1$:

    $\mu(\boldsymbol{q}^{t,k}) = \mathrm{MLP}_6(\boldsymbol{s}^t)$

    $\boldsymbol{q}^{t,k} \sim \mathcal{N}(\mu(\boldsymbol{q}^{t,k}), \Sigma)$

    $\mu(\boldsymbol{a}^{t,k}) = \mathrm{MLP}_4(\boldsymbol{s}^t, \boldsymbol{q}^{t,k})$

    $\boldsymbol{a}^{t,k} \sim \mathcal{N}(\mu(\boldsymbol{a}^{t,k}), \Sigma)$

  **else**:

    $\mu(\boldsymbol{q}^{t,k}) = \mathrm{MLP}_5(\boldsymbol{s}^t, \boldsymbol{q}^{t,k-1}, \boldsymbol{a}^{t,k-1})$

    $\boldsymbol{q}^{t,k} \sim \mathcal{N}(\mu(\boldsymbol{q}^{t,k}), \Sigma)$

    $\mu(\boldsymbol{a}^{t,k}) = \mathrm{MLP}_3(\boldsymbol{s}^t, \boldsymbol{q}^{l,k-1}, \boldsymbol{a}^{t,k-1}, \boldsymbol{q}^{t,k})$

    $\boldsymbol{a}^{l,k} \sim \mathcal{N}(\mu(\boldsymbol{a}^{l,k}), \Sigma)$

  **end if**

## Appendix A. III: The Detailed Prompts with Startup Success Ontology (SSO)

Assume you are an investment analysis expert. Please summarize the given private company's expert network call content from the following perspectives in no more than 200 words: Startup Characteristics (Product, Financial Health, Founding Team, Leadership, Intellectual Property);Market and Industry Dynamics (Market Size, Competitive Landscape, Economic and Regulatory Environment);Investor Characteristics (VC Expertise, Investment Strategy, Governance Role);Buyer Characteristics (Strategic Fit, Buyer's reputation for successful cultural and operational integration);Relationships Between Factors. If there is no description associated with the above feature, output the None value.

## Appendix A. IV Figures

**Figure A.1: Topic Composition Across Industries**

This figure shows the distribution of discussion topics across industries. T1–T10 correspond to the ten most salient discussion topics identified in Table 2, representing the dominant thematic categories extracted from expert network call transcripts. Each horizontal bar represents one industry, with color segments indicating the proportion of transcripts devoted to each topic within that industry. The figure highlights industry-specific emphases-for instance, software and IT services exhibit stronger focus on technology-related topics, while investment and financial services concentrate on market and pricing discussions. The proportions are normalized within industries to sum to 100%, allowing comparison of related topic salience across sectors.

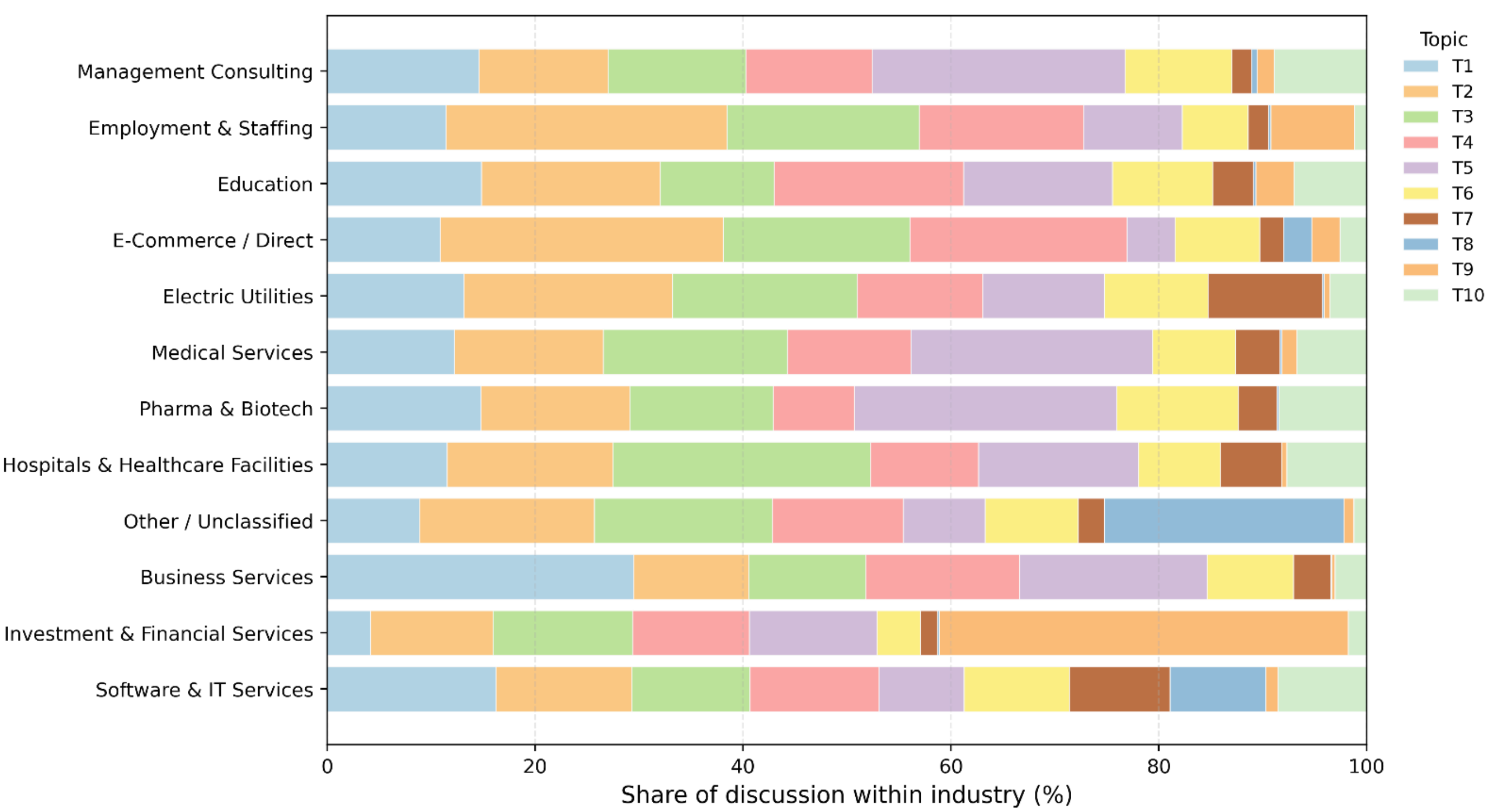

**Figure A.2: Sentiment by Topic**

This figure plots the mean sentiment scores across discussion topics identified from expert–client consultation calls. Each point represents the average sentiment for a given topic, and horizontal lines indicate the variation across topics. Topics are ordered by their mean sentiment score, which ranges from −2 (strongly negative) to +2 (strongly positive). Higher values reflect more optimistic or favorable discussions, while lower values indicate concerns or negative outlooks.

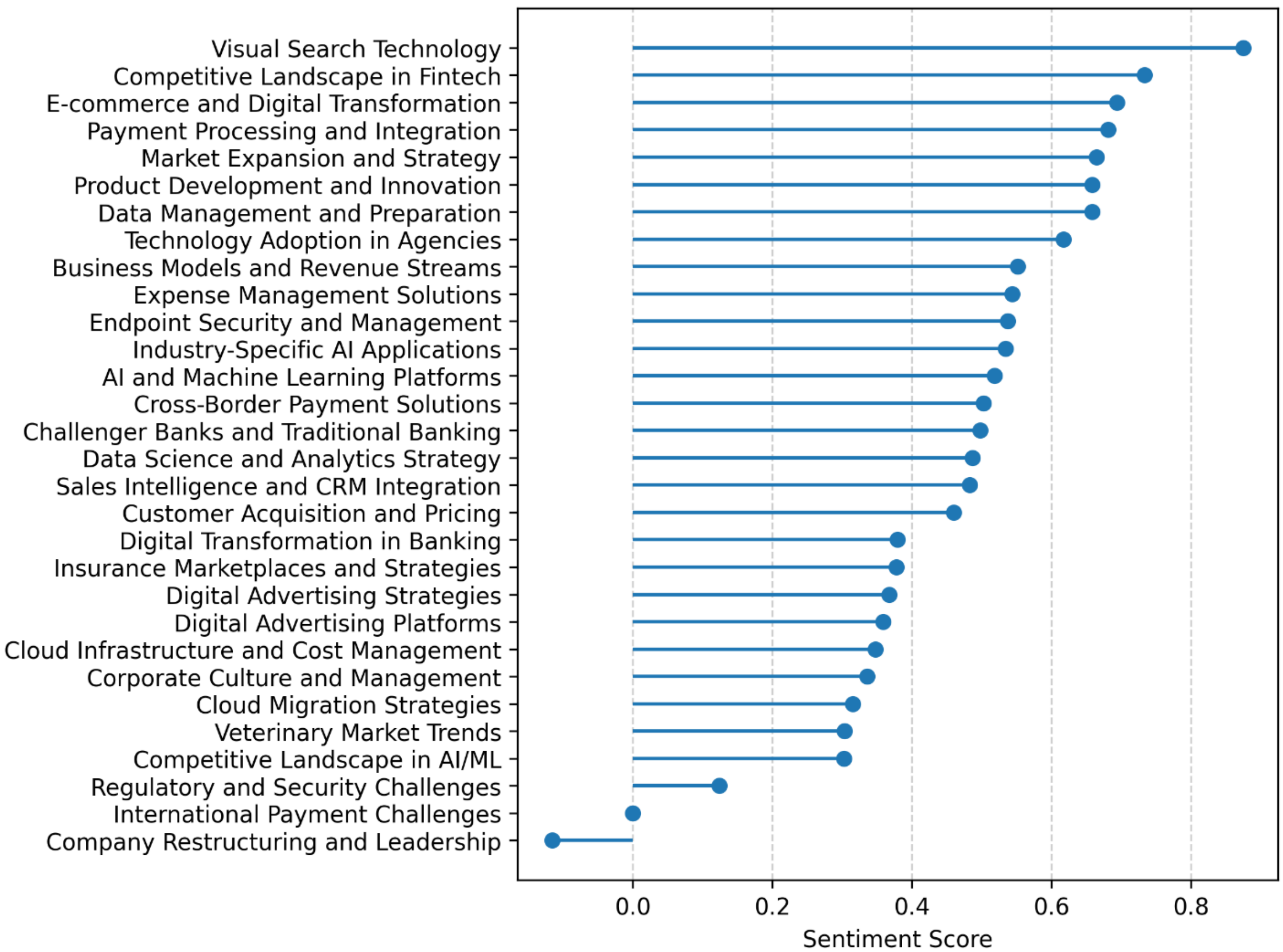

**Figure A.3: Expert Network Call Information Improves Prediction of Investment Outcomes**

This figure reports the receiver operating characteristic (ROC) curves for models predicting investment outcomes using firm characteristics alone and in combination with expert network call information. The horizontal axis shows the false positive rate, and the vertical axis shows the true positive rate. The baseline model ("fv") uses firm-level characteristics, while the augmented model ("fv + text") additionally incorporates information extracted from expert network call transcripts. The curve for the augmented model lies uniformly above that of the baseline model, indicating improved classification performance. The increase in area under the curve (AUC) demonstrates that expert network call information provides incremental predictive content beyond firm fundamentals. This result suggests that expert network calls contain economically meaningful signals that are not captured by traditional firm characteristics.

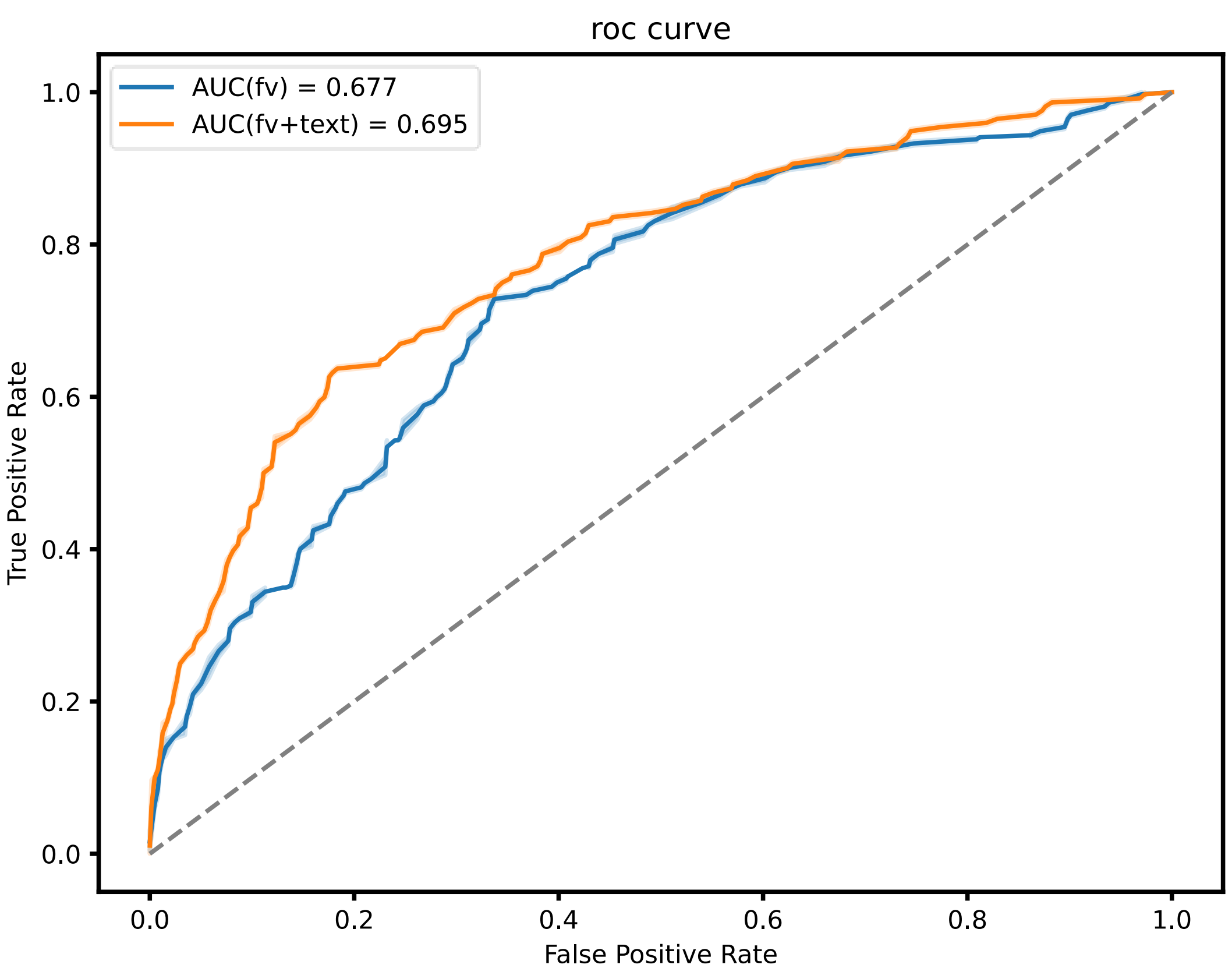

**Figure A.4: Robustness of Investment Performance to Payoff and Cost Assumptions**

This figure examines the robustness of the model's economic performance to alternative payoff and cost assumptions. Each panel presents a heatmap of differences in portfolio outcomes between our model and benchmark approaches across a grid of parameter values. The horizontal axis varies the investment cost (IC), while the vertical axis varies the payoff from successful investments ($FIV_{TP}$). Panels A and B report differences relative to the baseline model ("HAT") in return on investment (ROI) and multiple of invested capital (MOIC), respectively. Panels C and D report the corresponding differences relative to the SVM benchmark. Color intensity reflects the magnitude of performance differences, with warmer colors indicating regions where our model outperforms the benchmark.

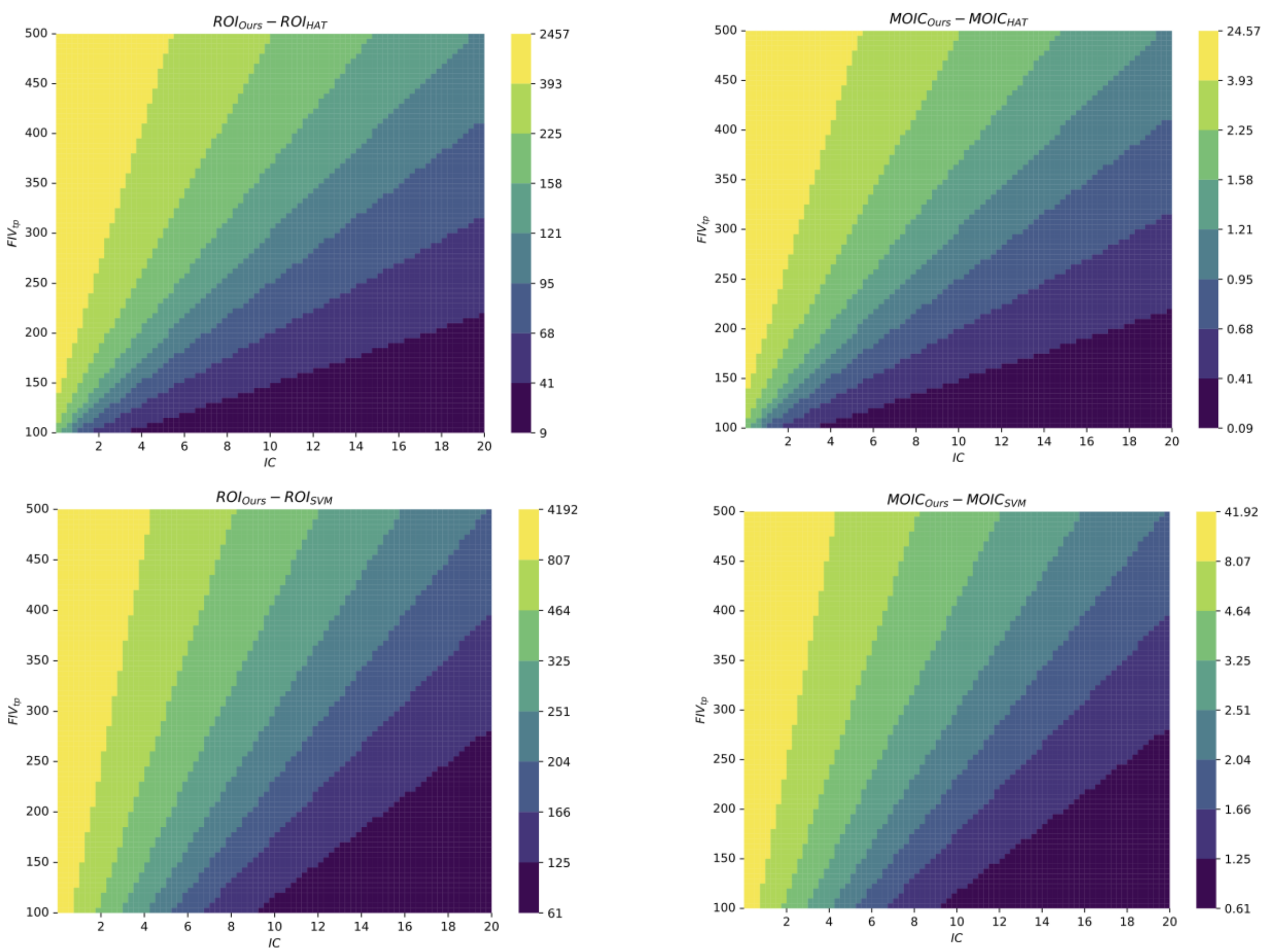

## Appendix A. V Tables

**Table A.1: Notation**

This table summarizes the notation used throughout the paper. Symbols describe key entities, model components, and data representations in the LLM-Bayesian framework.

| Notation | Description |
|---|---|
| $c^t$ | The $t$-th expert consultation call of company. |
| $\boldsymbol{q}^{t,k}, \boldsymbol{a}^{t,k}$ | Representations of the summarized $k$-th question–answer pair in call $c^t$. |
| $K^t$ | The total number of question–answer pairs in call $c^t$. |
| $\boldsymbol{e}$ | External structured information (e.g., company profile, or funding information). |
| $f_{\boldsymbol{\theta}}(\cdot)$ | The prediction model parameterized by $\boldsymbol{\theta}$. |
| y | The observed binary success outcome. |
| $r^t$ | Investors' belief at the time of its $t$-th call. |
| $r^{1:T}$ | Investor's belief over time $T$. |
| $\tilde{r}^t$ | Sampled instance of variational distribution for $r^t$. |
| $\boldsymbol{s}^t$ | Firm's state at time $t$. |
| $\boldsymbol{s}^{1:T}$ | Firm's state over time $T$. |
| $\tilde{\boldsymbol{s}}^t$ | Sampled instance of variational distribution for $\boldsymbol{s}^t$. |
| $\boldsymbol{A}^t, \boldsymbol{Q}^t$ | Extracted vectors of questions and answers in the $t$-th expert consultation call respectively. |
| $\boldsymbol{A}^{1:T}, \boldsymbol{Q}^{1:T}$ | Extracted vectors of questions and answers of all expert consultation call respectively. |
| $\mathcal{L}^q$ | Lower bound on the log-likelihood of the observed variables. |
| $\boldsymbol{\Phi}^m, \boldsymbol{\Theta}^m$ | Model parameters at the $m$-th round learning process. |

**Table A.2: Variable Descriptions.**

| Variable | Description |
|---|---|
| | OUTCOME VARIABLE |
| Success | True (= 1) if a startup had an initial public offering, received funding, or has been acquired. False (=0) otherwise |
| | PREDICTORS |
| Fundamental variable (FV) | |
| Age | Time since the startup has been founded (in months) |
| Founders count | Number of founders of the startup |
| Number of investment rounds | Number of investment rounds |
| Raised funding | Total raised funding (in million USD) |
| Investor count | Overall number of investors that invested in the startup |
| Active products count | Number of active products |
| IT spending | The approximate amount a company spends on IT |
| Trademark class | The class assigned to the trademark |
| Number of Calls in last 24 month | Number of Calls in last 24 months (24 dimension vector) |
| Head quarters location | Head quarters location (one-hot vector) |
| Expert network call transcript | |
| Expert network call embedding | Semantic representation of the expert–client conversation encoded via a large language model (MPNet) |
| Expert types | |
| Competitor | Competitors provide insights grounded in their direct rivalry with the focal firm. |
| Customer | Customers represent the demand side of the focal firm. |
| Former exec | Former executives and employees provide a rare insider's view of the focal firm. |
| Industry cons | Industry consultants offer a broader vantage point of focal firm. |
| Partner | Partners, including distributors, suppliers, and strategic collaborators, illuminate the ecosystem dynamics that underpin a firm's business model. |
| Model-Implied Belief and Uncertainty | |
| Investor Success Belief ($p_{it}$) | The investor's posterior belief about firm $i$'s likelihood of success after expert network call $t$. This measure is generated by the LLM-Bayesian framework and reflects the updated probability that the firm will achieve a successful outcome, conditional on all information available up to call $t$. |
| Uncertainty ($U_{it}$) | The investor's posterior uncertainty about firm $i$'s success probability after expert network call $t$. This measure captures the dispersion or entropy of the belief distribution inferred by the model, with higher values indicating greater uncertainty about firm prospects and lower values indicating more precise beliefs. |
| ΔUncertainty(%) | The percentage reduction in investor uncertainty following expert network call $t$, defined as $\Delta\text{Uncertainty}_{it} = 100 \times \left(U_{i,t-1} - U_{it}\right)$. Positive values indicate that uncertainty decreases after the call, reflecting learning and belief updating. |
| Other Variable | |
| Opacity | A standardized measure of the firm's information environment complexity, constructed as the average of standardized firm youth and product scope, and rescaled to have mean zero and unit variance. |
| Operation Expert | An indicator variable equal to one if the expert has an operational role (e.g., customer, industry consultant, or former executive), and zero otherwise. |
| Technology Adoption Topic | An indicator for calls that focus on technology adoption and implementation issues (Topic 1, 2, 3, and 4) |

**Table A.3: Sentiment Scale**

This table presents the five-point sentiment scale used in the assignment prompt during our LLM-based topic modeling methodology. The scale ranges from −2 (Strongly Negative) to +2 (Strongly Positive), reflecting the tone and confidence expressed in expert–client discussions.

| Score | Description |
|---|---|
| +2 | Strongly Positive — clear enthusiasm, strong benefits, high confidence |
| +1 | Positive — generally favorable, more pros than cons |
| 0 | Neutral — balanced, factual tone |
| −1 | Negative — noticeable concerns or mild skepticism |
| −2 | Strongly Negative — severe criticism or major problems |

**Table A.4: Sensitivity Analysis**

This table evaluates the robustness of the LLM-Bayesian framework's predictive and economic performance to alternative hyperparameter settings. Panels vary the *dropout* probability (Panel A), the learning rate (Panel B), the dimension of latent status (Panel C), and the weight of the constraint term (Panel D). Performance is evaluated using predictive metrics, *Accuracy*, *Precision*, *F1-score*, *Weighted F1*, *Macro F1*, and *AUC*. Economic metrics include *ROI* (*Return on Investment*) and *MOIC* (*Multiple on Invested Capital*), reflecting realized investor performance under model-guided portfolio allocation. Bolded values indicate optimal settings.

| | | Accuracy | Precision | F1-score | Weighted F1 | Macro F1 | AUC | ROI | MOIC |
|---|---|---|---|---|---|---|---|---|---|
| Panel A: Dropout rate | | | | | | | | | |
| | 0.1 | 0.773 | 0.498 | 0.586 | 0.786 | 0.715 | 0.751 | 713.205 | 8.132 |
| | **0.15** | 0.777 | 0.505 | **0.590** | **0.789** | **0.718** | **0.753** | **723.350** | **8.230** |
| | 0.2 | 0.771 | 0.495 | 0.581 | 0.783 | 0.712 | 0.747 | 696.989 | 7.970 |
| | 0.3 | 0.773 | 0.497 | 0.574 | 0.784 | 0.709 | 0.739 | 646.668 | 7.467 |
| | 0.35 | **0.780** | **0.510** | 0.569 | 0.788 | 0.711 | 0.732 | 592.047 | 6.921 |
| Panel B: Learning rate | | | | | | | | | |
| | 1.00E-04 | 0.771 | 0.494 | 0.573 | 0.782 | 0.708 | 0.739 | 650.493 | 7.505 |
| | 5.00E-05 | 0.776 | 0.503 | 0.578 | 0.787 | 0.713 | 0.742 | 657.987 | 7.580 |
| $lr$ | **1.00E-05** | **0.777** | **0.505** | **0.590** | **0.789** | **0.718** | **0.753** | **723.350** | **8.230** |
| | 5.00E-06 | 0.777 | 0.504 | 0.582 | 0.788 | 0.715 | 0.746 | 680.894 | 7.809 |
| | 1.00E-06 | 0.775 | 0.503 | 0.580 | 0.787 | 0.714 | 0.744 | 672.922 | 7.729 |
| Panel C: Dimension of latent status $\boldsymbol{s}^l$ | | | | | | | | | |
| | 128 | 0.769 | 0.492 | 0.571 | 0.781 | 0.707 | 0.738 | 642.907 | 7.429 |
| | 256 | 0.772 | 0.497 | 0.574 | 0.783 | 0.709 | 0.740 | 652.348 | 7.523 |
| $dl$ | **512** | 0.777 | 0.505 | **0.590** | **0.789** | **0.718** | **0.753** | **723.350** | **8.230** |
| | 768 | **0.779** | **0.508** | 0.583 | **0.789** | 0.716 | 0.745 | 675.129 | 7.751 |
| | 1024 | 0.777 | 0.505 | 0.583 | 0.788 | 0.715 | 0.746 | 682.630 | 7.826 |
| Panel D: the weight of the constraint term $w$ | | | | | | | | | |
| | 1.00E-03 | 0.771 | 0.495 | 0.580 | 0.783 | 0.711 | 0.746 | 687.002 | 7.870 |
| | 5.00E-03 | 0.774 | 0.500 | 0.583 | 0.786 | 0.714 | 0.747 | 694.375 | 7.944 |
| $w$ | **1.00E-04** | **0.777** | **0.505** | **0.590** | **0.789** | **0.718** | **0.753** | **723.350** | **8.230** |
| | 5.00E-05 | 0.775 | 0.502 | 0.586 | 0.787 | 0.716 | 0.750 | 708.523 | 8.085 |
| | 1.00E-05 | 0.771 | 0.495 | 0.581 | 0.783 | 0.712 | 0.747 | 697.223 | 7.972 |

**Table A.5: Summary Statistics of Model-Implied Beliefs and Uncertainty**

This table presents descriptive statistics for Model-Implied Beliefs and Uncertainty, including *ΔUncertainty(%)*, *Uncertainty*, and *Investor Success Belief*. All variable definitions are reported in Table A.2.

| Variable | N | Mean | SD | p25 | p50 | p75 |
|---|---|---|---|---|---|---|
| ΔUncertainty(%) | 4255 | 0 | 0.303 | -0.001 | 0 | 0.001 |
| Uncertainty | 6130 | 0.688 | 0.008 | 0.685 | 0.691 | 0.693 |
| Investor Success Belief | 6130 | 0.492 | 0.052 | 0.455 | 0.494 | 0.523 |

**Table A.6: Expert Network Calls and Long-Run Performance**

The dependent variable is the success measure based on outcomes over the eight quarters following our deal analysis period. This measure captures the full spectrum of potential outcomes: bankruptcy (-1), no material change (0), stage progression (1), and the combination of stage progression with successful exit (2). *Call* indicates if an expert consultation occurred at t. *Positive Sentiment* captures whether the call had positive sentiment. *Technology Adoption Topic* is an indicator for calls that focus on technology adoption and implementation issues (Topic 1, 2, 3, and 4) and *Technology Adoption Topic Positive* captures whether *Technology Adoption Topic* had positive sentiment. *Regulatory and Security Challenges Topic Negative* captures whether *Regulatory and Security Challenges Topic* had negative sentiment. Column (2) reports the results for the subsample of firms that receive at least one call within our sample. Standard errors are clustered by firm. *, **, *** denote significance at 10%, 5%, 1% levels respectively. All variable definitions are reported in Table A.2.

| | (1)<br>Future Success | (2)<br>Future Success |
|---|---|---|
| Call | -0.051*** | -0.004 |
| | (-11.34) | (-0.99) |
| Overall Sentiment | -0.011*** | -0.010*** |
| | (-2.72) | (-2.64) |
| Technology Adoption Topic Positive | 0.005 | 0.004 |
| | (1.14) | (1.02) |
| Regulatory and Security Challenges Topic Negative | -0.014** | -0.014*** |
| | (-2.24) | (-2.72) |
| Observations | 1,836,626 | 24,706 |
| Adjusted R-squared | 0.856 | 0.951 |
| Firm FE | Y | Y |
| Quarter FE | Y | Y |